\documentclass[twocolumn,showpacs,preprintnumbers,amsmath,amssymb]{revtex4}
\usepackage{epsfig}
\usepackage{dcolumn}
\usepackage{bm}

\usepackage{cancel}
\usepackage{bigstrut}

\newcommand{\gp}{{g^\prime}}
\usepackage{subfigure}

\newcommand{\pmi}{\phi_{\rm min}}

\newcommand{\pdp}{ {\Phi^\dagger\Phi}}

\newcommand{\remove}[1]{}

\def\be{\begin{equation}}
\def\ee{\end{equation}}

\newcommand{\beq}{\begin{equation}}
\newcommand{\eeq}{\end{equation}}
\newcommand{\beqa}{\begin{eqnarray}}
\newcommand{\eeqa}{\end{eqnarray}}

\renewcommand{\pl}{\partial}

\newcommand{\vv}{{\bf v}}

\newcommand{\vx}{{\bf x}}

\renewcommand{\vr}{{\bf r}}

\newcommand{\tchi}{{\tilde{\chi}}}

\newcommand{\tg}{{\tilde{g}}}

\newcommand{\tPhi}{{\tilde{\Phi}}}
\newcommand{\tPsi}{{\tilde{\Psi}}}

\newcommand{\cM}{{\cal M}}

\newcommand{\bea}{\begin{array}}
\newcommand{\ea}{\end{array}}

\begin{document}

\title{Supersymmetric chameleons and ultra-local models }

\author{Philippe Brax}
\affiliation{Institut de Physique Th\'eorique,\\
Universit\'e Paris-Saclay CEA, CNRS, F-91191 Gif-sur-Yvette, C\'edex, France\\}
\author{Luca Alberto Rizzo}
\affiliation{Institut de Physique Th\'eorique,\\
Universit\'e Paris-Saclay CEA, CNRS, F-91191 Gif-sur-Yvette, C\'edex, France\\}

\author{Patrick Valageas}
\affiliation{Institut de Physique Th\'eorique,\\
Universit\'e Paris-Saclay CEA, CNRS, F-91191 Gif-sur-Yvette, C\'edex, France\\}

\vspace{.2 cm}

\date{\today}
\vspace{.2 cm}

\begin{abstract}
Super-chameleon models where all types of matter belong to three secluded sectors, i.e.
the dark, supersymmetry breaking and matter sectors, are shown to be dynamically
equivalent to ultra-local models of modified gravity. In the dark sector, comprising both
dark matter and dark energy, the interaction range between the dark energy field and
dark matter is constrained to be extremely short, i.e. shorter than the inverse gravitino
mass set by supersymmetry breaking. This realises an extreme version of chameleon
screening of the dark energy interaction.
On the other hand, the baryonic matter sector decouples from the dark energy in
a Damour-Polyakov way. These two mechanisms preclude the existence of any modification
of gravity locally in the Solar System due to the presence of the super-chameleon field.
On larger scales, the super-chameleon can have effects on the growth of structure and
the number of dark matter halos. It can also affect the dynamics of galaxies where the
fifth force interaction that it induces can have the same order of magnitude as
Newton's interaction.

\end{abstract}

 \vskip2pc

\maketitle

\section{Introduction}
\label{sec:Introduction}

Dark energy \cite{Riess:1998cb,Perlmutter:1998hx,Copeland:2006wr}  is still as mysterious now as it was when the first observations of its existence
appeared more than fifteen years ago. Moreover it has been realized over the last ten years
that very often dark energy and a modification of gravity on large scales are intimately
connected \cite{Joyce:2014kja}. This is the case for models as diverse as $f(R)$ theories \cite{Hu:2007nk} or Galileons \cite{Nicolis:2008in}.
These models utilise a scalar field as the simplest way of going beyond a mere cosmological
constant. Such theories where the dynamical equations of motion are of second order have
been classified \cite{Horndeski:1974wa}. Their dynamics depend on the coupling of the scalar degree of freedom to
matter. In the most general case \cite{Bekenstein:1992pj}, this coupling can be either conformal or disformal  with
different physical consequences. For conformal couplings, the resulting scalar-induced fifth
force needs to to be screened locally. This appears to be feasible in only a few ways:
chameleon \cite{Khoury:2003rn}, Damour-Polyakov \cite{Damour:1994zq}, K-mouflage \cite{Babichev:2009ee,Brax:2014a} and Vainshtein \cite{Vainshtein:1972sx}.
Another mechanism, associated with the ultra-local models introduced in a companion
paper \cite{Brax:2016vpd}, arises from the absence of kinetic terms and the locality of the theory.
We will find  in this paper that this case can be related, outside the Compton wavelength,
to the chameleon models with a large mass.
For disformal couplings, no fifth force is present in (quasi)-static situations \cite{Brax:2012hm} precluding the
need for a specific screening mechanism.

All these theories involve non-linearities, either in the potential or kinetic terms and as such
appear as low-energy effective field theories. In particular, the issue of the structure of the
radiative corrections to the bare Lagrangian is a thorny one, only alleviated in some cases by
non-renormalisation theorems, e.g. for Galileons \cite{Nicolis:2008in} or K-mouflage \cite{deRham:2014wfa}. For this reason, and because
of its radiative stability, supersymmetry might be a promising setting for dark energy models \cite{Brax:1999gp,Brax:2009kd}.
In this paper, we will consider the super-chameleon models \cite{Brax:2013yja,Brax:2012mq} where the chameleon model is
embedded in a supersymmetric setting. This requires the existence of three separate sectors.
The dark sector where both dark matter and dark energy live. The matter sector which should
include the standard model of particle physics and finally a supersymmetry breaking sector
which shifts the masses of the matter superpartners compared to their standard model
counterparts. The analysis of this model was already presented in \cite{Brax:2013yja,Brax:2012mq}. Here we recall the
salient features and emphasize two facts. First of all, the interaction between dark matter
particles mediated by dark energy is of extremely short range, shorter than the inverse
gravitino mass. Nevertheless, dark matter will see its dynamics modified, i.e. a
modification of gravity, on very large scales where collective phenomena for the
coarse-grained dark matter fluid can be present. Second, we also recall that ordinary matter
decouples from dark energy due to the Damour-Polyakov mechanism leading to no
modification of gravity in the Solar System.

In this paper we point out that on cosmological and astrophysical scales these
super-chameleon models can be identified to the ultra-local models introduced in
a companion paper  \cite{Brax:2016vpd}. These ultra-local models correspond to modified source models \cite{Carroll:2006jn} where the coupling to matter has a magnitude of order $\vert \ln A\vert \lesssim 10^{-6}$ to guarantee that the contribution of modified gravity to Newton's potential is at most of order one. Ultra-local models are such that the value of the dark energy
field depends algebraically on the local dark matter density. This leads to a certain number of
important properties. First, the growth of structure in the linear to quasi-linear regime has an
instability at short scales which is eventually tamed by the absence of fifth forces on short
distances like the Solar System. This screening mechanism is quite different from the
usual screening mechanisms encountered in other modified-gravity scenarios as it directly
follows from the locality of the fifth-force interaction.
The intermediate region between the very large and very small scales is not amenable to our
analysis and would require numerical simulations which go beyond our analysis,
although we present a thermodynamic approach to investigate the fifth-force non-linear
regime.
We find that the number of intermediate dark matter halos is affected by the presence of the
super-chameleon. This is all the more true for galactic size and mass halos where the
fifth force is of the same magnitude as Newton's force. A more complete analysis would require numerical simulations which are left for future work.

This paper is organized as follows.
In section~\ref{sec:SUSY-chameleons} we describe the supersymmetric chameleon models
and the dark and baryonic sectors.
Next, in section~\ref{sec:Dilaton} we show that these models can be identified with
 ultra-local models introduced in a companion paper, over the scales that are relevant for
cosmological purposes.
We describe the background dynamics and the growth of large-scale structures in
section~\ref{sec:ultra-local-dynamics}, considering both linear perturbation theory and
the spherical collapse dynamics.
In section~\ref{sec:halos} we estimate the magnitude of the fifth force within spherical
halos and on cluster and galaxy scales.
In section~\ref{sechistory-cosmo} we use a thermodynamic approach to investigate the
non-linear fifth-force regime for the cosmological structures that turn non-linear at high
redshift and for the cores of dark matter halos.
We briefly investigate the dependence on the parameter $\alpha$ of our results in
section~\ref{sec:x-dependence} and we conclude in section~\ref{sec:conclusion}.

\section{Supersymmetric chameleons}
\label{sec:SUSY-chameleons}

\subsection{Super-chameleons}
\label{sec:super-chameleons}

The nature of the dark part of the Universe, i.e. dark matter and dark energy, is still unknown.
It is not ruled out that both types of dark elements belong to a secluded sector of the ultimate
theory of physics describing all the interactions of the Universe. In this paper, we will use a
supersymmetric setting at low energy and assume that the theory comprises three sectors
with only gravitational interaction between each other. We will assume that the standard
model of particles to which baryons belong is one of them. We will also add a supersymmetry
breaking sector $\cancel{\mathrm{SG}}$ and a dark sector comprising both the dark energy field, which will turn out to
be a supersymmetric version of a chameleon dark energy model, and dark matter
represented by fermions in separate superfields from the super-chameleon one. For details about supersymmetry and its relation to cosmology, see for instance \cite{Binetruy:2006ad}.

Baryons are  introduced in  a secluded sector defined by the
K\"{a}hler potential $K_M$ and the superpotential $W_M$. This is the matter  sector which
complements the dark sector and the supersymmetry breaking one. Assuming no direct
interaction between the super-chameleon $\Phi$ and matter, we take for the total K\"ahler
potential which governs the kinetic terms of the model
\begin{equation}
K = K(\Phi\Phi^\dagger) + K_{\cancel{\mathrm{SG}}} + K_M
\end{equation}
and similarly for the superpotential which is responsible for the interactions between the fields
\begin{equation}
W = W(\Phi) + W_{\cancel{\mathrm{SG}}} + W_M.
\end{equation}
The kinetic terms for the complex scalar fields $\phi^i$ of the model obtained as the scalar
components of the superfields $\Phi^i$ are given by
\be
{\cal L}_{\rm kin} = - K_{i\bar j} \, \partial_\mu \phi^i \partial^\mu\bar \phi^{\bar j}
\ee
where we have defined
\be
K_{i\bar j} = \frac{\partial^2 K}{\partial \Phi^i\partial\bar \Phi^{\bar j}}
\equiv \partial_i \bar \partial_{\bar j} K
\ee
and its matrix inverse such that $K^{i\bar j} K_{k\bar j}= \delta^i_{ k}$.
The scalar potential obtained from the $F$-terms of the superfields is given by
\be
V_{\rm F} =  K^{i\bar j} \partial_i W \bar \partial_{\bar j} \bar W ,
\ee
where $\bar W$ is the complex conjugate of $W$. This is the only term in the scalar potential
when the fields are not charged under gauge groups.

We will also need to add a $D$-term potential to the scalar potential when some extra fields
in the dark sector are charged under a gauge symmetry. We will also consider the corrections
due to supergravity induced by the presence of the supersymmetry breaking sector.
This will be dealt with in the corresponding sections.

\subsection{The supersymmetric model}
\label{sec:SUSY-model}

We consider supersymmetric models where the scalar potential and the coupling to Cold Dark
Matter (CDM) arise from a particular choice of the K\"{a}hler potential for the dark energy
superfield $\Phi$ which is non-canonical whilst the dark matter superfields $\Phi_\pm$
have a canonical normalisation
\begin{equation}
 K(\Phi\Phi^\dagger)=\frac{\Lambda_1^2}{2} \left(\frac{\pdp}{\Lambda_1^2}\right)^\gamma
+ \Phi_+^\dagger\Phi_+ + \Phi_-^\dagger\Phi_-  .
\label{Kahler-gamma-def}
\end{equation}
The self-interacting part of the superpotential is
\begin{equation}
W = \frac{\gamma}{\sqrt{2}\omega} \left(\frac{\Phi^\omega}{\Lambda_0^{\omega-3}}\right)
+ \frac{1}{\sqrt{2}} \left(\frac{\Phi^\gamma}{\Lambda_2^{\gamma-3}}\right)  ,
\;\;\; 0 < \omega < \gamma ,
\label{W-superpotential-def}
\end{equation}
where $\Phi$ contains a complex scalar $\phi$ whose modulus acts as  super-chameleon
and $\Phi_\pm$ are chiral superfields containing dark matter fermions $\psi_\pm$.
Defining the super-chameleon field as $\phi(x)=|\phi|e^{i\theta}$ and identifying
$\phi\equiv|\phi|$, one can minimise the potential over the angular field $\theta$ and after
introducing the new scales
\beq
\Lambda = \Lambda_2 \left(\frac{\Lambda_1}{\Lambda_2}\right)^{(\gamma-1)/2} , \;\;\;
\phi_{\rm min} = \Lambda_2
\left( \frac{\Lambda_2}{\Lambda_0} \right)^{(\omega-3)/(\gamma-\omega)} ,
\label{eq:scales}
\eeq
the scalar potential becomes
\be\label{eq:Fpot}
V_{\rm F}(\phi) = K^{\Phi\Phi^{\dagger}} \left\vert \frac{d W}{d\Phi} \right\vert^2
= \Lambda^4 \left[ 1- \left(\frac{\phi_{\rm min}}{\phi}\right)^\frac{n}{2} \right]^2 ,
\ee
with
\beq
n = 2 (\gamma - \omega)  \;\;\; \mbox{for} \;\;\; n \geq 2 , \;\;\; \gamma \geq \omega+1 .
\label{n-def}
\eeq
When $\phi\ll\pmi$ equation (\ref{eq:Fpot}) reduces to the Ratra-Peebles potential \cite{Ratra:1987rm}
\begin{equation}
\phi \ll \phi_{\rm min} : \;\;\; V_{\rm F}(\phi) \approx \Lambda^4 \left(\frac{\pmi}{\phi}\right)^n ,
\end{equation}
which has been well studied in the context of dark energy
and used to define chameleons. This is the reason why this model is called super-chameleon.
At larger field values the potential has a minimum at $\phi=\pmi$ where
$V_{\rm F}(\pmi)=0$ and $d W/d \phi=0$.
Supersymmetry is therefore broken whenever $\phi\ne\pmi$ and restored at the minimum
where the supersymmetric minimum always has a vanishing energy (this follows from the
supersymmetry algebra).
Then, a new mechanism must be introduced in order to have a non-vanishing cosmological
constant at the minimum of the potential.

\subsection{The Fayet-Iliopoulos mechanism}
\label{sec:Fayet-Iliopoulos}

An effective cosmological constant can be implemented by introducing two new scalars
$\Pi_\pm=\pi_\pm+\ldots$ with charges $\pm q$ under a local $\mathrm{U}(1)$ gauge
symmetry in the dark sector.
These have the canonical K\"{a}hler potential
\begin{equation}
K(\Pi_\pm) = \Pi_+^\dagger e^{2qX} \Pi_+ + \Pi_-^\dagger e^{-2qX} \Pi_- ,
\;\;\; q > 0 ,
\end{equation}
where $X$ is the $\mathrm{U}(1)$ vector multiplet containing the $\mathrm{U}(1)$
gauge field $A_\mu$.
They are chosen to couple to the super-chameleon via the superpotential
\begin{equation}
W_\pi = g^\prime \Phi \Pi_+ \Pi_-
\end{equation}
where $g^\prime = {\cal O}(1)$ is a coupling constant.
This construction gives rise to new terms in the scalar potential.
The first contribution is the D-term potential coming from the fact that the $\Pi_\pm$
fields are charged
\begin{equation}
V_{\rm D} = \frac{1}{2} \left( q \pi_+^2 - q \pi_-^2 - \xi^2 \right)^2 ,
\label{VD-pi}
\end{equation}
where we have included a Fayet-Illiopoulos term $\xi^2$ which will later play the role of the
cosmological constant. The second part of the new scalar potential is far more complicated
with the addition of these new fields but when $\langle\pi_-\rangle=0$ it simplifies
and the sum of both terms yields
\begin{equation}\label{eq:D-term}
V(\pi_+) = \frac{1}{2} \left( q \pi_+^2 - \xi^2 \right)^2 + {g^{\prime}}^2 \phi^2 \pi_+^2 ;
\quad \langle \pi_- \rangle = 0  ,
\end{equation}
where  we have put $\pi_+=|\pi_+|$. It can be shown \cite{Brax:2013yja} that $\langle \pi_- \rangle=0$ minimises
the whole potential so we only consider the effects of the new term $V(\pi_+)$.
In particular, the mass of the charged scalar $\pi_+$ is
\be
m_{\pi_+}^2 = 2 \gp^2 \phi^2 - 2 q \xi^2 .
\ee
At early times the super-chameleon is small ($\phi \ll \phi_{\rm min}$) and this mass
is negative. The $\mathrm{U}(1)$ symmetry is therefore broken ($\langle \pi_+ \rangle \ne 0$).
However, as the cosmological field evolves towards its minimum this mass  increases until
it reaches zero, restoring the symmetry so that $\langle \pi_+ \rangle = 0$.
Minimising (\ref{eq:D-term}) with respect to $\pi_+$ one finds
\beqa
&& \phi < \frac{\sqrt{q}}{g^\prime} \xi : \;\;\; V_{\rm min} = - \frac{m_{\pi_+}^4}{8 q^2}
+ \frac{\xi^4}{2} ,   \\
&& \phi > \frac{\sqrt{q}}{g^\prime} \xi : \;\;\; V_{\rm min} = \frac{\xi^4}{2} .
\label{xi4-cosmological-constant}
\eeqa
Therefore, at late times we recover the present-day dark energy density by taking
\beq
\xi^4 = 2 \bar\rho_{\rm de0} ,
\label{Fayet-constant}
\eeq
which gives $\xi \sim 10^{-3} {\rm eV}$.
This mechanism requires that $\phi_{\rm min}> \sqrt{q} \xi/g^\prime$, which imposes
restrictions on the parameter space,
\beq
\Lambda_2 \left( \frac{\Lambda_2}{\Lambda_0} \right)^{(\omega-3)/(\gamma-\omega)}
> \frac{\sqrt{q}}{g^\prime} \left( 2 \bar\rho_{\rm de0} \right)^{1/4} .
\eeq

\subsection{The coupling to Cold Dark Matter}
\label{sec:CDM-coupling}

Dark energy in the form of $\Phi$ is coupled to dark matter.
The coupling function between the two dark sides of the model is found by considering
the interaction of $\Phi$ and $\Phi_\pm$
\begin{equation}
W_{\rm int} = m \left[ 1 + \frac{g \, \Phi^\sigma}{m \,\Lambda_3^{\sigma-1}} \right]
\Phi_+\Phi_-  , \;\;\; \sigma > 0 ,
\label{W-int-CDM}
\end{equation}
which gives a super-chameleon dependent mass to the dark matter fermions
\begin{equation}
\mathcal{L} \supset \frac{\partial^2 W_{\rm int}}{\partial\Phi_+ \partial\Phi_-} \psi_+\psi_-  .
\end{equation}
When the dark matter condenses to a finite density, $\rho=m\langle\psi_+\psi_-\rangle$, this
term gives a density-dependent contribution to the scalar potential
\begin{equation}
\mathcal{L}\supset A(\phi)\rho ,
\end{equation}
from which one can read off the coupling function
\begin{equation}
\label{eq:AmodB}
A(\phi) = 1 + \frac{g \, \phi^{\sigma}}{m \, \Lambda_3^{\sigma-1}} .
\end{equation}
This function reappears in the form of the conformal coupling between dark matter and dark
energy considered as a scalar-tensor theory

\subsection{The normalised dark-energy scalar field $\varphi$}
\label{sec:normalized-dynamics}

Because $K_{\phi\bar\phi}\ne 1$ the field $\phi$ is not canonically normalised, since the
kinetic term in the Lagrangian reads
\begin{equation}
\mathcal{L}_{\rm kin} = - K_{\phi\bar\phi} \, \partial_\mu\phi \partial^\mu\bar\phi
= - \frac{\gamma^2}{2} \left( \frac{| \phi |}{\Lambda_1} \right)^{2(\gamma-1)}
\partial_\mu\phi \partial^\mu\bar\phi .
\end{equation}
The normalised field is then easily defined by
\begin{equation}
\label{eq:vphi}
\varphi = \Lambda_1 \left( \frac{\phi}{\Lambda_1} \right)^\gamma ,
 \end{equation}
and the coupling function (\ref{eq:AmodB}) becomes
\begin{equation}
A(\varphi) = 1 + \alpha \left( \frac{\varphi}{\varphi_{\rm min}} \right)^{\sigma/\gamma}
\;\;\; \mbox{with} \;\;\; \alpha \equiv \frac{g \phi_{\rm min}^{\sigma}}{m \Lambda_3^{\sigma-1}} ,
\label{eq:x}
\end{equation}
and
\beq
\varphi_{\rm min} = \Lambda_1 \left( \frac{\phi_{\rm min}}{\Lambda_1} \right)^\gamma
= \Lambda_1 \left( \frac{\Lambda_2}{\Lambda_1} \right)^{\gamma}
\left( \frac{\Lambda_2}{\Lambda_0} \right)^{\gamma(\omega-3)/(\gamma-\omega)} ,
\eeq
while the effective potential $V_{\rm F}(\varphi) +\rho (A(\varphi)-1)$ is
\begin{equation}
V_{\rm eff}(\varphi) = \Lambda^4 \left[ \left( \frac{\varphi_{\rm min}}{\varphi}
\right)^{n/2\gamma} - 1 \right]^2 + \alpha \, \rho \left( \frac{\varphi}{\varphi_{\rm min}}
\right)^{\sigma/\gamma}  .
\label{eq:bveffa}
\end{equation}
Notice that the effective potential in this model coincides with the one obtained in
a scalar tensor theory with the potential $V_{\rm F}(\varphi)$ and the coupling function
$A(\varphi)$. We will exploit this fact below.
Since we require the cosmology to remain close to the $\Lambda$-CDM scenario,
i.e. the fifth force must not be much greater than Newtonian gravity, within this
framework we can infer that the coupling function $A(\varphi)$ must remain close
to unity. This provides the constraint
\beq
\alpha \ll 1
\label{A-x-small}
\eeq
on the parameter combination $\alpha$ of Eq.(\ref{eq:x}).

The dynamics of the model can be determined by minimizing the effective potential.
This  leads to the minimum $\varphi$ of the theory in the presence of matter (CDM)
\begin{equation}
\label{eq:phimineq}
\left(\frac{\varphi_{\rm min}}{\varphi}\right)^{(n+\sigma)/\gamma} -
\left(\frac{\varphi_{\rm min}}{\varphi}\right)^{(n+2\sigma)/2\gamma}
= \frac{\rho}{\rho_\infty}  ,
\end{equation}
where we have defined the energy density
\begin{equation}
\label{eq:rhoinf}
\rho_\infty = \frac{n}{\alpha \sigma} \Lambda^4 = \bar\rho_0 (1+z_{\infty})^3 , \;\;\;
\mbox{and} \;\;\; 0 < \varphi \leq \varphi_{\rm min} \, ,
\end{equation}
where $z_\infty$ is the redshift below which the field becomes close to  its supersymmetric
minimum $\varphi_{\rm min}$\footnote{Although the vacuum energy due to $V_D$ lifts
the vanishing energy density of a true supersymmetric minimum and therefore
supersymmetry is broken by $\xi$, we will still refer to $\varphi_{\rm min}$ as the
supersymmetric minimum as it minimises $V_F$.}.

As in scalar tensor theories, such as dilaton models or $f(R)$ theories,
it is convenient to introduce the coupling $\beta(\varphi)$ defined by
\beqa
&& \hspace{-1cm} \beta(\varphi) = M_{\rm Pl} \frac{d\ln A}{d\varphi}
\label{beta-chameleon-def} \\
&& \hspace{-0.8cm} = \frac{\alpha \sigma}{\gamma} \, \frac{M_{\rm Pl}}{\varphi_{\rm min}}
\left[ 1 + \alpha \left(\frac{\varphi}{\varphi_{\rm min}}\right)^{\sigma/\gamma} \right]^{-1}
\left( \frac{\varphi}{\varphi_{\rm min}} \right)^{\sigma/\gamma-1}
\label{eq:betaphi}
\eeqa
and the effective mass $m^2_{\rm eff} = \partial^2 V_{\rm eff}/\partial \varphi^2$
at the minimum of the effective potential,
\beqa
&& \hspace{-1.cm} m^2_{\rm eff}(\varphi) = \frac{\alpha\sigma}{\gamma}
\frac{\rho_{\infty}}{\varphi^2} \left(\frac{\varphi}{\varphi_{\rm min}}\right)^{\sigma/\gamma}
\left[ \frac{n}{\gamma} \left(\frac{\varphi_{\rm min}}{\varphi}\right)^{(n+\sigma)/\gamma}
\right. \nonumber \\
&& \left. - \frac{n}{2\gamma} \left(\frac{\varphi_{\rm min}}{\varphi}\right)^{(n+2\sigma)/2\gamma}
+ \frac{\sigma}{\gamma} \frac{\rho}{\rho_{\infty}} \right] ,
\label{m2eff-def}
\eeqa
where we used Eq.(\ref{eq:phimineq}).
The quasi-static approximation (\ref{eq:rhoinf}) applies if $m_{\rm eff}^2 \gg H^2$.
This holds for redshifts $z \leq z_{\infty}$ provided
\beq
\frac{\alpha\rho_{\infty}}{\varphi_{\rm min}^2} \gg H^2_{\infty} , \;\;\; \mbox{whence} \;\;\;
\left( \frac{\varphi_{\rm min}}{M_{\rm Pl}} \right)^2 \ll \alpha ,
\label{quasi-static-1}
\eeq
where in the second inequality we assumed $z_{\infty} \leq z_{\rm eq}$.
At higher redshifts, $m_{\rm eff}(z)$ grows at least as fast as $H(z)$ in both the matter and
radiation eras if we have
\beq
\mbox{matter era:} \;\; \sigma  \leq 2\gamma , \;\;
\mbox{radiation era:} \;\; \sigma \leq \gamma + \omega/2 .
\label{delta-meff-bounds}
\eeq

\subsection{Supersymmetry breaking}
\label{sec:SUSY-breaking}

Supersymmetry is  broken by values much larger than the energy density of CDM.
This is achieved in a dedicated sector of the theory which we do not need to specify here.
Gravitational interactions lead to a correction to the scalar  potential coming from
supersymmetry breaking \cite{Brax:2012mq}
\begin{equation}
\Delta V_{\cancel{\mathrm{SG}}} = \frac{m_{3/2}^2 \left| K_{\Phi} \right|^2}
{K_{\Phi\Phi^\dagger}} \sim \frac{m_{3/2}^2 \phi^{2\gamma}}{\Lambda_1^{2\gamma-2}}  ,
\end{equation}
where $m_{3/2}$ is the gravitino mass.
This competes with the density dependent term in the effective potential (\ref{eq:bveffa}).
This correction does not upset the dynamics of the model as long as
\begin{equation}
\label{eq:sugraphi}
\left(\frac{\varphi_{\rm min}}{M_{\rm Pl}}\right)^2 \ll
\frac{\alpha \rho_\infty}{M_{\rm Pl}^2 m_{3/2}^2} .
\end{equation}
This is typically much more stringent than the quasi-static condition (\ref{quasi-static-1}).
Using Eq.(\ref{m2eff-def}) this can also be shown to correspond to a condition on the mass
of the scalar field $\varphi$ at the supersymmetric minimum, for $z \leq z_{\infty}$,
\begin{equation}
\label{eq:minfbound}
m_{\rm eff}^2(\varphi_{\rm min}) \sim \frac{\alpha \rho_\infty}{\varphi_{\rm min}^2}
\gg m_{3/2}^2  .
\end{equation}
As the gravitino mass is always greater than $10^{-5}$  eV in realistic models of supersymmetry
breaking \cite{Fayet:1986zc}, we deduce that the range of the scalar interaction mediated by $\varphi$ is very
small, at most at the  cm level. Because the scalar interaction has such a short
range, we call these models ultra-local.
In fact, we shall see below that they can be related to the so-called``ultra-local models''
introduced in the companion paper \cite{Brax:2016vpd}.

\subsection{Coupling to baryons}
\label{coupli-bary}

We consider that matter fermions $\psi$ belong to a  superfield $\Phi_M$.
The mass of the canonically normalised matter fermions becomes
\begin{equation}
m_\psi= e^{K(\Phi,\Phi^\dagger)/2M_{\rm Pl}^2} \; m_\psi^{(0)} ,
\end{equation}
where $m_\psi^{(0)}$ is the bare mass of the baryons $\frac{\partial^2 W}{\partial \Phi_M^2}$.
The exponential prefactor is at the origin of the coupling function between the
matter fields and the super-chameleon in the Einstein frame. This leads to the identification of
coupling function in the matter sector
\begin{equation}
A_M(\varphi) = e^{\varphi^2/2M_{\rm Pl}^2}
\end{equation}
for the canonically normalised super-chameleon, and the coupling to baryons
\begin{equation}
\beta_M(\varphi) = \frac{\varphi}{M_{\rm Pl}} ,
\end{equation}
which is the coupling of a dilaton to matter.
As long as $\varphi_{\rm min} \ll M_{\rm Pl}$, which is already required to suppress the
supergravity corrections to the scalar potential, the coupling to baryons is negligible.
Hence this model describes a scenario where dark energy essentially couples to dark matter
and decouples from ordinary matter.

\section{The supersymmetric chameleon as an ultra-local model}
\label{sec:Dilaton}

\subsection{Definition of ultra-local models}
\label{sec:definition-ultra-local}

We define ultra-local scalar field models by the action \cite{Brax:2016vpd}
\beqa
S & = & \int d^4 x \; \sqrt{-\tg} \left[ \frac{\tilde{M}_{\rm Pl}^2}{2} \tilde{R}
+ \tilde{\cal L}_{\varphi}(\varphi) \right] \nonumber \\
&& + \int d^4 x \; \sqrt{-g} \, {\cal L}_{\rm m}(\psi^{(i)}_{\rm m} , g_{\mu\nu}) ,
\label{S-def}
\eeqa
where the dark matter fields $\psi^{(i)}_{\rm m}$ follow the Jordan-frame metric
$g_{\mu\nu}$, with determinant $g$, which is related to the Einstein-frame metric
$\tg_{\mu\nu}$ by
\beq
g_{\mu\nu} = A^2(\varphi) \tg_{\mu\nu} .
\label{conformal-g-tg}
\eeq
We explicitly take no coupling between baryons and the scalar field to make possible
the equivalence with the supersymmetric chameleon models.
In this paper we restrict ourselves to large cosmological scales, which are dominated
by the dark matter, and we neglect the impact of baryons.
Ultra-local models are defined by the property that their scalar-field kinetic term
is negligible,
\beq
\tilde{\cal L}_{\varphi}(\varphi) = - V(\varphi) .
\label{L-phi-def}
\eeq
Introducing the characteristic energy scale $\cM^4$ of the potential and the dimensionless
field $\tilde\chi$ as
\beq
\tilde\chi \equiv - \frac{V(\varphi)}{\cM^4} , \;\;\; \mbox{and} \;\;\;
A(\tilde\chi) \equiv A(\varphi) ,
\label{tchi-def}
\eeq
these models are fully specified by a single function, $A(\tilde\chi)$, which is
defined from the initial potential $V(\varphi)$ and coupling function $A(\varphi)$
through Eq.(\ref{tchi-def}).
In other words, because the kinetic term is negligible there appears a degeneracy
between the potential $V(\varphi)$ and the coupling function $A(\varphi)$.
The change of variable (\ref{tchi-def}) absorbs this degeneracy and we are left
with a single free function $A(\tilde\chi)$.

\subsection{Cosmological background of ultra-local models}
\label{sec:background-ultra-local}

Because the matter fields follow the geodesics set by the Jordan frame and
satisfy the usual conservation equations in this frame, we mostly work in the
Jordan frame.
We introduce the time dependent coupling
\be
\epsilon_2(t) \equiv \frac{d\ln\bar{A}}{d\ln a} ,
\label{eps2-def}
\ee
such that, as shown in the companion paper, the Friedmann equation reads as
\beq
3 M_{\rm Pl}^2 {\cal H}^2 = (1-\epsilon_2)^{-2} a^2 ( \bar{\rho} + \bar{\rho}_{\rm rad}
+ \bar{\rho}_{\tchi} ) ,
\label{Friedmann-J}
\eeq
where $\tau$ is the conformal time, ${\cal H}$ the conformal Hubble expansion rate,
and the Jordan-frame Planck mass is
\beq
M_{\rm Pl}^2(t) = \bar{A}^{-2}(t) \, \tilde M_{\rm Pl}^2 ,
\label{Planck-J}
\eeq
while $\bar\rho$, $\bar\rho_{\rm rad}$ and $\bar\rho_{\tilde\chi}$ are the matter,
radiation and scalar field energy densities.
In particular, the background matter and radiation densities evolve as usual as
\beq
\bar\rho = \frac{\bar\rho_0}{a^3} , \;\,\,\,
\bar\rho_{\rm rad} = \frac{\bar\rho_{\rm rad 0}}{a^4} ,
\label{rho-a-J}
\eeq
while the scalar field energy density is given by
\beq
\bar\rho_{\tilde\chi} = - \bar{A}^{-4} {\cal M}^4 \bar{\tilde\chi} ,
\label{rho-chi-def}
\eeq
and the equation of motion of the background scalar field is
\beq
\cM^4 = \bar{A}^4 \bar{\rho} \frac{d\ln\bar{A}}{d\bar{\tchi}} \;\;\;
\mbox{hence} \;\;\;
\frac{d\bar{\tchi}}{d\tau} = \bar{A}^4 \frac{\bar{\rho}}{\cM^4} \epsilon_2
{\cal H} .
\label{KG-J-background}
\eeq
It is convenient to write the Friedmann equation (\ref{Friedmann-J}) in a more
standard form by introducing the effective dark energy density $\bar\rho_{\rm de}$
defined by
\beq
3 M_{\rm Pl}^2 {\cal H}^2 = a^2 ( \bar{\rho} + \bar{\rho}_{\rm rad}
+ \bar{\rho}_{\rm de} ) ,
\label{Friedmann-J-de}
\eeq
which gives
\beq
\bar{\rho}_{\rm de} = \bar\rho_{\tchi} + \frac{2\epsilon_2-\epsilon_2^2}{(1-\epsilon_2)^2}
( \bar{\rho} + \bar{\rho}_{\rm rad} + \bar{\rho}_{\tchi} ) .
\label{rho-de-def}
\eeq

\subsection{Cosmological perturbations of ultra-local models}
\label{sec:bperturbations-ultra-local}

We write the Newtonian gauge metric as
\beq
d s^2= a^2 [ - (1+2\Phi) d\tau^2 + (1-2\Psi) d\vx^2 ] ,
\label{metric-J}
\eeq
so that the Einstein- and Jordan-frame metric potentials are related by
\beq
1+2\Phi = \frac{A^2}{\bar{A}^2} (1+2\tPhi) , \;\;\;
1-2\Psi = \frac{A^2}{\bar{A}^2} (1-2\tPsi) ,
\label{Phi-J-Phi-E}
\eeq
while the Jordan-frame Newtonian potential is defined by
\beq
\frac{\nabla^2}{a^2} \Psi_{\rm N} \equiv \frac{\delta\rho+\delta\rho_{\tchi}}
{2 M_{\rm Pl}^2} .
\label{tPsi-N-J}
\eeq
Because we wish the deviations of $\Phi$ and $\Psi$ from the Newtonian potential
$\Psi_{\rm N}$ to remain modest, and we typically have
$| \Psi_{\rm N} | \lesssim 10^{-5}$ for cosmological and astrophysical structures,
we require $| \delta\ln A | \lesssim 10^{-5}$ and
$| \delta\rho_{\tilde\chi} | \lesssim | \delta\rho |$.
This first constraint is fulfilled by choosing coupling functions $A(\tilde\chi)$
that are bounded and deviate from unity by less than $10^{-5}$, which reads as
\beq
| \ln A(\tilde\chi) | \lesssim 10^{-5} ,
\label{bound-lnA}
\eeq
while the second constraint will follow naturally because the characteristic scalar field
energy density is the dark energy density today.
Then, we can linearize Eq.(\ref{Phi-J-Phi-E}) in $\delta\ln A$. This leads to
\beq
\Phi = \Psi_{\rm N} + \delta\ln A , \;\;\; \Psi = \Psi_{\rm N} - \delta\ln A ,
\label{Phi-Psi-J}
\eeq
while the dark energy density fluctuations read as
\beq
\delta\rho_{\tilde\chi} = - {\cal M}^4 \delta\tilde\chi .
\label{delta-rho-chi}
\eeq
In Eq.(\ref{delta-rho-chi}) and in the following we use the characteristic property
(\ref{bound-lnA}) of ultra-local models to write $A \simeq 1$ wherever this approximation
is valid within a $10^{-5}$ accuracy (the only place where deviations of $A$ from unity
are important is for the computation of the fifth force through the gradient $\nabla\ln A$).

In general configurations including perturbations, the equation of motion of the scalar field
reads as
\beq
\frac{d\ln A}{d\tchi} = \frac{\cM^4}{\rho} .
\label{KG-pert-J}
\eeq
The dark matter component obeys  the continuity and Euler equations
\beq
\frac{\pl\rho}{\pl\tau} + (\vv\cdot\nabla) \rho + ( 3 {\cal H}+\nabla\cdot\vv) \rho = 0 ,
\label{continuity-J}
\eeq
and
\beq
\frac{\pl\vv}{\pl\tau} + (\vv\cdot\nabla)\vv + {\cal H} \vv = -\nabla\Phi .
\label{Euler-J}
\eeq
From Eq.\eqref{Phi-Psi-J} we have $\nabla \Phi = \nabla \Psi_{\rm N} + \nabla \ln A$,
and then the scalar field equation \eqref{KG-pert-J} gives
\beq
\nabla\ln A = \frac{\cM^4}{\rho} \nabla\tchi .
\label{grad-lnA}
\eeq
Thus in terms of matter dynamics, the scalar field appears via the modification of the Poisson
equation \eqref{tPsi-N-J}, because of the additional source associated to the scalar field and
the time dependent Planck mass, and via the appearance of the ``new'' term (\ref{grad-lnA})
in the Euler equation \eqref{Euler-J}, which is due to the spatial variation of $\ln A$.

On large scales  we may linearize the equations of motion.
Expanding the coupling function $A(\tchi)$ as
\beq
\ln A(\tchi) = \ln\bar{A} + \sum_{n=1}^{\infty} \frac{\beta_n(t)}{n!} (\delta\tchi)^n ,
\label{beta-n-def}
\eeq
the scalar field equation (\ref{KG-pert-J}) gives at the background and linear
orders
\beq
\beta_1 = \frac{\cM^4}{\bar\rho} , \;\;\;
\delta\tchi = - \frac{\beta_1}{\beta_2} \delta .
\label{dchi-linear}
\eeq
Defining
\beq
\epsilon_1(t) \equiv \frac{\beta_1}{\beta_2} \frac{\cM^4}{\bar\rho}
= \frac{\beta_1^2}{\beta_2} ,
\label{eps1-def}
\eeq
we have for the linear matter density contrast $\delta$
\beq
\frac{\partial\delta^2}{\partial\tau^2} + {\cal H} \frac{\partial\delta}{\partial\tau}
+ \epsilon_1 c^2 \nabla^2 \delta = \frac{\bar\rho a^2}{2 M_{\rm Pl}^2} (1+\epsilon_1) \delta ,
\label{linear-delta-real}
\eeq
which also reads in Fourier space as
\begin{equation}
\frac{\partial^2\delta}{\partial\tau^2} + {\cal H} \frac{\partial\delta}{\partial\tau}
- \frac{3}{2} \Omega_{\rm m}(\tau) {\cal H}^2 \left[ 1+\epsilon(k,\tau)\right] \delta =0 ,
\label{pertlin-ultra-local}
\end{equation}
where $\epsilon(k,\tau)$, which corresponds to the deviation from the $\Lambda$-CDM
cosmology, is given by
\beq
\epsilon(k,\tau) = \epsilon_1(\tau) \left[ 1 + \frac{2}{3\Omega_{\rm m}}
\frac{c^2k^2}{a^2H^2} \right] .
\label{eps-def}
\eeq
The $k$-dependent term dominates when $ck/aH>1$, i.e. on sub-horizon
scales. Moreover, we have $(ck/aH)^2 \sim 10^7$ today at scales of about
$1 \, h^{-1}$Mpc. Therefore, we must have
\beq
| \epsilon_1 | \lesssim 10^{-7}
\label{epsilon1-bound1}
\eeq
to ensure that the growth of large-scale structures is not too significantly modified.
This small value does not require introducing additional small parameters as it
will follow from the constraint (\ref{bound-lnA}), which already leads
to the introduction of a small parameter $\alpha \lesssim 10^{-5}$ that gives the amplitude
of the coupling function $\ln A$.

The quantity $\epsilon_2$ introduced in Eq.(\ref{eps2-def}) is related to the quantity
$\epsilon_1$ defined in Eq.(\ref{eps1-def}) by
\beq
\epsilon_2 = 3 \epsilon_1 , \;\; \mbox{hence} \;\; | \epsilon_2 | \lesssim 10^{-7} .
\label{epsilon2-bound1}
\eeq
This implies that at the background level the ultra-local model behaves like the
$\Lambda$-CDM cosmology, see Eqs.(\ref{rho-chi-def})-(\ref{rho-de-def}),
as the scalar field and dark energy densities coincide and are almost constant at low $z$,
within an accuracy of $10^{-6}$.

\subsection{Super-chameleon identification}
\label{sec:super-chameleon-identification}

Super-chameleon models are such that the mass of the scalar field is so large that the kinetic
terms are negligible. They behave like ultra-local models on distances
$r \gtrsim m_{\rm eff}^{-1}$. It is only on very short distances, which are negligible on
astrophysical and cosmological scales, that the kinetic terms play a role.
The identification with an ultra-local model is therefore valid on scales
\be
\frac{k}{a} \lesssim m_{\rm eff} ; \;\;\;
\mbox{this includes the range} \;\; \frac{k}{a} \lesssim m_{3/2}  \ll m_{\rm eff} ,
\ee
where we used Eq.(\ref{eq:minfbound}).
Even as early as $a_{\rm BBN} \sim 10^{-10}$, the model is equivalent to an ultra-local
model on comoving scales larger than 10 km, well below the distances of interest in the
growth of cosmological structures. As a result, for all practical purposes
super-chameleon models can be identified with ultra-local models.
Thus, the coupling function $A(\varphi)$ and the potential $V(\varphi)$ defined in
Eqs.(\ref{conformal-g-tg})-(\ref{L-phi-def}) for the ultra-local model can be read from
the effective potential (\ref{eq:bveffa}) of the super-chameleon model, to which we must add
the cosmological constant contribution (\ref{xi4-cosmological-constant}).
Using the mapping (\ref{tchi-def}) in terms of the dimensionless field $\tilde\chi$
this yields
\beq
A(\tilde\chi) = 1 + \alpha \left( \frac{\varphi}{\varphi_{\rm min}} \right)^{\sigma/\gamma}
\eeq
and
\beq
- {\cal M}^4 \tilde\chi = V = \Lambda^4 \left[ \left( \frac{\varphi_{\rm min}}{\varphi}
\right)^{n/2\gamma} - 1 \right]^2 + \frac{\xi^4}{2} .
\label{V-chi-def}
\eeq
We have seen in Eq.(\ref{Fayet-constant}) that $\xi^4=2\bar\rho_{\rm de0}$ to recover
the cosmological constant associated with the current expansion of the Universe.
We can also take ${\cal M}^4=\bar\rho_{\rm de0}$ without loss of generality,
as this only sets the choice of normalization of $\tilde\chi$.
To simplify the model we also take $\Lambda^4=\bar\rho_{\rm de0}$, which avoids
introducing another scale. This gives
\be
{\cal M}^4 = \Lambda^4=\bar\rho_{\rm de0} : \;\;\;
\tilde\chi= -1 - \left[ \left(\frac{\varphi_{\rm min}}{\varphi}\right)^{n/2\gamma} - 1 \right]^2
\label{chi-phi-rel}
\ee
and
\be
A(\tilde\chi)= 1 +  \alpha \left( 1 + \sqrt{-1-\tilde\chi} \right)^{-2\sigma/n}
\;\; \mbox{with} \;\; \tilde\chi \leq -1 ,
\label{A-chi}
\ee
which is the expression of the coupling function in terms of the ultra local scalar field.
The comparison with the supersymmetric model can be completed by verifying that the
cosmological perturbations also obey the same dynamics.

The coupling of dark energy to dark matter implies that the growth of the density contrast of
CDM is modified \cite{Brax:2004qh,Brax:2005ew,Brax:2012gr} and the linear density contrast
$\delta = \delta\rho/\rho$ of the super-chameleon model in the conformal Newtonian Gauge
evolves on sub-horizon scales according to
\begin{equation}
\label{eq:pertfull}
\frac{\partial\delta}{\partial\tau^2} + {\cal H} \frac{\partial\delta}{\partial\tau}
- \frac{3}{2} \Omega_{\rm m}(\tau) {\cal H}^2
\left( 1 + \frac{2\beta^2(\varphi)}{1 + \frac{m_{\rm eff}^2 a^2}{k^2}} \right) \delta = 0  .
\end{equation}
Physically, the last term in (\ref{eq:pertfull}) corresponds to a scale dependent enhancement
of Newton's constant. As the mass of the scalar field is always very large compared to
astrophysical wave numbers, we can simplify (\ref{eq:pertfull}) to find
\begin{equation}
\label{eq:pertlin}
\frac{\partial\delta}{\partial\tau^2} + {\cal H} \frac{\partial\delta}{\partial\tau}
- \frac{3}{2} \Omega_{\rm m}(\tau) {\cal H}^2
\left( 1 + \frac{2 k^2 \beta^2(\varphi)}{m_{\rm eff}^2 a^2} \right) \delta = 0
\end{equation}
for $k/a \ll m_{\rm eff}$.
This equation is the same as the equation (\ref{pertlin-ultra-local}) obtained for the ultra-local
models, on sub-horizon scales where we can neglect the unit factor in Eq.(\ref{eps-def}).
Indeed, the chameleon coupling $\beta(\varphi)$ defined in Eq.(\ref{beta-chameleon-def}),
$\beta = M_{\rm Pl} d\ln A/d\varphi$, and the ultra-local coupling $\beta_1(\tilde\chi)$
defined in Eq.(\ref{beta-n-def}), $\beta_1 = d\ln A/d\tilde\chi$, are related by
\beq
\beta = \beta_1 M_{\rm Pl} \frac{d\tilde\chi}{d\varphi} .
\label{beta-beta1}
\eeq
From the identification (\ref{V-chi-def}) we can write the effective chameleon potential
of Eq.(\ref{eq:bveffa}) as
\beq
V_{\rm eff}(\varphi) = - {\cal M}^4 \tilde\chi + \rho (A-1) - \bar\rho_{\rm de0} ,
\label{Veff-chi-no-cosmo-constant}
\eeq
where we explicitly subtract the cosmological constant.
Then, the quasi-static equation (\ref{eq:phimineq}) for $\varphi$, which corresponds
to the minimum of the potential $\partial V_{\rm eff}/\partial\varphi = 0$,
yields $\beta_1={\cal M}^4/\rho$, where we used Eq.(\ref{beta-beta1}) and $A \simeq 1$,
and we recover the ultra-local equation of motion (\ref{KG-pert-J})-(\ref{dchi-linear}).
Next, from the definition of the chameleon effective mass,
$m_{\rm eff}^2 = \partial^2 V_{\rm eff}/\partial \varphi^2$, we obtain using
Eq.(\ref{Veff-chi-no-cosmo-constant}) and the result $\beta_1={\cal M}^4/\rho$,
\beq
m_{\rm eff}^2(\varphi) = \frac{\rho \beta_2 \beta^2}{M_{\rm Pl}^2 \beta_1^2} ,
\label{m-eff-beta1-beta2}
\eeq
where the ultra-local factor $\beta_2=d^2\ln A/d\tilde\chi^2 = d\beta_1/d\tilde\chi$
was introduced in Eq.(\ref{beta-n-def}).
This gives $2\beta^2/m_{\rm eff}^2=2M_{\rm Pl}^2\beta_1^2/\rho\beta_2$ and
we find that Eq.(\ref{eq:pertlin}) coincides with Eq.(\ref{pertlin-ultra-local})
over the range $H \ll k/a \ll m_{\rm eff}$,
using the second expression (\ref{eps1-def}) for $\epsilon_1(t)$.

This identification of the super-chameleon model with the ultra-local model
shows that on cosmological scales, $H \ll k/a \ll m_{\rm eff}$, the dynamics
is set by the single function $A(\tilde\chi)$ obtained in Eq.(\ref{A-chi}).
This implies that structure formation is only sensitive to two combinations of the parameters
introduced in the supersymmetric chameleon setting, namely the exponent ratio
$\sigma/n$ and the ratio $\Lambda^4/\bar\rho_{\rm de0}$ (which we set to unity in this paper),
in addition to the cosmological constant $\xi^4/2=\bar\rho_{\rm de0}$.
Conversely, there is a wide model degeneracy and the same coupling function
(\ref{A-chi}) corresponds to many different chameleon models.

We can note here that in the context of usual chameleon models such as $f(R)$ theories,
where $\beta \sim 1$, having a very large effective mass $m_{\rm eff}^2$,
with $m_{\rm eff}^{-1} \ll 10^{-4} {\rm mm}$, would lead to negligible departure
from the $\Lambda$-CDM cosmology for the formation of large scale structures,
as seen from Eq.(\ref{eq:pertlin}). This is not the case for the super-chameleon
models studied in this paper because the coupling $\beta$ is also very large and
much greater than unity. Indeed, from Eq.(\ref{eq:betaphi}) we have
$\beta \sim \alpha M_{\rm Pl}/\varphi_{\rm min} \gg 1$, whereas from
Eq.(\ref{m2eff-def}) we have $m_{\rm eff}^2 \sim \alpha \rho_{\infty}/\varphi_{\rm min}^2$.
This yields
\beq
\frac{\beta^2}{m_{\rm eff}^2} \sim \frac{\alpha^2 M_{\rm Pl}^2}{\Lambda^4} ,
\label{beta-m2}
\eeq
and $\beta^2 k^2/m_{\rm eff}^2 a^2$ can be of order unity on kpc to Mpc scales,
even with $\alpha \ll 1$, as we typically have $\Lambda^4 \sim M_{\rm Pl}^2 H_0^2$.

\subsection{Example of models}
\label{sec:examples}

It is interesting to consider templates for ultra-local models coming from super-chameleons.

A good set of models can be obtained for instance by taking the cut-off of the theory
$\Lambda_1=M_{\rm Pl}$ in the K\"{a}hler potential (\ref{Kahler-gamma-def}).
To obtain $\Lambda^4=\bar\rho_{\rm de0}$ as in Eq.(\ref{chi-phi-rel}) this requires
the non-renormalised scale in the superpotential $W$ of Eq.(\ref{W-superpotential-def})
to be $\Lambda_2 = M_{\rm Pl} (\bar\rho_{\rm de0}/M_{\rm Pl}^4)^{1/(6-2\gamma)}$.
A simple choice for the exponents $\omega$ and $\gamma$ is $\omega=1$ and $\gamma=2$,
which gives $n=2$ and the K\"ahler potential becomes
\begin{equation}
K(\Phi\Phi^\dagger) = \frac{M_{\rm Pl}^2}{2} \left( \frac{\pdp}{M_{\rm Pl}^2} \right)^2
+ \Phi_+^\dagger \Phi_+ + \Phi_-^\dagger \Phi_-
\end{equation}
while the self-interacting part of the superpotential is
\begin{equation}
W = \sqrt{2} \Lambda_0^2 {\Phi} + \sqrt{\frac{3 \Omega_{\rm de0}}{2}} H_0 {\Phi^2} ,
\end{equation}
which contains a linear term and a mass term, with
$\Lambda_2=\sqrt{3\Omega_{\rm de0}} H_0$.
Both $\Lambda_0$ and $H_0$ are protected by supersymmetry under renormalisation.

The supersymmetric minimum $\phi_{\rm min}$ of Eq.(\ref{eq:scales}) becomes
\beq
\phi_{\rm min} = \frac{\Lambda_0^2}{\Lambda_2} .
\eeq
Requiring that $\phi_{\rm min} > \sqrt{q} \xi/g^\prime$ to recover the late cosmological
constant behavior (\ref{xi4-cosmological-constant}) and using Eq.(\ref{Fayet-constant})
we obtain the lower bound on $\Lambda_0$
\beq
\Lambda_0^2 \gtrsim M_{\rm Pl}^2 \left( \frac{H_0}{M_{\rm Pl}} \right)^{3/2} .
\label{lower-bound-Lambda0}
\eeq
The normalized chameleon field $\varphi$ of Eq.(\ref{eq:vphi}) reads as
\beq
\frac{\varphi}{M_{\rm Pl}} = \frac{\phi^2}{M_{\rm Pl}^2} , \;\;\;\;\;
\frac{\varphi_{\rm min}}{M_{\rm Pl}} =
\frac{\Lambda_0^4}{3\Omega_{\rm de0} M_{\rm Pl}^2 H_0^2} ,
\eeq
while the characteristic density $\rho_{\infty}$ of Eq.(\ref{eq:rhoinf}) is
\beq
\rho_{\infty} = \frac{2}{\alpha\sigma} \bar\rho_{\rm de0}  \sim
\frac{\bar\rho_{\rm de0}}{\alpha} .
\eeq
We must also satisfy the constraint (\ref{eq:sugraphi}), which yields the upper bound
on $\Lambda_0$
\beq
\Lambda_0^2 \ll M_{\rm Pl}^2 \left( \frac{H_0}{M_{\rm Pl}} \right)^{3/2}
\left( \frac{M_{\rm Pl}}{m_{3/2}} \right)^{1/2} .
\label{upper-bound-Lambda0}
\eeq
As we always have $m_{3/2}\ll M_{\rm Pl}$, the comparison of
Eq.(\ref{upper-bound-Lambda0}) with Eq.(\ref{lower-bound-Lambda0})
shows that the range of values for $\Lambda_0$ is fairly large.

The scales $m$ and $\Lambda_3$ of the dark matter interaction $W_{\rm int}$
in Eq.(\ref{W-int-CDM}) are only constrained through their combination with
$\phi_{\rm min}$ in the coupling parameter $\alpha$ of Eq.(\ref{eq:x}), which must be small
as noticed in Eq.(\ref{A-x-small}).
In fact, the identification with the ultra-local model and the study presented in the companion
paper shows that we must require $\alpha \lesssim 10^{-6}$ to keep the formation of
large cosmological structures close to the $\Lambda$-CDM behavior.
From Eq.(\ref{delta-meff-bounds}) the exponent $\sigma$ should satisfy $\sigma \leq 5/2$
if we wish to ensure that the quasi-static approximation remains valid up to arbitrarily
high redshifts, which gives $0< \sigma/n \leq 5/4$.
More generally, combining Eqs.(\ref{n-def}) and (\ref{delta-meff-bounds}) we have
\beq
0 < \frac{\sigma}{n} \leq \frac{\gamma+\omega/2}{2(\gamma-\omega)} \;\;\;
\mbox{hence} \;\;\; 0 < \frac{\sigma}{n} \leq \frac{3\gamma-1}{4} .
\eeq

It is interesting to obtain the characteristic scales of the coupling $\beta$ and effective
mass $m_{\rm eff}$ of these super-chameleon models.
Using the bounds (\ref{lower-bound-Lambda0}) and (\ref{upper-bound-Lambda0})
we obtain
\beq
\beta \sim \frac{\alpha M_{\rm Pl}^2 H_0^2}{\Lambda_0^4}  \;\;\; \mbox{hence} \;\;\;
\frac{\alpha m_{3/2}}{H_0} \ll \beta \lesssim \frac{\alpha M_{\rm Pl}}{H_0} ,
\eeq
and
\beq
m_{\rm eff}^2 \sim \frac{M_{\rm Pl}^4 H_0^5}{\Lambda_0^8} \;\;\; \mbox{hence} \;\;\;
m_{3/2}^2 \ll m_{\rm eff}^2 \lesssim M_{\rm Pl}^2 .
\eeq
We can check that both $\beta$ and $m_{\rm eff}$ are large in these super-chameleon
models.

As noticed above from Eq.(\ref{A-chi}), eventually we will study the super-chameleon models
of this type where the only parameters are $\alpha$, which will be chosen to be $10^{-6}$ or
lower, and $\zeta=\sigma/n$, of order unity.

\section{Ultra-local dynamics}
\label{sec:ultra-local-dynamics}

\subsection{Chameleon and ultra-local potentials and coupling functions}
\label{sec:background-cosmo}

\begin{figure*}
\begin{center}
\epsfxsize=5.8 cm \epsfysize=5.5 cm {\epsfbox{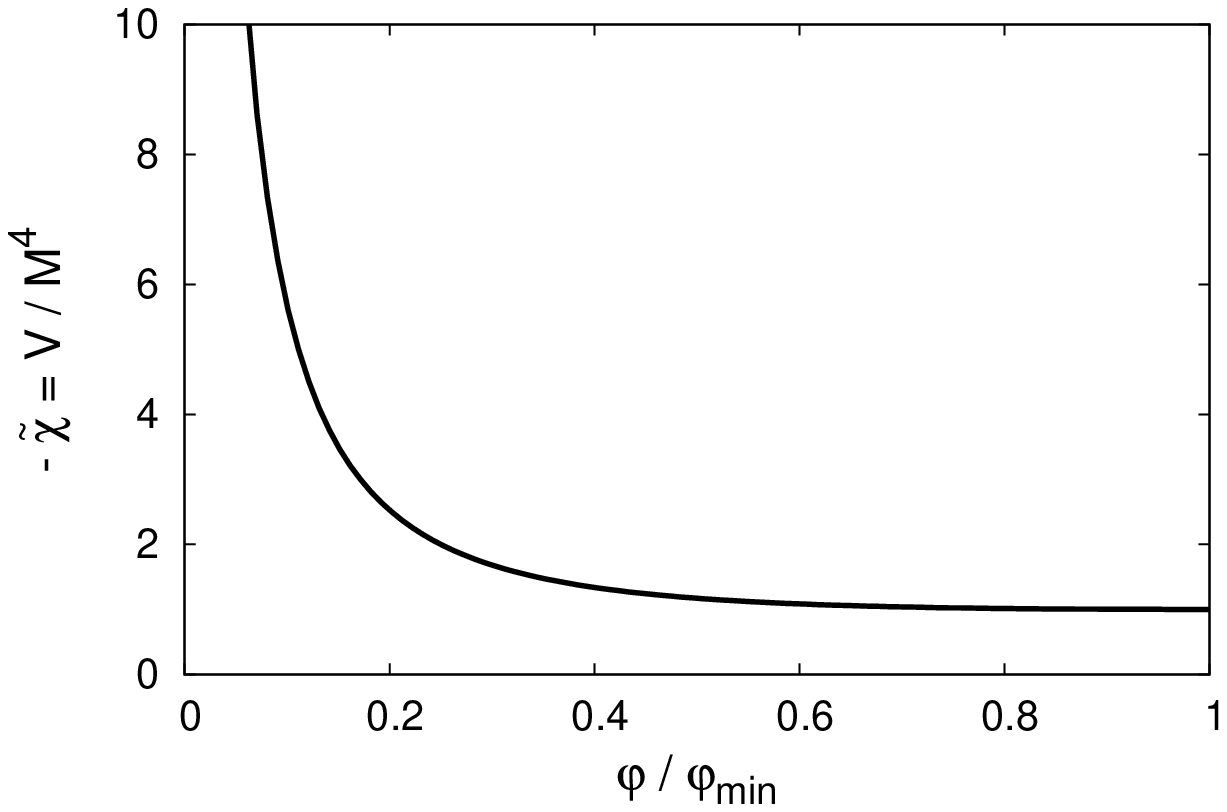}}
\epsfxsize=5.8 cm \epsfysize=5.5 cm {\epsfbox{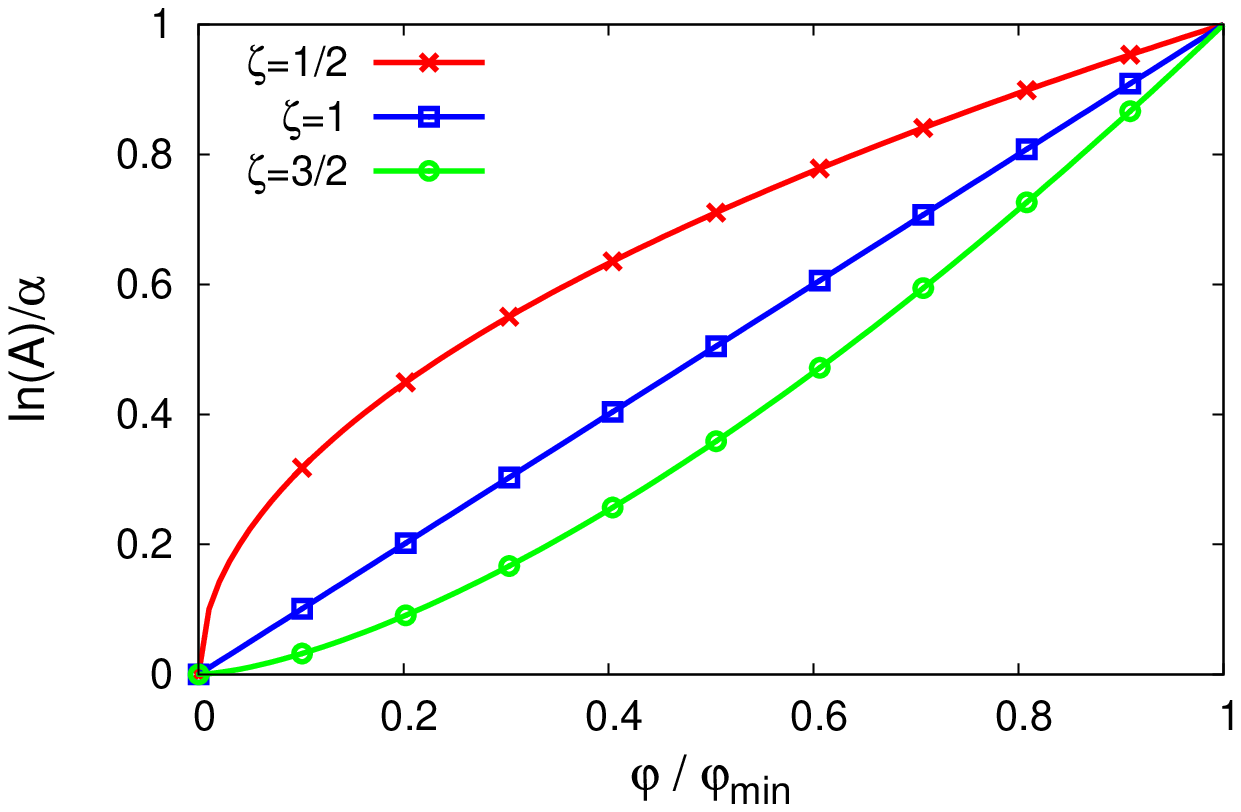}}
\epsfxsize=5.8 cm \epsfysize=5.5 cm {\epsfbox{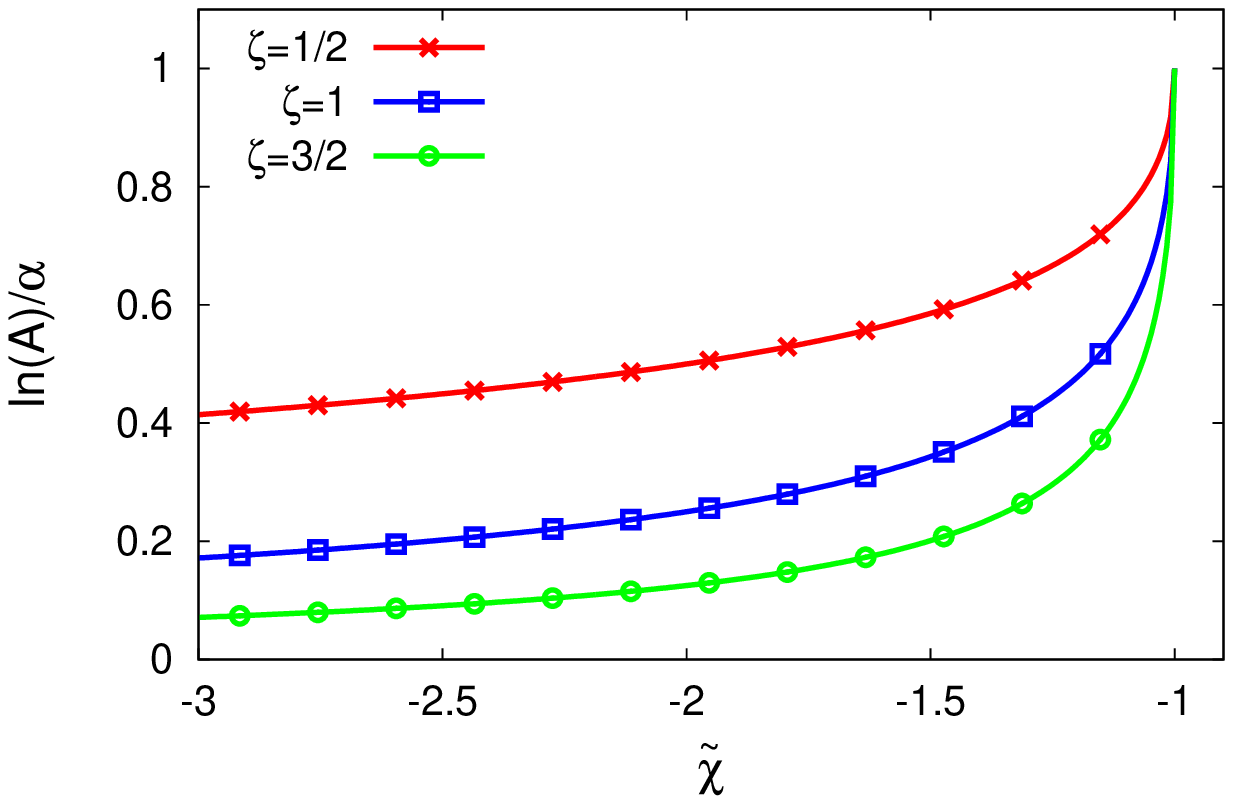}}
\end{center}
\caption{
{\it Left panel:} ultra-local scalar field or chameleon potential, $-\tilde\chi=V/{\cal M}^4$,
as a function of the chameleon scalar field $\varphi / \varphi_{\rm min}$,
as in Eq.(\ref{chi-phi-rel}) for $\gamma=2, n=2$.
{\it Middle panel:} coupling function $\ln A(\varphi)$ as a function of the chameleon scalar
field from Eq.(\ref{eq:x}), with $\gamma=2$, $\sigma=1,2,3$, which corresponds to
$\zeta=1/2,1,3/2$ with $n=2$.
{\it Right panel:} coupling function $\ln A(\tilde\chi)$ as a function of the ultra-local scalar
field $\tilde\chi$ from Eq.(\ref{A-def}), for $\zeta=1/2, 1,3/2$.
}
\label{fig:A-phi}
\end{figure*}

As the total variation of $A(\tilde\chi)$ is bounded by $\alpha \lesssim 10^{-6}$,
we can approximate Eq.\eqref{A-chi} as
\be
\ln A(\tilde\chi) = \alpha \left( 1 + \sqrt{-1-\tilde\chi} \right)^{-2\zeta} , \;\; \zeta>0 ,
\label{A-def}
\ee
where we defined $\zeta= \sigma/n$.
Equation (\ref{A-def}) fully defines the ultra-local model that corresponds to the
super-chameleon models considered in this paper.
For the numerical applications below we take $\alpha = 10^{-6}$ and $\zeta$ among
$\{1/2,1,3/2\}$. The first two choices can be obtained with $\sigma=1$ and $\sigma=2$
for the explicit super-chameleon model described in section~\ref{sec:examples}
with $\gamma=n=2$. The choice $\zeta=3/2$ requires a model with $\gamma \geq 7/3$
or corresponds to a model with $\gamma<7/3$ where the field $\varphi$ has not yet
reached the quasi-static equilibrium (\ref{eq:phimineq}) at very high redshift
(which is not very important as the dark energy and the fifth force do not play
a significant role at high redshifts far in the radiation era).

Using Eq.\eqref{A-def}, the equation for the evolution of the scalar field
\eqref{KG-pert-J} becomes
\be
\frac{\rho}{\rho_{\alpha}} = \frac{1}{\zeta} \sqrt{-1-\tilde\chi}
\left(1 + \sqrt{-1-\tilde\chi} \right)^{2\zeta + 1} ,
\label{KG-static}
\ee
where we introduced
\beq
\rho_{\alpha} = \frac{{\cal M}^4}{\alpha} = \frac{\bar\rho_{\rm de0}}{\alpha} .
\label{rho-alpha-def}
\eeq
This explicitly shows that, because of the small parameter $\alpha$, such models
introduce a second density scale $\rho_{\alpha} \gtrsim 10^6 \bar\rho_{\rm de0}$
in addition to the current dark energy density $\bar\rho_{\rm de0}$.

Eq.\eqref{KG-static} can be used to express $\tilde\chi$ as a function of the density in
the high- and low-density limits,
\be
\rho \gg \rho_{\alpha} : \;\;\; \tilde\chi(\rho) \sim
- \left( \frac{\zeta\rho}{\rho_{\alpha}} \right)^{1/(1+\zeta)} ,
\label{chi-rho-high}
\ee
\be
\rho \ll \rho_{\alpha} : \;\;\; \tilde\chi(\rho) \simeq
-1 - \left( \frac{\zeta\rho}{\rho_{\alpha}} \right)^2 .
\label{chi-rho-low}
\ee
At the background level, we switch from the high-density regime (\ref{chi-rho-high})
to the low-density regime (\ref{chi-rho-low}) at the redshift $z_{\alpha}$, with
\be
a_{\alpha} = \alpha^{1/3} \lesssim 0.01 , \;\;
z_{\alpha} = \alpha^{-1/3} \gtrsim 100 , \;\; \bar\rho(z_{\alpha}) = \rho_{\alpha} .
\label{zalpha-def}
\ee
Thus, together with the density scale $\rho_{\alpha}$ these ultra-local models also
select a particular redshift $z_{\alpha} \gtrsim 100$.
This is the redshift where the fifth force effects are the strongest, in terms of the
formation of cosmological structures, even though at the background level the scalar
field energy density only becomes dominant at low $z$ as a dark energy contribution.
Up to factors of order unity, the density $\rho_{\alpha}$ and redshift $z_{\alpha}$
also correspond to the density $\rho_{\infty}$ and redshift $z_{\infty}$ introduced
in Eq.(\ref{eq:rhoinf}), where the super-chameleon field $\varphi$ reaches the
supersymmetric minimum $\varphi_{\rm min}$
(we chose $\Lambda^4=\bar\rho_{\rm de0}$).
Thus, within this supersymmetric setting the density and redshift
$(\rho_{\alpha},z_{\alpha})$ obtain an additional physical meaning.

From Eqs.(\ref{chi-rho-high}) and (\ref{chi-rho-low}) we also obtain the behavior of
the coupling function $\ln A(\rho)$ in terms of the matter density,
\be
\rho \gg \rho_{\alpha} : \;\;\; \ln A(\rho) \sim \alpha
\left( \frac{\zeta\rho}{\rho_{\alpha}} \right)^{-\zeta/(1+\zeta)}  ,
\label{dlnAdlnrho-high-density}
\ee
\be
\rho \ll \rho_{\alpha} : \;\;\; \ln A(\rho) \simeq \alpha
\left( 1 - 2 \zeta^2 \frac{\rho}{\rho_{\alpha}} \right) .
\label{dlnAdlnrho-low-density}
\ee
As shown in the companion paper, the derived function $\ln A(\rho)$ is particularly
important when applied to static configurations and can be used to probe the existence
of a screening mechanism for this theory as we will show in sec.\ref{sec:screening-clusters}.

We show in Fig.~\ref{fig:A-phi} the characteristic functions that define the super-chameleon
models and the associated ultra-local models, for the choice of chameleon exponents
$\gamma=2, \omega=1, n=2$ for the K\"{a}hler potential $K$ and the superpotential $W$,
and $\sigma=1,2,3$ for the interaction potential $W_{\rm int}$.
This gives $\zeta=1/2,1,3/2$ for the ultra-local coupling function $\ln A(\tilde\chi)$.
The left panel shows the normalized chameleon potential $V/{\cal M}^4$, which is also
equal to the opposite of the ultra-local field $\tilde\chi$ from Eq.(\ref{V-chi-def}).
It is identical for the three models that we consider in the numerical computations
presented in this paper.
The middle panel shows the chameleon coupling function $\ln A(\varphi)$ for the
three choices for the exponent $\sigma$.
The right panel shows the ultra-local coupling function $\ln A(\tilde\chi)$ for the
corresponding three choices of the exponent $\zeta$.
In terms of the ultra-local model, or for the dynamics of cosmological perturbation
in the chameleon model over scales $H \ll k/a \ll m_{\rm eff}$, this function
$\ln A(\tilde\chi)$ fully defines the system.

In the right panel of Fig.~\ref{fig:A-phi} we show the coupling function $\ln A$ as a function
of the normalized scalar field $\varphi$ for different values of the parameter $\zeta$.
For all the models we have $|\bar{A}-1| \lesssim 10^{-6} \ll 1$ which means that we recover
the $\Lambda$-CDM cosmology at the background level to a $10^{-6}$ accuracy: in particular
as we increase $\zeta$ the coupling function becomes steeper making the effect of the presence
of the scalar field on the growth of structure more relevant, as we will demonstrate in
section~\ref{sec:cosmo-per}.

\subsection{Cosmological background and perturbations}
\label{sec:cosmo-per}

\begin{figure}
\begin{center}
\epsfxsize=8. cm \epsfysize=6 cm {\epsfbox{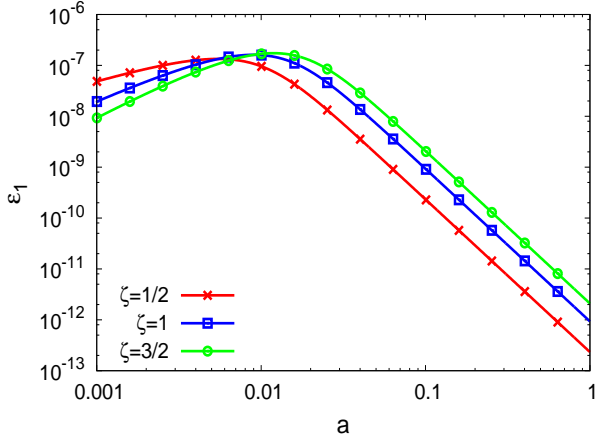}}
\end{center}
\caption{
Time evolution of the factor $\epsilon_1(a)$ as a function of the scale factor for
$\zeta=1/2,1,3/2$.
}
\label{fig:eps1}
\end{figure}

\begin{figure*}
\begin{center}
\epsfxsize=5.8 cm \epsfysize=5.5 cm {\epsfbox{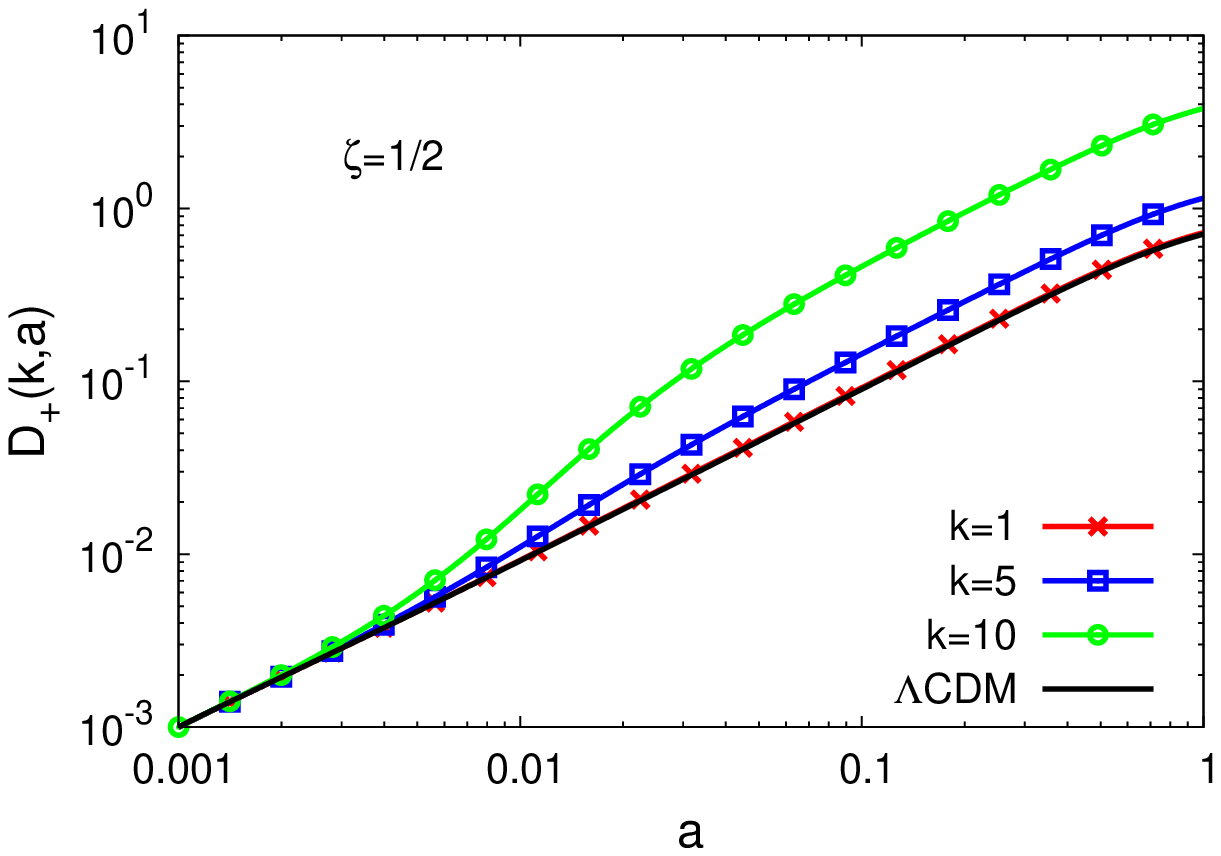}}
\epsfxsize=5.8 cm \epsfysize=5.5 cm {\epsfbox{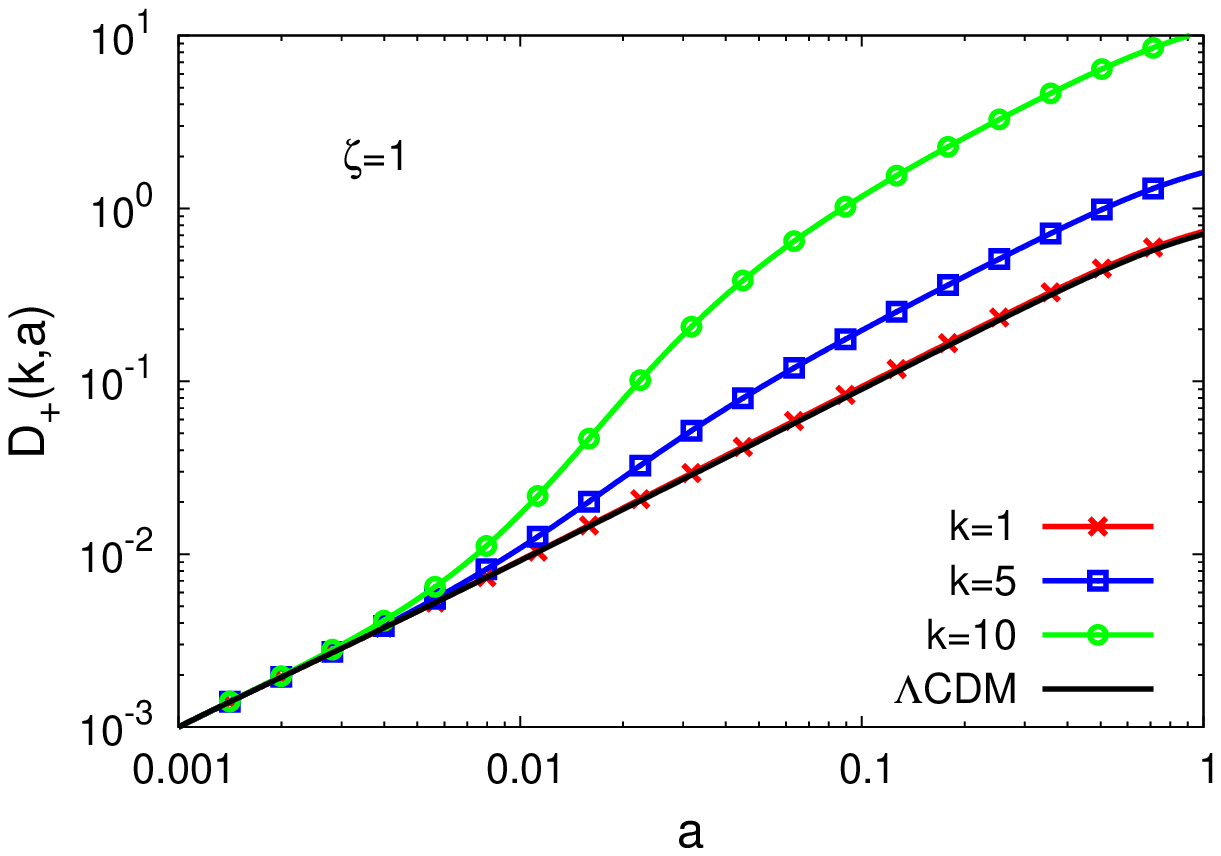}}
\epsfxsize=5.8 cm \epsfysize=5.5 cm {\epsfbox{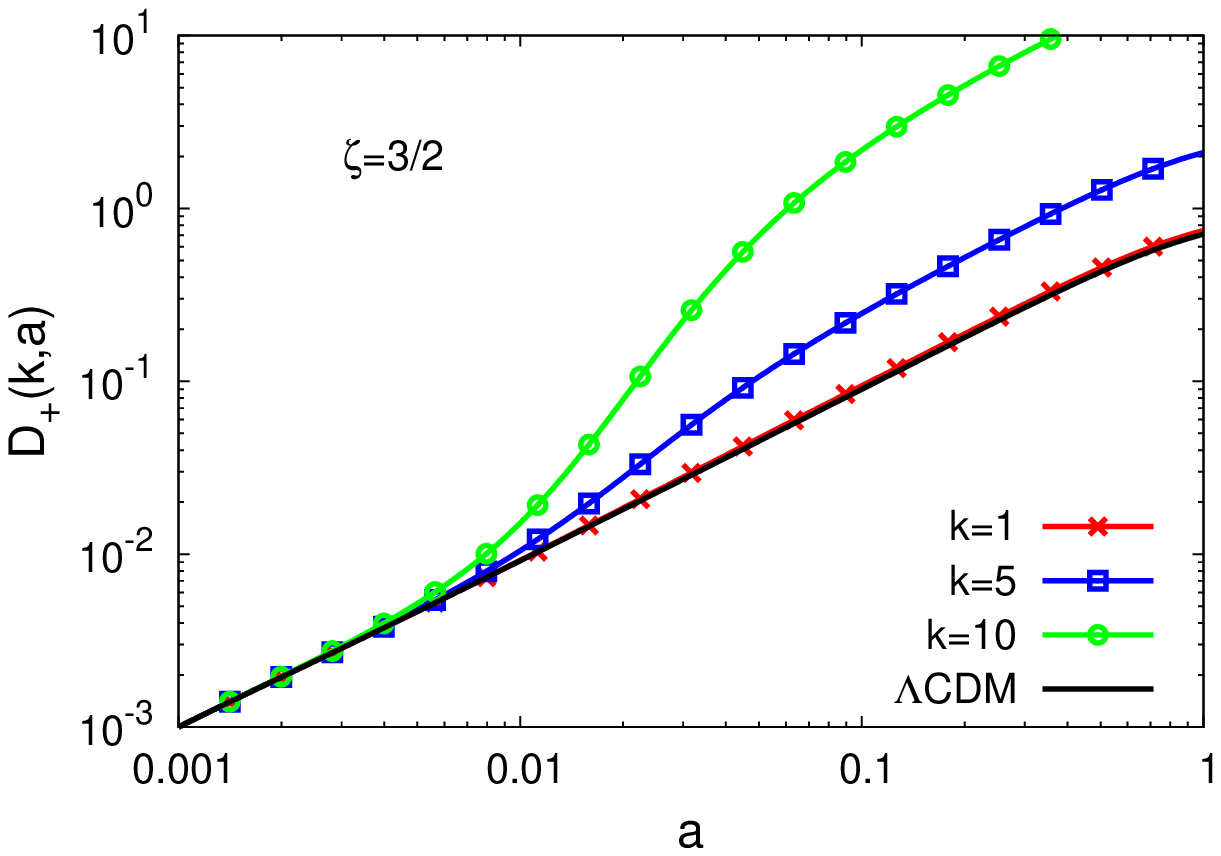}}
\end{center}
\caption{
Linear growing mode $D_+(k,a)$ for the models defined by Eq.\eqref{A-def}, as a function
of the scale factor for $k= 1, 5$ and $10 \, h {\rm Mpc}^{-1}$, and for the $\Lambda$-CDM
cosmology. We consider the cases $\zeta=1/2,1$ and $3/2$ (respectively left, center and
right panel).
}
\label{fig:Dlin_SUSY}
\end{figure*}

\begin{figure*}
\begin{center}
\epsfxsize=5.8 cm \epsfysize=5.5 cm {\epsfbox{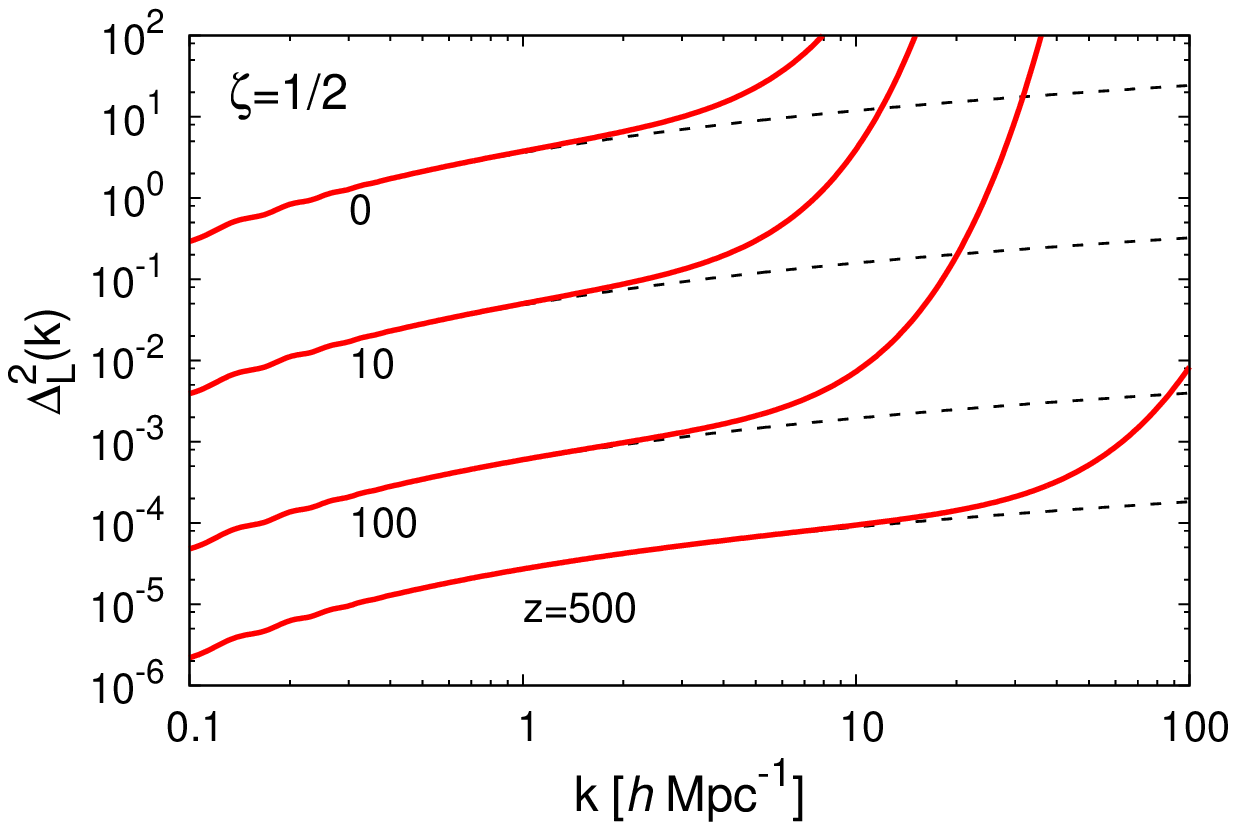}}
\epsfxsize=5.8 cm \epsfysize=5.5 cm {\epsfbox{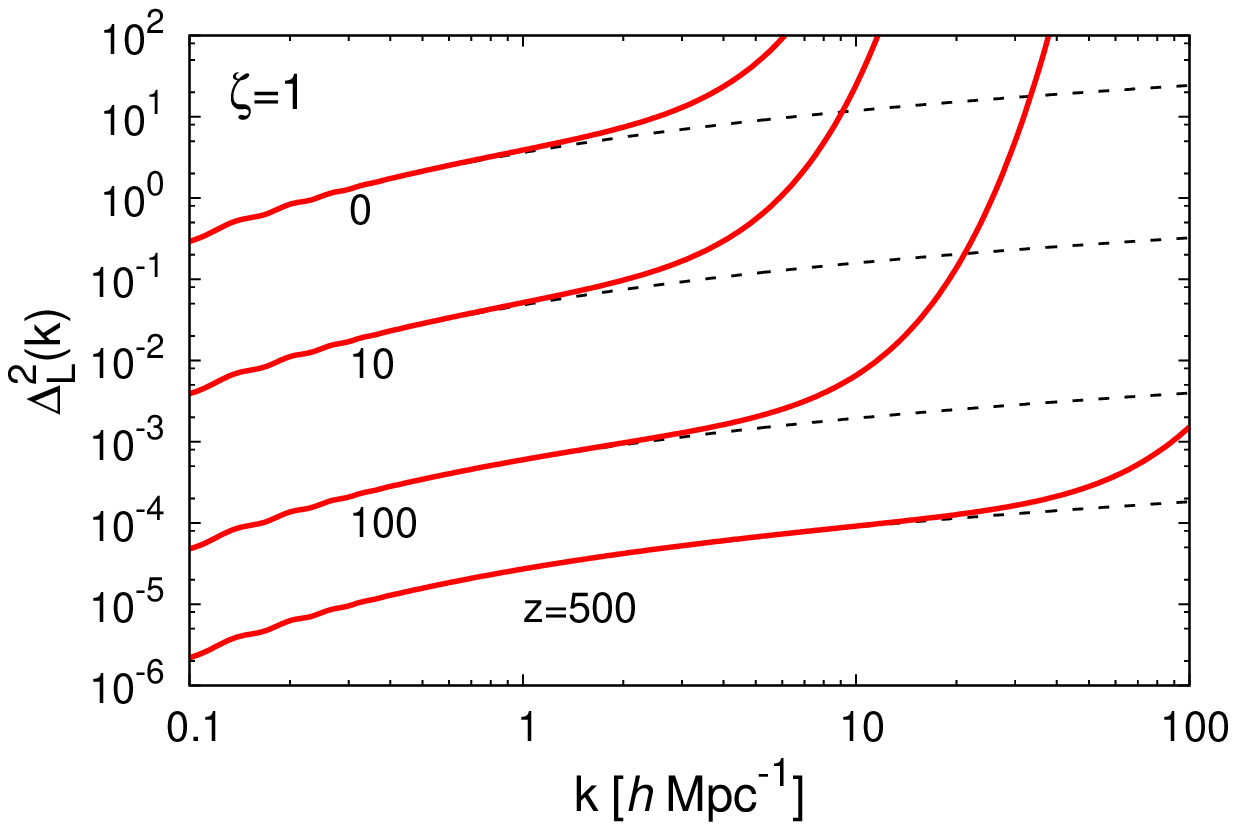}}
\epsfxsize=5.8 cm \epsfysize=5.5 cm {\epsfbox{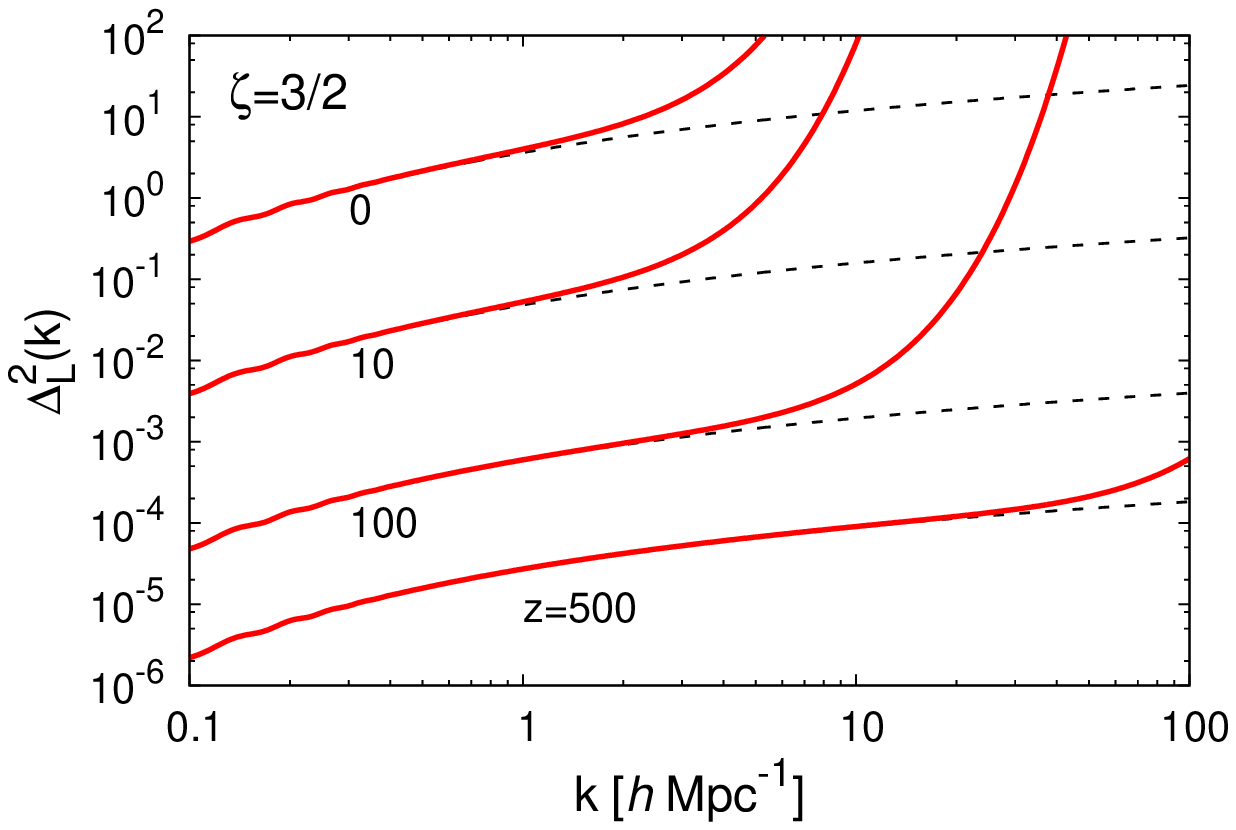}}
\end{center}
\caption{
Logarithmic linear power spectra $\Delta_L^2(k,z)$ at redshifts $z=0, 10, 100$ and $500$
(from top to bottom) at fixed $\zeta=1/2,1$ and $3/2$ (respectively left, center and
right panel).
}
\label{fig:Pk_SUSY}
\end{figure*}

For all the models we have $|\bar{A}-1| \lesssim 10^{-6} \ll 1$, which means that we recover
the $\Lambda$-CDM cosmology at the background level to a $10^{-6}$ accuracy.
Therefore, to distinguish such models from the $\Lambda$-CDM scenario we must
consider the dynamics of cosmological perturbations.
As we can see from Eq.\eqref{pertlin-ultra-local}, the linear growth $D_+(k,t)$ of the dark
matter density contrast is modified with respect to the $\Lambda$-CDM case only by the
presence of the factor $\epsilon(k,t)$, which for the models presented in the previous
sections is equal to
\be
\epsilon_1 = 2 \alpha \zeta \frac{\sqrt{-1-\tilde\chi}
\left( 1 + \sqrt{-1-\tilde\chi} \right)^{-2\zeta}} {1+ 2 ( \zeta  + 1 ) \sqrt{-1-\tilde\chi}} ,
\label{eps1-chi}
\ee
where we used the definition (\ref{eps1-def}).
From Eq.\eqref{chi-rho-high} and \eqref{chi-rho-low} we have the following simplified
expressions for $\epsilon_1$ as function of the density
\be
\rho \gg \rho_{\alpha} : \;\; \epsilon_1(\rho) \sim \frac{\alpha\zeta}{1+\zeta}
\left( \frac{\zeta\rho}{\rho_{\alpha}} \right)^{-\zeta/(1+\zeta)} ,
\label{eps1-high-rho-squared}
\ee
\be
\rho \ll \rho_{\alpha} : \;\; \epsilon_1(\rho) \sim 2 \alpha \zeta^2 \frac{\rho}{\rho_{\alpha}} .
\label{eps1-low-rho-costant}
\ee
This explicitly shows that $\epsilon_1$ decreases both at high and low densities and
peaks around $\rho_{\alpha}$.
This also gives the evolution of $\epsilon_1(t)$ as a function of the scale factor $a(t)$
using $\bar\rho=\bar\rho_0 a^{-3}$,
\be
a \ll a_{\alpha} = \alpha^{1/3} : \;\; \epsilon_1(a) \sim \alpha
\left( \frac{a}{a_{\alpha}} \right)^{3\zeta/(1+\zeta)} ,
\label{eps1-low-a}
\ee
\be
a \gg a_{\alpha} = \alpha^{1/3} : \;\; \epsilon_1(a) \sim \alpha
\left( \frac{a}{a_{\alpha}} \right)^{-3} ,
\label{eps1-high-a}
\ee
which peaks at the scale factor $a_{\alpha}$ that corresponds to $\bar\rho = \rho_{\alpha}$.
In Fig.~\ref{fig:eps1} we show the evolution of $\epsilon_1$, for $\zeta=1/2,1,3/2$,
as a function of the scale factor. It is always positive for these models leading to an
amplification of the Newtonian gravity. We can check that $\epsilon_1$ has a maximum
around $a_{\alpha}=\alpha^{1/3}$, which for this paper corresponds to a value of
$a_{\alpha} = 0.01$.
At low redshifts we recover the same decrease as $\epsilon_1 \propto a^{-3}$
of Eq.(\ref{eps1-high-a}), whereas at high redshift the decrease is stronger for higher
exponent $\zeta$, in agreement with Eq.(\ref{eps1-low-a}).
At its peak at $a_{\alpha}$, we have $\epsilon_1 \sim \alpha = 10^{-6}$, whereas today
we have $\epsilon_1 \sim \alpha^2=10^{-12}$.

As shown in the companion paper, the growth of structure is vastly enhanced by the presence
of the scalar field when $\epsilon(k,a) \gg 1$ in Eq.(\ref{pertlin-ultra-local}).
Because $\epsilon(k,a)$ grows as $k^2$ at high $k$, there exists a time dependent scale
$k_{\alpha}(a)$ such that for any scale smaller than the latter $D_+(k,a)$
deviates significantly from the $\Lambda$-CDM one.
This threshold $k_{\alpha}(a)$ can be computed from the condition
$\epsilon[k_{\alpha}(a),a]=1$ in Eq.\eqref{eps-def}, to obtain
\be
k_{\alpha}(a) = \frac{a H}{ c \sqrt \epsilon_1}\sim \frac {H_0}{c {\sqrt {\epsilon_1 a}}} ,
\label{kalpha-def}
\ee
where we used $H^2 \propto a^{-3}$ in the matter era.
Because $\epsilon_1$ decreases at both high and low redshifts, with a peak at
$a_{\alpha}$, the threshold $k_{\alpha}(a)$ is minimum at the scale factor $a_{\alpha}$,
\be
k_{\alpha}^{\rm min} = k_{\alpha}(a_{\alpha}) \sim \frac{H_0}{c \, \alpha^{2/3}}
\sim 3 h {\rm Mpc}^{-1} ,
\label{kmin}
\ee
Therefore, low wave numbers $k<k_{\alpha}^{\rm min}$ are never sensitive to the fifth
force whereas high wave numbers $k>k_{\alpha}^{\rm min}$ are sensitive to the
fifth force around $a_{\alpha}$. The range of scale factors $[a_-(k),a_+(k)]$
where a wave number $k$ feels the fifth force broadens at higher $k$.
From Eq.(\ref{kalpha-def}) we obtain
\be
k > k_{\rm min} : \;\;\;
a_-(k) \sim a_{\alpha} \left( \frac{k}{k_{\rm min}} \right)^{-(2 \zeta + 2)/(4 \zeta + 1)} ,
\label{amin-def}
\ee
\be
a_+(k)  \sim a_{\alpha} \frac{k}{k_{\rm min}} .
\label{amax-def}
\ee

In Fig.~\ref{fig:Dlin_SUSY} we show the evolution of the linear growing mode $D_+(k,a)$
obtained numerically solving Eq.\eqref{pertlin-ultra-local} at three different scales,
for the models considered in this paper.
In agreement with the discussion of Eq.(\ref{kmin}) above, low wave numbers
$k<k^{\rm min}_{\alpha}$ are never sensitive to the fifth force and follow the
$\Lambda$-CDM growth.
Higher wave numbers depart from the $\Lambda$-CDM behavior around
$a_{\alpha} \sim 0.01$ and show a faster growth over a limited time range
$[a_-,a_+]$, resuming the $\Lambda$-CDM growth at later times.
This transient speed-up increases with $k$.
This effect becomes stronger at higher $\zeta$ because of the higher amplitude of
$\epsilon_1$ found in Fig.~\ref{fig:eps1}.

The presence of the scalar field leads to a very steep increase of $D_+(k,a)$ at
$k \gg 1  \, h$ $\textrm{Mpc}^{-1}$ and so these scales enter the nonlinear regime
much earlier than in the $\Lambda$-CDM cosmology, at $z \sim z_{\alpha}$.
This can be seen in Fig.~\ref{fig:Pk_SUSY} where we plot the logarithmic linear
power spectrum $\Delta_L^2(k,z) = 4\pi k^3 P_L(k,z)$.

\subsection{Spherical collapse}
\label{sec:Spherical-collapse}

\begin{figure*}
\begin{center}
\epsfxsize=5.8 cm \epsfysize=5.5 cm {\epsfbox{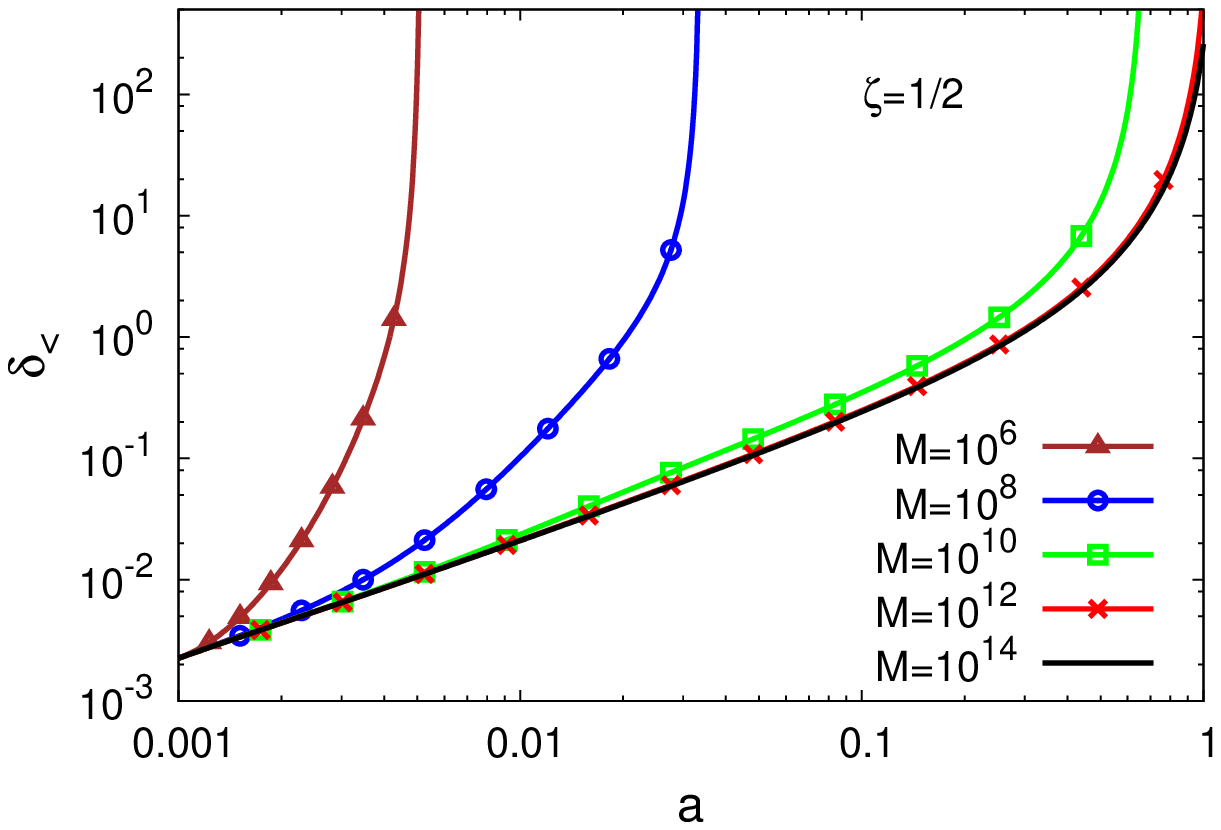}}
\epsfxsize=5.8 cm \epsfysize=5.5 cm {\epsfbox{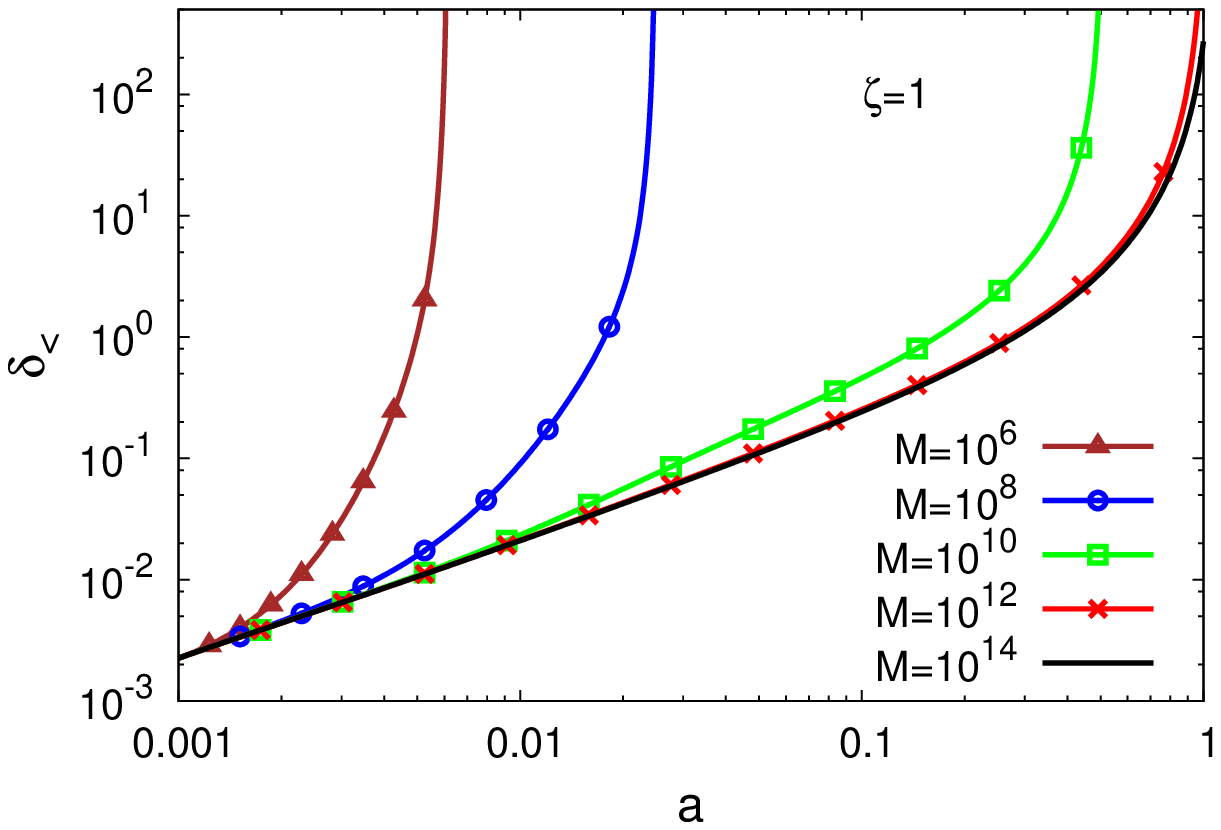}}
\epsfxsize=5.8 cm \epsfysize=5.5 cm {\epsfbox{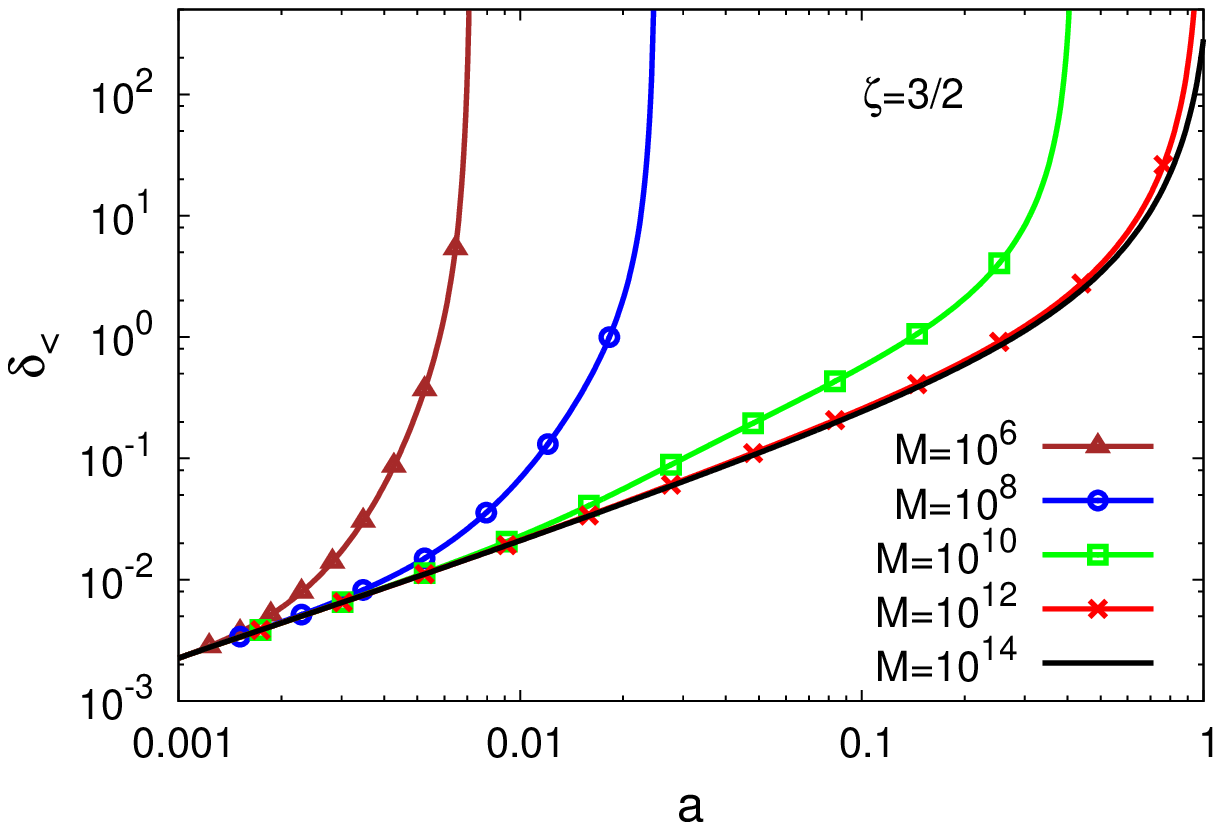}}
\end{center}
\caption{
Time evolution of the nonlinear density contrast $\delta(<r)$ given by the spherical
dynamics, as a function of the scale factor, for several masses (in units of
$h^{-1} M_{\odot}$) at fixed $\zeta=1/2,1,3/2$ (respectively left, center and right panel).
The initial condition corresponds to the same linear density contrast
$\delta_L^{\Lambda-{\rm CDM}}=1.6$ today, using the $\Lambda$-CDM growth factor.
}
\label{fig:delta_nl_masses}
\end{figure*}

\begin{figure*}
\begin{center}
\epsfxsize=5.8 cm \epsfysize=5.5 cm {\epsfbox{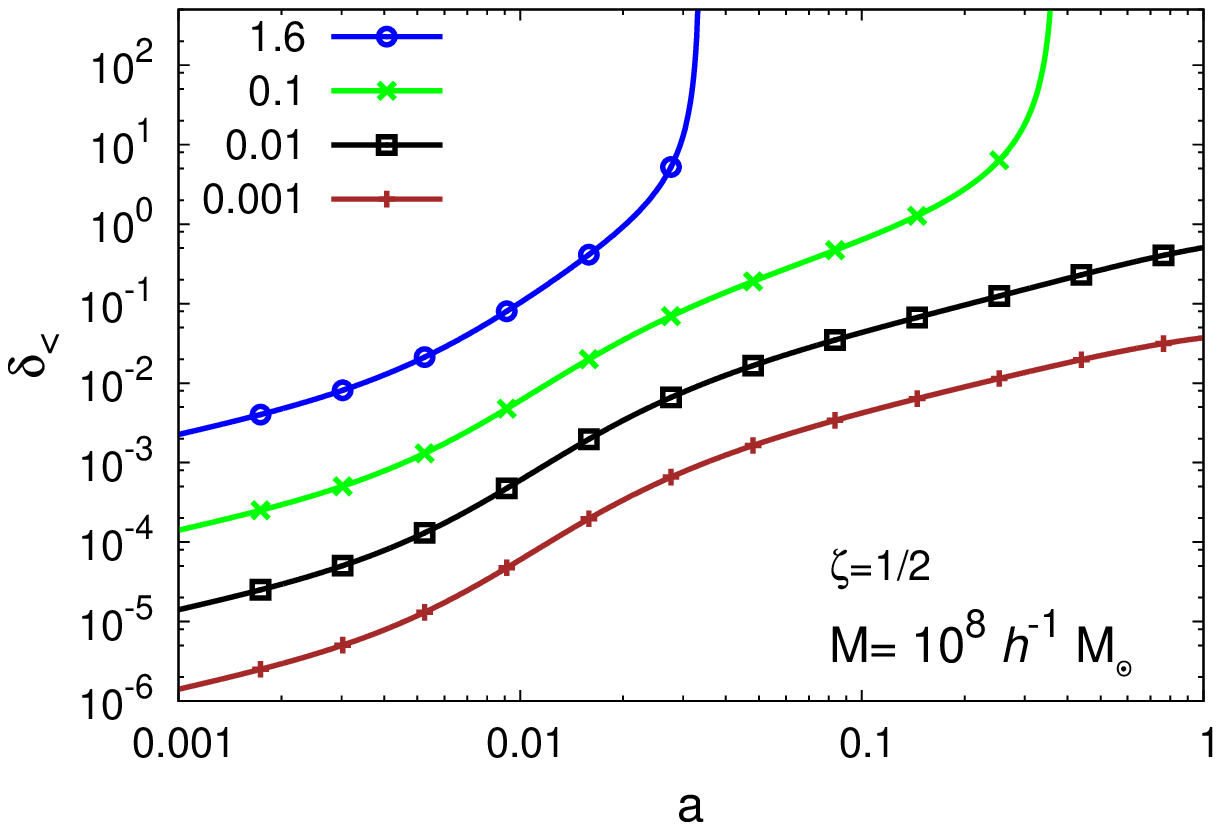}}
\epsfxsize=5.8 cm \epsfysize=5.5 cm {\epsfbox{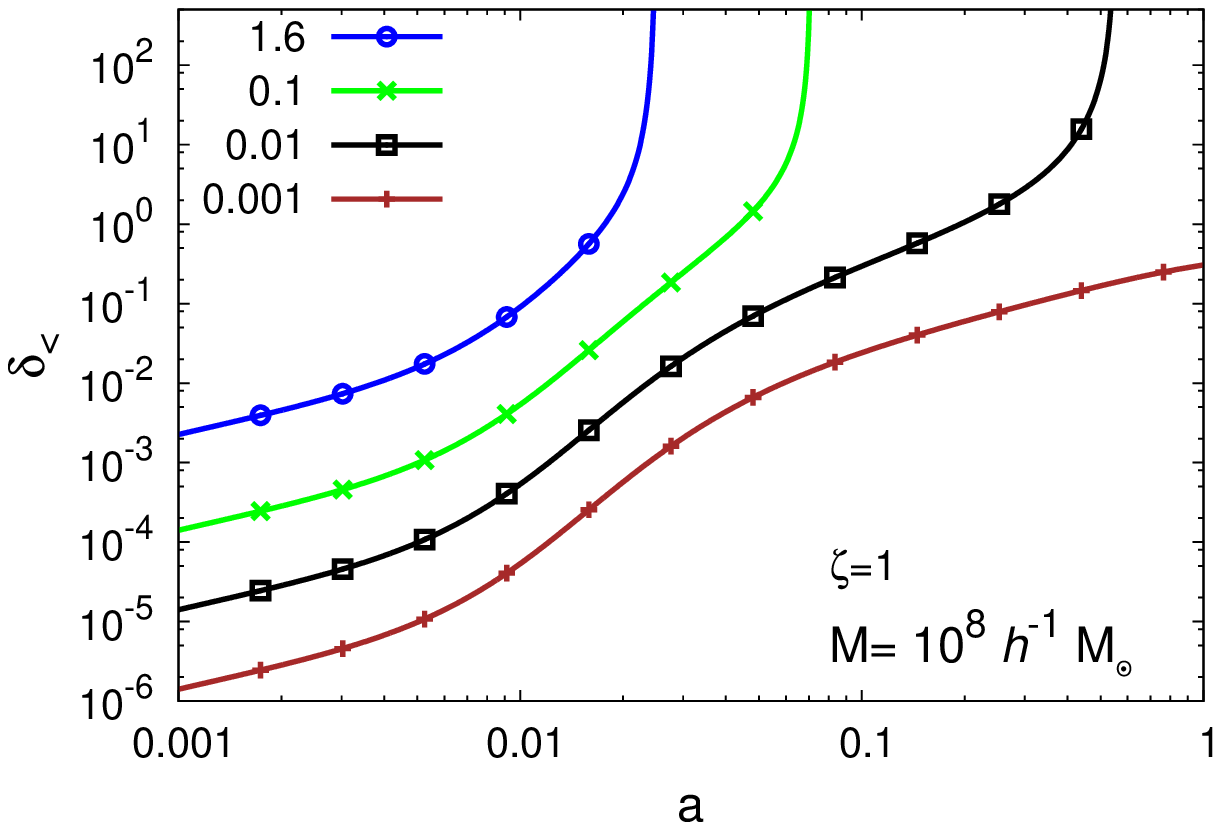}}
\epsfxsize=5.8 cm \epsfysize=5.5 cm {\epsfbox{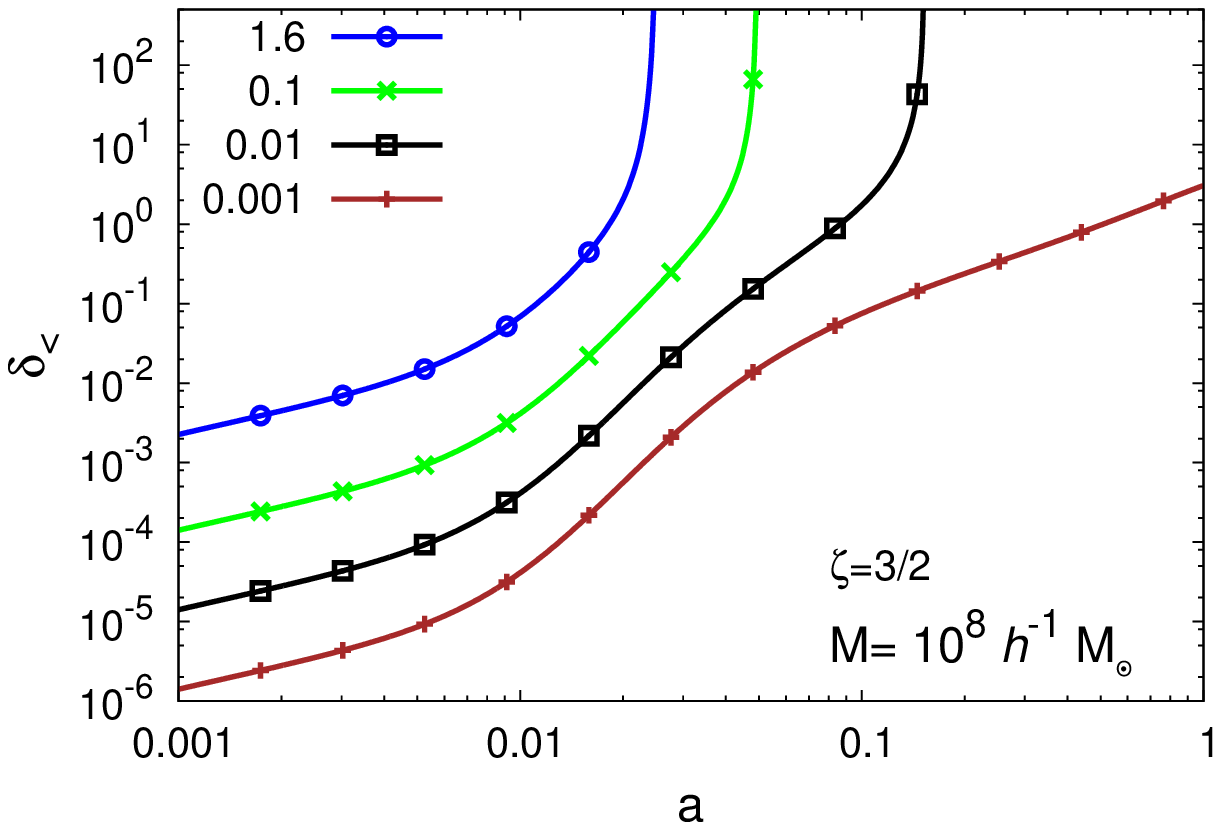}}
\end{center}
\caption{
Time evolution of the nonlinear density contrast $\delta(<r)$ given by the spherical dynamics,
as a function of the scale factor, for several values of the initial density contrast and for
a mass of $M=10^{8} h^{-1} M_{\odot}$ at fixed $\zeta=1/2,1,3/2$(respectively left, center
and right panel).
We show our results for different initial conditions, which correspond to
$\delta_L^{\Lambda-{\rm CDM}}=1.6, 0.1, 0.01$ and $0.001$ from top to bottom.
}
\label{fig:delta_nl_deltai}
\end{figure*}

On large scales where the baryonic pressure is negligible, the particle trajectories
$\vr(t)$ follow the equation of motion
\beq
\frac{d^2 \vr}{d t^2}  - \frac{1}{a} \frac{d^2 a}{d t^2} \vr =
- \nabla_{\vr} \left( \Psi_{\rm N} + \Psi_A \right) ,
\label{trajectory-Jordan}
\eeq
where $\vr=a\vx$ is the physical coordinate, $\Psi_{\rm N}$ the Newtonian potential
and $\Psi_A = c^2 \ln A$ the fifth-force potential.
To study the spherical collapse before shell crossing, it is convenient to label each shell
by its Lagrangian radius $q$ or enclosed mass $M$, and to introduce its
normalized radius $y(t)$ by
\beq
y(t) = \frac{r(t)}{a(t) q} \;\;\; \mbox{with} \;\;\;
q = \left(\frac{3M}{4\pi\bar\rho_0}\right)^{1/3} , \;\;\; y(t=0) = 1 .
\label{y-def-Jordan}
\eeq
In particular, the matter density contrast within radius $r(t)$ reads as
\beq
1+ \delta_{<}(r) = y(t)^{-3} .
\label{deltaR-def}
\eeq
The equation of motion becomes
\beqa
&& \frac{d^2 y}{d(\ln a)^2} + \left( 2+\frac{1}{H^2} \frac{d H}{d t} \right)
\frac{d y}{d\ln a} + \frac{\Omega_{\rm m}}{2} y (y^{-3} - 1) = \nonumber \\
&& - y \left( \frac{c}{Hr} \right)^2 \frac{d\ln A}{d\ln\rho}
\frac{r}{1+\delta} \frac{\partial\delta}{\partial r} .
\label{y-lna-1}
\eeqa
The fifth force introduces a coupling as it depends on the density profile, through the local
density $\rho(r) = \bar\rho (1+\delta(r))$.

In the following, we use the density profile defined by
\beqa
\lefteqn{ \delta(x') = \frac{\delta_{<}(x)}{\sigma^2_{x}} \int_V \frac{d \vx''}{V} \,
\xi_L(\vx',\vx'') } \nonumber \\
&& = \frac{\delta_{<}(x)}{\sigma^2_{x}} \int^{+\infty}_{0} \frac{d k}{k} \,
\Delta^2_{L}(k) \tilde{W}(k x) \frac{\sin(kx')}{kx'} .  \;\;\;
\label{density-profile-ansatz}
\eeqa
Here $x(t)=a(t) r(t)$ is the comoving radius of the spherical shell of mass $M$ that we are
interested in while $x'$ is any radius along the profile; $\xi_L$ and $\Delta^2_L$ are
the linear correlation function and logarithmic power spectrum of the matter density
contrast, $\sigma^2_{x}=\langle \delta_{L<}(x)^2\rangle$ its variance within radius $x$,
which defines a sphere of volume $V$; and
$\tilde{W}(k x) =  3 [ \sin(kx) - kx \, \cos(kx) ]/(kx)^3$ the Fourier transform of the 3D top hat
of radius $x$.
The profile (\ref{density-profile-ansatz}) is the typical profile around a density fluctuation
at scale $x$ in the initial Gaussian field and provides a convenient ansatz (here we use the
initial linear power spectrum or its $\Lambda$-CDM amplified value at the redshift of interest). 

We show in Fig.~\ref{fig:delta_nl_masses} the time evolution of the nonlinear density contrast
$\delta_<(r)$ within a shell of mass $M$, given by the spherical dynamics (\ref{y-lna-1}),
for different values of the mass $M$, fixing the initial linear density contrast
so that $\delta_L^{\Lambda-{\rm CDM}}=1.6$ today (the initial condition is set at high
redshift before the onset of the fifth force and it is common to all models and the
$\Lambda$-CDM cosmology; as usual it is convenient to describe this initial condition
by its value today using the $\Lambda$-CDM linear growth factor).
In agreement with what we found by studying the evolution of linear perturbations,
we can see that at large masses, $M \gtrsim 10^{12} h^{-1} M_{\odot}$, the evolution
of $\delta_<(r)$ closely follows the $\Lambda$-CDM one, whereas the collapse of small
masses is strongly accelerated around $a_{\alpha}$. This faster growth occurs earlier
for smaller mass, as $a_-(k)$ decreases on smaller scales.

We show in Fig.~\ref{fig:delta_nl_deltai} the spherical dynamics for a fixed value of the mass
$M=10^8 h^{-1} M_{\odot}$ and several initial density contrasts. The acceleration of
the growth of structure due to the presence of the scalar field makes halos collapse before
$a=1$, even starting from $\delta_L^{\Lambda-{\rm CDM}} \simeq 0.1$.
In agreement with previous figures, the acceleration of the collapse occurs around
$a_{\alpha}$. For sufficiently high initial conditions this leads to a collapse at high redshift
around $z_{\alpha}$. For lower initial conditions the dynamics is still in the linear regime
after the fifth force has vanished, at low redshift, but with a higher amplitude than in the
$\Lambda$-CDM cosmology and a higher final collapse redshift.
Again, we can see that the effect of the fifth force increases with $\zeta$.

We show in the upper panel of  Fig.~\ref{fig:halo-mass} the linear density contrast threshold,
measured by $\delta_{L}^{\Lambda \rm -CDM}$ (i.e., the extrapolation up to $z=0$ of the
linear initial density contrast by the $\Lambda$-CDM growth rate), required to reach a
nonlinear density contrast $\delta_{<}=200$ today.
In  agreement with Figs.~\ref{fig:delta_nl_masses} and \ref{fig:delta_nl_deltai},
at large mass we recover the $\Lambda$-CDM linear density threshold,
$\delta_{L}^{\Lambda \rm -CDM} \simeq 1.6$, whereas at small mass we obtain a much
smaller linear density threshold, $\delta_{L}^{\Lambda \rm -CDM} \ll 1$, because of the
acceleration of the collapse by the fifth force.
Again, at small masses the threshold $\delta_{L}$ becomes smaller for larger exponent
$\zeta$ as the effect of the fifth force increases.

\subsection{Halo mass function}
\label{sec:Halo-mass-function}

\begin{figure}
\begin{center}
\epsfxsize=8. cm \epsfysize=6 cm {\epsfbox{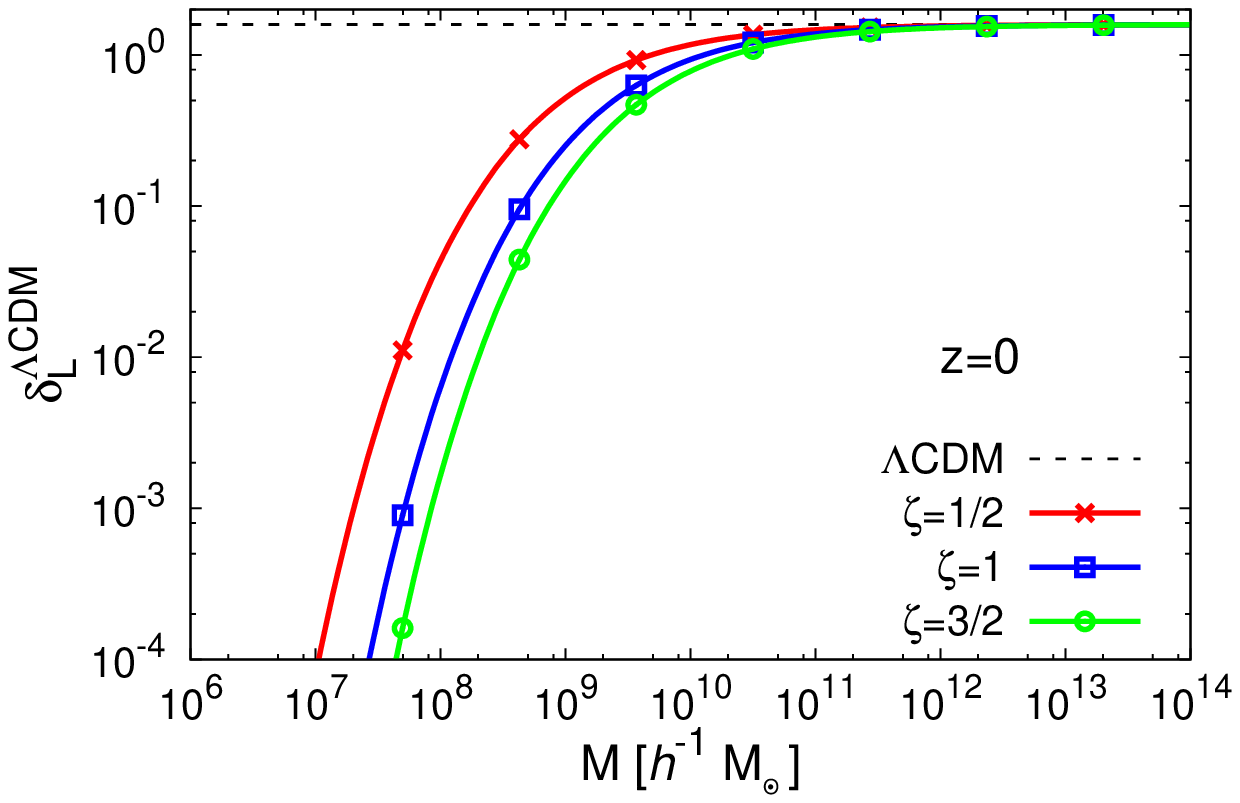}}
\epsfxsize=8. cm \epsfysize=6 cm {\epsfbox{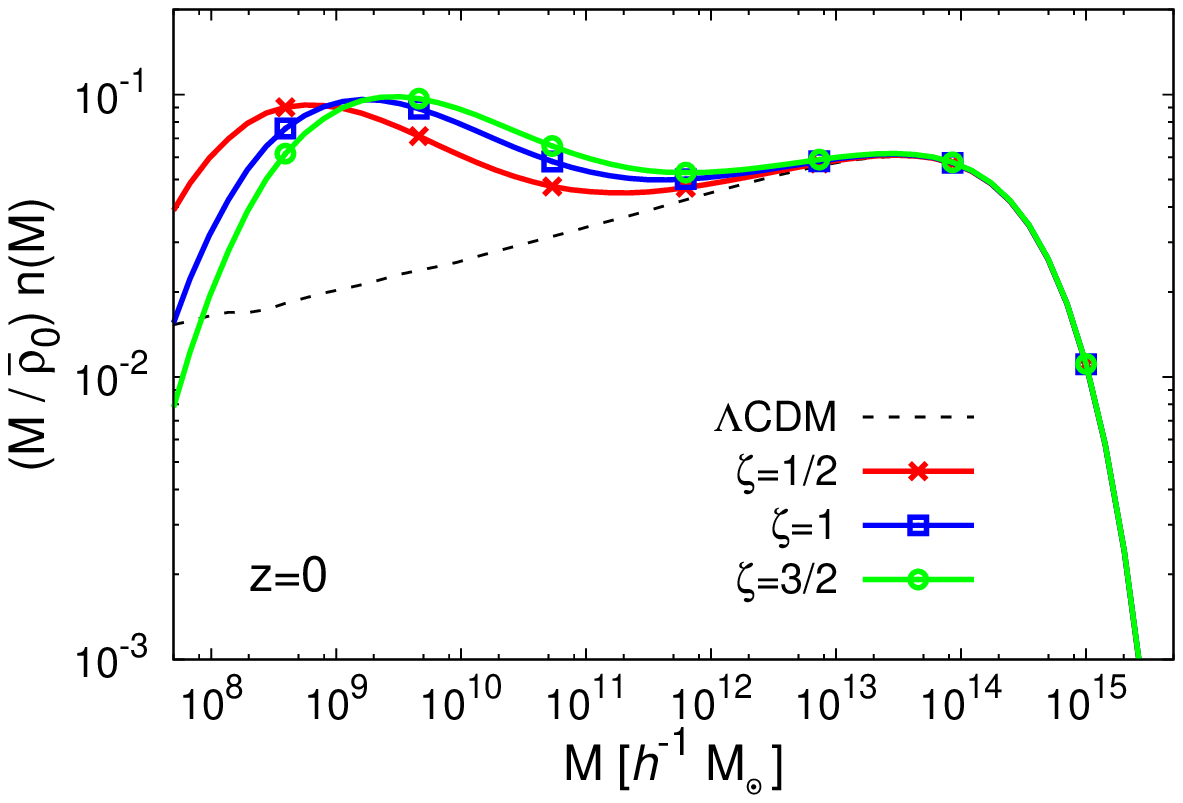}}
\end{center}
\caption{
{\it Upper panel:} Initial linear density contrast, as measured by
$\delta_{L}^{\Lambda \rm -CDM}$, that gives rise to a nonlinear density contrast
$\delta_{<}=200$ at $z=0$, as a function of the halo mass $M$ for fixed $\zeta=1/2,1$
and $3/2$.
{\it Lower panel:} Halo mass function at $z=0$ for fixed $\zeta=1/2,1$ and $3/2$, and
for the $\Lambda$CDM cosmology.
}
\label{fig:halo-mass}
\end{figure}

As for the $\Lambda$-CDM cosmology, we write the comoving halo mass function as
\beq
n(M) \frac{d M}{M} = \frac{\bar\rho_0}{M} f(\nu) \frac{d\nu}{\nu} ,
\label{nM-def}
\eeq
where the scaling variable $\nu(M)$ is defined as
\beq
\nu(M) = \frac{\delta_L^{\Lambda- \rm CDM}(M)}{\sigma^{\Lambda- \rm CDM}(M)} ,
\label{nu-def}
\eeq
and $\delta_L^{\Lambda- \rm CDM}(M)$ is again the initial linear density contrast
(extrapolated up to $z=0$ by the $\Lambda$-CDM linear growth factor) that is required
to build a collapsed halo (which we define here by a nonlinear density contrast of 200 with
respect to the mean density of the Universe) and $\sigma^{\Lambda-\rm CDM}$
its variance.
The variable $\nu$ measures whether such an initial condition corresponds to a  rare and
very high overdensity in the initial Gaussian field ($\nu \gg 1$) or to a typical fluctuation
($\nu \lesssim 1$).
In the Press-Schechter approach, we have $f(\nu) = \sqrt{2/\pi} \nu e^{-\nu^2/2}$.
Here we use the same function as in \cite{Press1974}.
Then, the impact of the modified gravity only arises through the linear threshold
$\delta_L^{\Lambda \rm CDM}(M)$, as we assume the same initial matter density power
spectrum as for the $\Lambda$-CDM reference at high redshift.

The threshold $\delta_L^{\Lambda- \rm CDM}(M)$ was shown in the upper panel
of Fig.~\ref{fig:halo-mass}.
We show the mass function in the lower panel of Fig.~\ref{fig:halo-mass}.
Once again, we can notice that at large mass all the mass functions are close to the
$\Lambda$-CDM prediction whereas at smaller masses,
$M \sim 10^8-10^{10} h^{-1} M_{\odot}$, they are higher.
This is because the fifth force has no effect on very large scales and accelerates
the formation of structures on small scales.
At lower mass, $M \lesssim 10^7 h^{-1} M_{\odot}$, the mass function becomes smaller
than in the $\Lambda$-CDM cosmology, because both mass functions are normalized
to unity (the sum over all halos cannot give more matter than the mean matter density).

At large masses, $M > 10^{12}  h^{-1} M_{\odot}$, where the formation of large-scale
structures remains close to the $\Lambda$-CDM case, with only a modest acceleration, and
the mass function is dominated by the Gaussian tail $\sim e^{-\nu^2/2}$, we can expect that
the results obtained are robust, since in this regime the shape of the halo mass function is
dominated by the exponential tail $e^{-\nu^2/2}$.
At low masses, $M < 10^{12}  h^{-1} M_{\odot}$, where the history of gravitational clustering
is significantly different from the $\Lambda$-CDM scenario, as a large range of masses have
collapsed together before a redshift of $100$, and the halo mass function is no longer
dominated by its universal Gaussian tail, these results are unlikely to be accurate.
Nevertheless, we can still expect the halo mass function to be significantly higher than in the
$\Lambda$-CDM case for masses $M \sim 10^8 - 10^{11}  h^{-1} M_{\odot}$,
although it is difficult to predict the maximum deviation and the transition to a negative
deviation at very low masses.

\section{Astrophysical effects}
\label{sec:halos}

\subsection{Screening within spherical halos}
\label{sec:screening-clusters}

\subsubsection{Radial profiles}
\label{sec:radial-profile-eta}

We first consider here how the ratio of the fifth force to Newtonian gravity behaves within
spherical halos with a mean density profile such as the Navarro-Frenk-White (NFW) \cite{Navarro:1996}
density profile.
In particular, we wish to find the conditions for the fifth force not to diverge at the center
of the halos and to remain modest at all radii, to be consistent with observations of
X-ray clusters.
Within spherical halos, the Newtonian force reads as
\beq
F_{\rm N} = - \frac{{\cal G}_{\rm N} M(<r)}{r^2} = - \frac{\Omega_{\rm m}}{2}
\Delta(<r) r H^2 ,
\label{FN-Delta}
\eeq
where $\Delta(<r)$ is the mean overdensity within radius $r$.
The fifth force reads
\beq
F_A = - c^2 \frac{d\ln A}{d r} = - \frac{c^2}{r} \frac{d\ln A}{d\ln\rho}
\frac{d\ln\rho}{d\ln r} .
\eeq
We can also use $F_{\rm N}$ and $F_A$ to define characteristic velocity scales,
\beq
F_{\rm N} = - \frac{v_{\rm N}^2(r)}{r} , \;\;\; F_A = - \frac{c_s^2(r)}{r} ,
\eeq
with
\beq
v_{\rm N}^2 = \frac{{\cal G}_{\rm N} M(<r)}{r} , \;\;\;
c_s^2 = c^2 \frac{d\ln A}{d\ln r} ,
\label{vN-cs-def}
\eeq
where $v_{\rm N}$ is the Newtonian circular velocity.
Therefore, the ratio of the fifth force to the Newtonian force is
\beq
\eta \equiv \frac{F_A}{F_{\rm N}} = \frac{c_s^2}{v_{\rm N}^2}
 = \frac{2}{\Omega_{\rm m} \Delta(<r)}
\left( \frac{c}{r H} \right)^{\!2} \frac{d\ln A}{d\ln\rho} \frac{d\ln\rho}{d\ln r} .
\label{eta-def-Delta}
\eeq
From Eq.(\ref{dlnAdlnrho-low-density}), we have at moderate densities,
$\rho \sim \bar\rho(z)$,
\beq
\rho \ll \rho_{\alpha} :  \;\;\; | \eta | \sim \frac{\alpha^2 \zeta^2}{a^3}
\left( \frac{c}{r H} \right)^2 .
\label{eta-low-density}
\eeq
Thus, in the late Universe  the ratio $\eta$ is suppressed by a factor $\alpha^2$ so that
$\eta$ only reaches unity at $r \sim 3 h^{-1} {\rm kpc}$,
i.e. at galaxy scales (see also Sec.~\ref{sec:galaxies} below).
At higher densities, we obtain from Eq.(\ref{dlnAdlnrho-high-density})
\beq
\rho \gg \rho_{\alpha} :  | \eta | \sim \frac{\alpha^2 \zeta}{a^3}
\left( \frac{a^3}{\alpha \zeta \Delta} \right)^{(1+2\zeta)/(1+1\zeta)}
\left( \frac{c}{r H} \right)^2 .
\label{eta-high-density}
\eeq

\begin{figure}
\begin{center}
\epsfxsize=8 cm \epsfysize=6 cm {\epsfbox{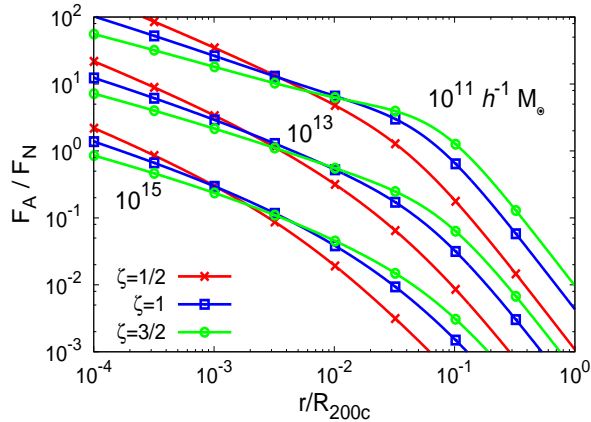}}
\end{center}
\caption{Ratio $\eta = F_{A} / F_{N}$ as a function of the radius $r$, within spherical halos
with an NFW profile. We display the cases of halo  masses $M=10^{11}, 10^{13}$ and
$10^{15} h^{-1} M_{\odot}$ from top to bottom, for the ultra-local model exponent
$\zeta=1/2,1$ and $3/2$.
}
\label{fig:ratio-forces}
\end{figure}

We plot the ratio $\eta$ for several halo masses, with an NFW
density profile in Fig.~\ref{fig:ratio-forces}.
In agreement with the results obtained in previous sections, we can see that the
fifth force is more important for smaller halos, which also correspond to smaller scales.
For a power-law density profile, of exponent $\gamma_p>0$ and
critical radius $r_{\alpha}$,
\beq
\rho(r) = \rho_{\alpha} \left( \frac{r}{r_{\alpha}} \right)^{-\gamma_p} ,
\label{gamma-p-def}
\eeq
we have
\be
r < r_{\alpha} , \;\;\; \eta \sim r^{\gamma_p ( 1+2\zeta)/(1+\zeta) -2} .
\label{eta-small-r-gammap}
\ee
If we consider halos with a mean NFW density profile,
which has $\gamma_p=1$, we find that $\eta \sim r^{-1/(1+\zeta)}$ and
the relative importance of the fifth force does not vanish at the center for the models,
whatever the value of the exponent $\zeta$, in agreement with Fig.~\ref{fig:ratio-forces}.
This suggests that these models would lead to significant modifications in the cluster dynamics
with respect to the $\Lambda$-CDM model and so would be ruled out by the observations,
which show a good agreement with the $\Lambda$-CDM cosmology.
However, as we can see from Fig.~\ref{fig:ratio-forces}, for typical cluster masses
$\eta$ only becomes of the order of unity far within the virial radius,
$r \lesssim 0.01 R_{200c}$ for $M \gtrsim 10^{13} h^{-1} M_{\odot}$.
Because at these scales clusters have significant substructures the approximation of
a smooth profile is not any more correct.
Then, deeper analyses are needed to unravel the dynamics of  clusters of galaxies
considering the ultra-local behaviour of the theory. We leave these analysis for future studies
when we may need to use numerical simulations and to estimate the observational accuracy
of the measured halo profiles. On the other hand, we will perform a thermodynamic analysis of the system in \ref{sechistory-cosmo} where we find
that for large enough clusters, the mean density approximation is valid.

\subsubsection{Clusters of galaxies}
\label{sec:clusters}

We now estimate the fifth force to Newtonian gravity ratio $\eta$ on a global scale,
for clusters and for galaxies.
In contrast with the companion paper, we do not need to study the Solar System, the Earth or
the laboratory, because within the supersymmetric setting considered in this paper
baryons do not couple to the fifth force. Therefore, astrophysical systems which are dominated
by baryons do not feel the effect of the fifth force and we automatically recover the
General Relativity or Newtonian dynamics in these systems.

We have seen in Eq.(\ref{eta-def-Delta}) that $\eta = c_s^2/v_{\rm N}^2$, whence
$\eta \sim (c/v_{\rm N})^2 | d\ln A/d\ln\rho|$ if we take $d\ln\rho/d\ln r \sim 1$.
From Eq.(\ref{dlnAdlnrho-low-density}), we also have at moderate densities below
$\rho_{\alpha} \sim 10^6 \bar\rho_0$, $d\ln A/d\ln\rho \sim - \alpha^2 \Delta$ at
redshift $z=0$. This gives
\beq
z=0 : \;\;\; \eta \sim \left( \frac{\alpha c}{v_{\rm N}} \right)^2 \; \Delta .
\eeq
For clusters of galaxies, with $\Delta \sim 10^3$ and $v_{\rm N} \sim 500$ km/s,
this yields
\beq
\mbox{clusters:} \;\;\; \eta \sim (10^4 \, \alpha)^2 \ll 1 .
\eeq
Therefore, the fifth force is negligible on cluster scales.
However, as seen in Fig.~\ref{fig:ratio-forces}, this is no longer the case far inside the cluster,
where the characteristic scales are smaller and the density greater, which gives rise to a greater fifth force.

\subsubsection{Galaxies}
\label{sec:galaxies}

We now consider a typical galaxy, such as the Milky Way, with $\Delta \sim 10^6$,
which is at the upper limit of the regime $\rho \lesssim \rho_{\alpha}$,
and $v_{\rm N} \sim 200$ km/s. This gives
\beq
\mbox{galaxies:} \;\;\; \eta \sim (10^6 \, \alpha)^2 \sim 1 .
\eeq
Thus, the fifth force is of the same order as the Newtonian gravity on galaxy scales.
This suggests that interesting phenomena could occur in this regime and that galaxies could provide a useful probe of such models, as we can see from Fig.~\ref{fig:ratio-forces}
for low-mass halos $M \lesssim 10^{11} h^{-1} M_{\odot}$.

\subsection{Fifth-force dominated regime}
\label{sec:fifth-force-regime}

\begin{figure}
\begin{center}
\epsfxsize=9 cm \epsfysize=6 cm {\epsfbox{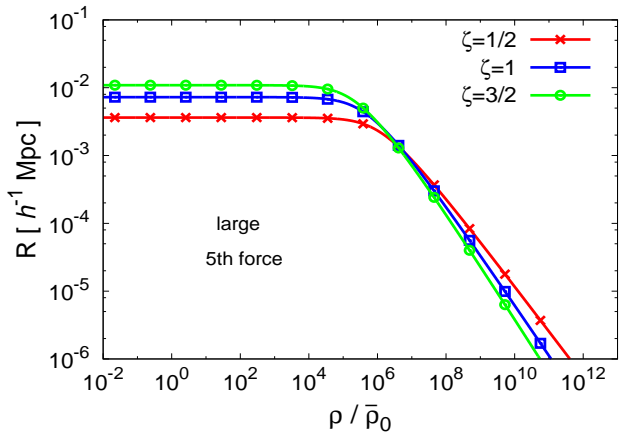}}
\end{center}
\caption{
Domain in the density-radius plane where the fifth force is greater than Newtonian gravity
(bottom left area below the curves), for the ultra-local exponents $\zeta=1/2,1$ and $3/2$.
}
\label{fig:eta_R_rho}
\end{figure}

It is useful to reformulate the analysis presented above for clusters and galaxies
and to determine the domain of length, density and mass scales where the fifth force
is dominant.
Taking $d\ln\rho/d\ln r \sim 1$, we write for structures of typical radius $R$, density $\rho$
and mass $M=4\pi\rho R^3/3$,
\beq
| \eta | \sim \frac{2}{\Omega_{\rm m 0}} \frac{\bar\rho_0}{\rho}
\left( \frac{c}{R H_0} \right)^{\!2} \left| \frac{d\ln A}{d\ln\rho} \right| .
\label{eta-R-rho}
\eeq
Then, the fifth force is greater than Newtonian gravity if we have
\beq
|\eta| \geq 1 : \;\;\; R^2 \leq R_\eta^2\equiv  \left( \frac{c}{H_0} \right)^{\!2}
\frac{2}{\Omega_{\rm m 0}} \frac{\bar\rho_0}{\rho} \left | \frac{d\ln A}{d\ln\rho} \right | .
\label{R-rho}
\eeq

At low densities, using Eq.(\ref{dlnAdlnrho-low-density}) we obtain
\beq
\rho \ll \rho_{\alpha} : \;\; R_{\eta}(\rho) \sim R_{\alpha} \;\;
\mbox{with} \;\; R_{\alpha} \equiv \frac{\alpha \, \zeta \, c}{H_0} .
\label{Reta-low-density}
\eeq
Thus, at low densities we obtain a constant radius threshold, of order
$R_{\alpha} \sim 3 h^{-1} \,{\rm kpc}$ for $\alpha=10^{-6}$, which grows with $\zeta$
as we can check in Fig.~\ref{fig:eta_R_rho}. At high densities, we have the behaviour
\be
\rho \gg \rho_{\alpha} : \;\; R_\eta\sim R_{\alpha} \left( \frac{\rho}{\rho_{\alpha}}
\right)^{-(2 \zeta + 1)/(2\zeta + 2)} .
\label{Reta-high-density}
\ee
Thus, at high densities the upper boundary of the fifth-force domain decreases and
the fifth force becomes negligible in the center of halos with sufficiently steep profiles,
as seen in Eq.(\ref{eta-small-r-gammap}).

\begin{figure}
\begin{center}
\epsfxsize=9 cm \epsfysize=6 cm {\epsfbox{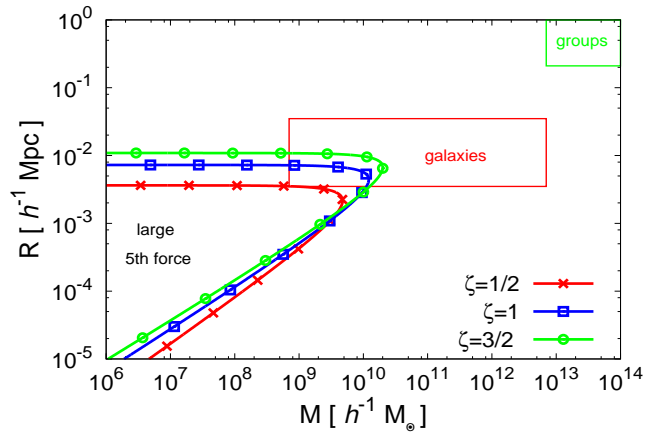}}
\end{center}
\caption{
Domain in the mass-radius plane where the fifth force is greater than Newtonian. 
The horizontal axis is the typical mass $M$ of the structure and the vertical axis its typical 
radius $R$.
The rectangles show the typical scales of galaxies and groups of galaxies.
}
\label{fig:eta_R_M}
\end{figure}

To facilitate the comparison with astrophysical structures, it is convenient to display
the fifth-force domain (\ref{R-rho}) in the mass-radius plane $(M,R)$.
This is shown in Fig.~\ref{fig:eta_R_M}, as the curve $R_{\eta}(\rho)$ provides
a parametric definition of the boundary $R_{\eta}(M)$, defining the mass
of the structure as $M=4\pi\rho R^3/3$.
We obtain a triangular domain, with a constant-radius upper branch and a lower branch
that goes towards small radius and mass with a slope that depends on $\zeta$.
The upper branch corresponds to the regime (\ref{Reta-low-density}), with
\beq
R_\eta \sim R_{\alpha} \;\; \mbox{ for } \;\; M < M_{\alpha} ,
\label{R-M-upper}
\eeq
and
\beq
M_{\alpha} \equiv \alpha^2 \zeta^3 \bar\rho_0 \left( \frac{c}{H_0} \right)^3 .
\label{Mainf-def}
\eeq
For $\alpha=10^{-6}$ this yields $M_{\alpha} \sim 10^{10} M_{\odot}$.
The lower branch corresponds to the regime (\ref{Reta-high-density}), which yields
for $M<M_\alpha$
\be
R \sim R_\alpha \left(\frac{M}{M_{\alpha}}\right)^{(2 \zeta + 1)/(4 \zeta +1)} .
\label{R_R_infy}
\ee
We also show in Fig.~\ref{fig:eta_R_M} the regions in this $(M,R)$-plane occupied by
groups and clusters of galaxies and by galaxies.
We only show astrophysical objects whose dynamics is due to the presence of dark matter
since for this model the coupling of the scalar field with baryons is negligible, as shown in
section~\ref{coupli-bary}. In agreement with section~\ref{sec:screening-clusters},
we find that the fifth force is negligible for clusters and groups (at their global scale), while it is
of the same order as Newtonian gravity for galaxies.
Therefore galaxies may provide strong constraints on the models considered in this paper.

\section{History and properties of the formation of cosmological structures}
\label{sechistory-cosmo}

To study the evolution of cosmological perturbations in the previous sections, either
through linear theory or the spherical collapse, we assumed that the density field remains
smooth and that the fifth force on cosmological scale $x$ is set by the density gradient on
the same scale. However, the ultra-local fifth force is directly sensitive to the
local density gradient, $\nabla\ln A = (d\ln A/d\ln\rho) \nabla\rho$, in contrast with the
Newtonian force which involves an average over scale $x$,
$F_{\rm N} \propto \int d\vx' \rho(\vx')/|\vx-\vx'|^2$.
Moreover, smaller scales are increasingly unstable because of the $k^2$ factor in the
factor $\epsilon(k,\tau)$ in Eq.(\ref{eps-def}) that amplifies the gravitational attraction
in the linear evolution equation (\ref{pertlin-ultra-local}).
This could invalidate the analysis presented above as small scales could develop strong
instabilities.
This would lead to a fragmentation of the system down to very small scales so that the
local density gradient, hence the fifth force, is nowhere related to cosmological scale
gradients.
This would in turn lead to an effective screening mechanism as isolated overdensities
no longer interact.
Note that this mechanism, due to the ultra-local character of the theory, is the key
to the screening of the fifth force in local environments, such as in the Solar System,
which is required in the theories studied in the companion paper where both the baryons
and the dark matter feel the fifth force.
In the supersymmetric setting considered in this paper, we do not need to invoke this
mechanism to ensure that the theory is consistent with Solar System tests as the baryons
do not feel the fifth force. However, it could still play a role in case it leads to a fragmentation
of the dark matter density field at high redshift, when the fifth force is dominant, and
makes a ``mean field'' approach inadequate.

As described in details in the companion paper \cite{Brax:2016vpd}, we can investigate this issue by using
a thermodynamic approach, which allows us to go beyond perturbation theory and
spherical dynamics.
Thus, we assume that at high redshift, when the fifth force is dominant, regions that
collapse and turn non-linear because of the fifth-force interaction relax towards the
thermodynamic equilibrium.
Then, if this equilibrium is strongly inhomogeneous the mean field approach used in the
previous sections breaks down, whereas if this equilibrium is homogeneous we can conclude
that the system does not develop strong small-scale inhomogeneities and the previous
analysis is correct.

\subsection{Cosmological non-linear transition}
\label{sec:cosmological-transition}

\begin{figure}
\begin{center}
\epsfxsize=8 cm \epsfysize=6 cm {\epsfbox{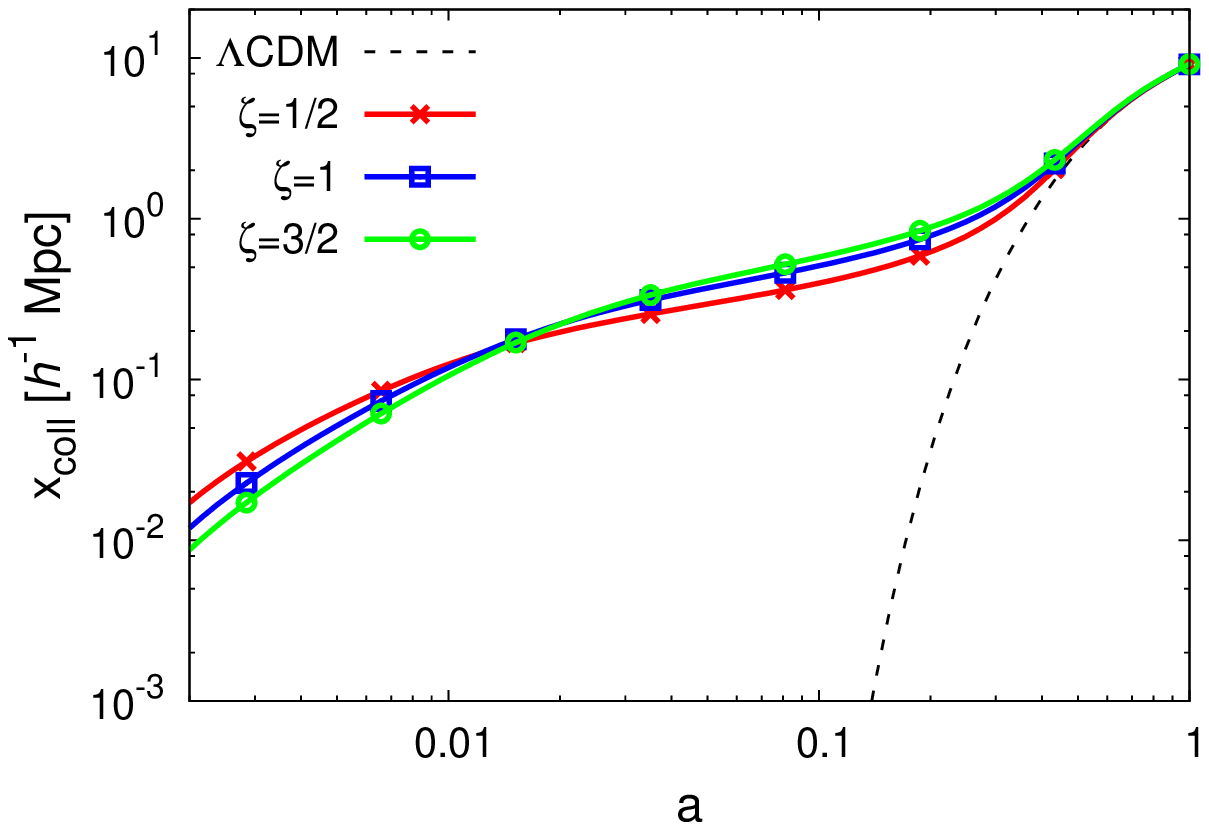}}
\epsfxsize=8 cm \epsfysize=6 cm {\epsfbox{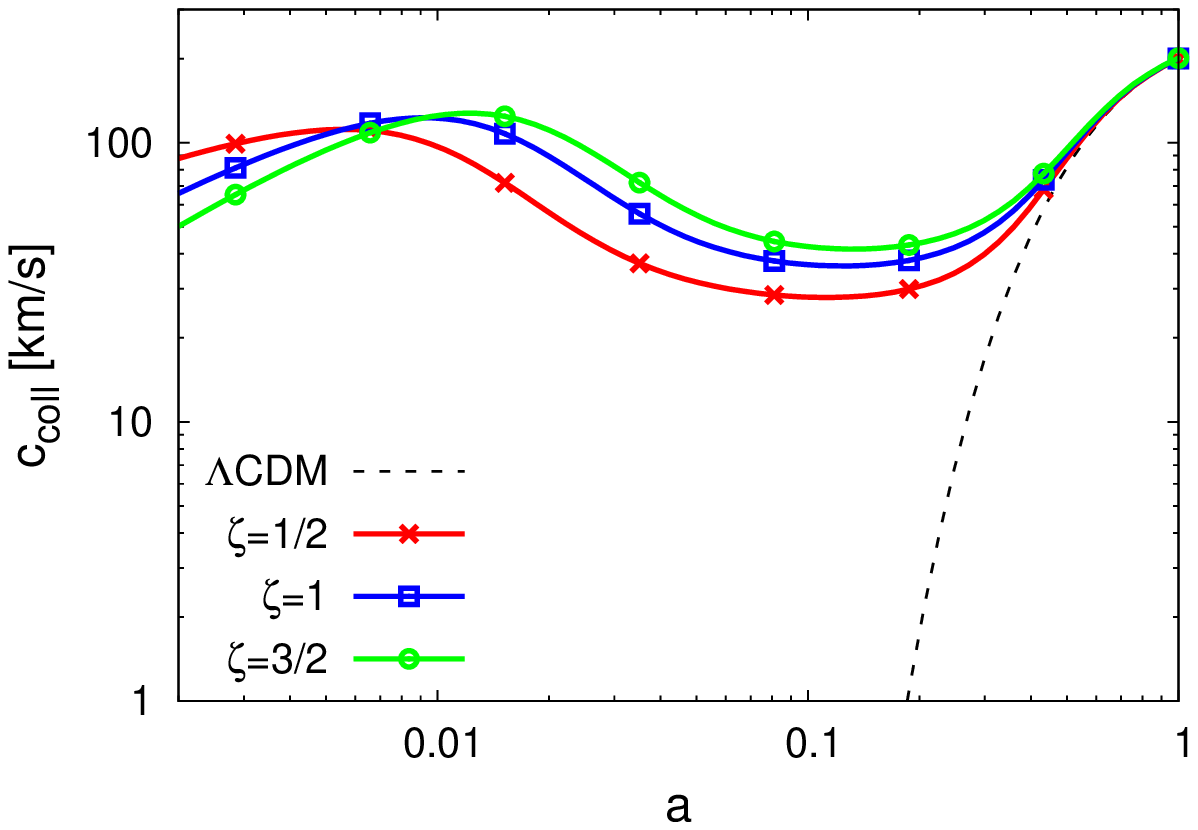}}
\end{center}
\caption{
{\it Upper panel:} collapse radius $x_{\rm coll}(z)$ (in comoving coordinates)
as a function of the scale factor, for the ultra-local models and the $\Lambda$-CDM
cosmology.
{\it Lower panel:} collapse velocity scale $c_{\rm coll}(z)$.
}
\label{fig:radius-velocity}
\end{figure}

We first study in this section the evolution with redshift of the comoving cosmological
scales $x_{\rm coll}(z)$ that enter the non-linear regime, which we define by
\beq
\Delta^2_L(\pi/x_{\rm coll},z) = 1.5
\label{xcoll-def}
\eeq
where $\Delta^2_L$ is the logarithmic linear power spectrum.
(The factor $1.5$ is chosen so that we obtain $x_{\rm coll} \simeq 8 h^{-1}$Mpc
at $z=0$ in the $\Lambda$-CDM scenario.)
As seen in the upper panel in Fig.~\ref{fig:radius-velocity}, $x_{\rm coll}(z)$ is
much greater than in the $\Lambda$-CDM cosmology at high redshift because the
fifth force amplifies the growth of structure.
After $a_{\alpha}$ the fifth force fastly decreases, as seen in Fig.~\ref{fig:eps1}.
This leads to the plateau for $x_{\rm coll}(z)$ over
$a_{\alpha} \leq a \leq a_{\Lambda \rm -CDM}$, with $a_{\alpha}=\alpha^{1/3} \sim 0.01$
associated with the peak of the fifth force and $a_{\Lambda \rm -CDM} \simeq 0.2$
associated with the convergence to the $\Lambda$-CDM prediction for
$x_{\rm coll}(z)$.
The reason why $a_{\alpha} \ll a_{\Lambda \rm -CDM}$ is that after $a_{\alpha}$
the fast decrease of the fifth force implies that structure formation due to the fifth force
stalls, and we need to wait until $a_{\Lambda \rm -CDM}$ for Newtonian gravity
to take over at the scale $x_{\rm coll}(z_{\alpha})$, because at $a_{\alpha}$ Newtonian
gravity was much weaker than the fifth force at the comoving scale $x_{\rm coll}(z_{\alpha})$.
Thus, from the point of view of cosmological structure formation, we have three
eras.
For $a<a_{\rm \alpha}$, the non-linear transition $x_{\rm coll}(z)$ of the cosmological
density field is due to the fifth force and occurs at scales much greater than in the
$\Lambda$-CDM scenario.
For $a_{\rm \alpha} < a < a_{\Lambda \rm -CDM}$, structure formation stalls as
the fifth force decreases and Newtonian gravity is still weak on these scales.
For $a_{\Lambda \rm -CDM} < a$, we recover the growth predicted by the
$\Lambda$-CDM cosmology, due to Newtonian gravity.

For the thermodynamic analysis presented in the next section we also need the initial
kinetic energy or typical velocity of the collapsing domains. Thus, we define the effective
velocity $c_{\rm coll}(z)$ by
\beq
c^2_{\rm coll}(z) = c_s^2 + c_{\rm N}^2 ,
\label{c-coll-def}
\eeq
with
\beq
c_s^2 = \epsilon_1 \, c^2 , \;\;\;
c_{\rm N}^2 = (1+\epsilon_1) \frac{3\Omega_{\rm m}}{2\pi^2} (H a x_{\rm coll})^2 .
\label{cs-cN-def}
\eeq
The term $c_s^2$ comes from the pressure-like term $\epsilon_1 c^2 \nabla^2\delta$
in Eq.(\ref{linear-delta-real}) while the term $c_{\rm N}^2$ comes from the right-hand
side in Eq.(\ref{linear-delta-real}), associated with Newtonian gravity (amplified by the
negligible factor $\epsilon_1$).
In the case of the $\Lambda$-CDM cosmology we only have
$c_{\rm coll}^{\Lambda\rm -CDM} = c_{\rm N}^{\Lambda\rm -CDM}$ as there is no fifth-force
pressure-like term.
As seen in the lower panel in Fig.~\ref{fig:radius-velocity}, at high redshift,
$a < a_{\Lambda \rm -CDM}$, we have $c_{\rm coll} \gg c_{\rm coll}^{\Lambda\rm -CDM}$,
whereas at low redshift, $a_{\Lambda \rm -CDM} < a$, we have
$c_{\rm coll} \simeq c_{\rm coll}^{\Lambda\rm -CDM}$ as we recover the
$\Lambda$-CDM behavior.
Between $a_{\alpha}$ and $a_{\Lambda \rm -CDM}$ the velocity scale first
decreases until $a_{c_s/c_{\rm N}} \simeq 0.1$ with the decline of the fifth force,
as $c_{\rm coll} \simeq c_s$, and next grows again with Newtonian gravity as
$c_{\rm coll} \simeq c_{\rm N}$.

This history singles out a characteristic mass and velocity scale, associated
with the plateau found in Fig.~\ref{fig:radius-velocity} over $0.02 \lesssim a \lesssim 0.2$.
This yields
\beqa
&& x_* \sim 0.355 \; h^{-1} {\rm Mpc} , \;\;\;
M_* \sim 2 \times 10^{10} \; h^{-1} M_{\odot} , \nonumber \\
&& c_* \sim 50 \; {\rm km/s} .
\label{x*-M*-c*-def}
\eeqa
As in Fig.~\ref{fig:eta_R_M}, we recover galaxy scales, more precisely here the scales
associated with small galaxies.
It is tempting to wonder whether this could help alleviate some of the problems
encountered on galaxy scales by the standard $\Lambda$-CDM scenario.
However, this would require detailed numerical studies that are beyond the scope of this
paper.

\subsection{Thermodynamic equilibrium on cosmological scales}
\label{sec:thermo-equilibrium-cosmo}

\begin{figure*}
\begin{center}
\epsfxsize=5.8 cm \epsfysize=5.5 cm {\epsfbox{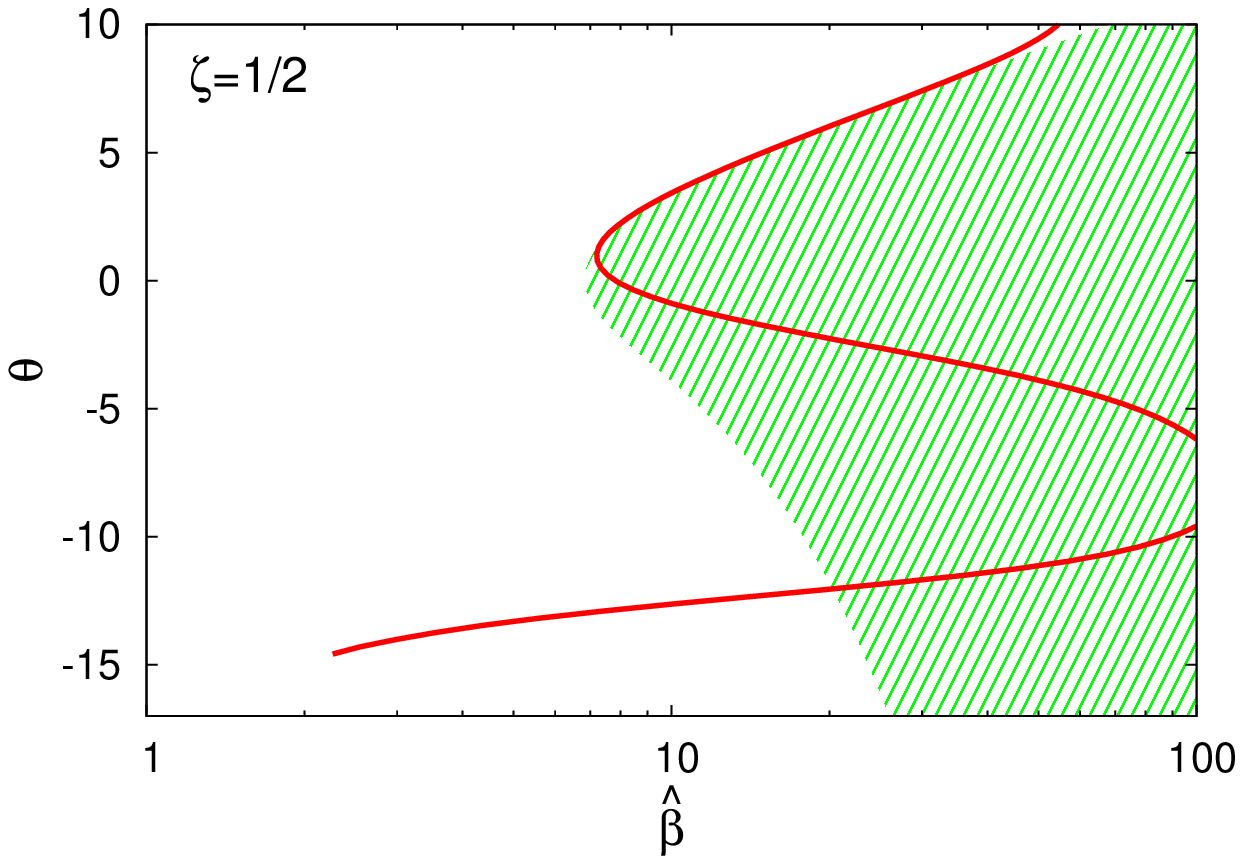}}
\epsfxsize=5.8 cm \epsfysize=5.5 cm {\epsfbox{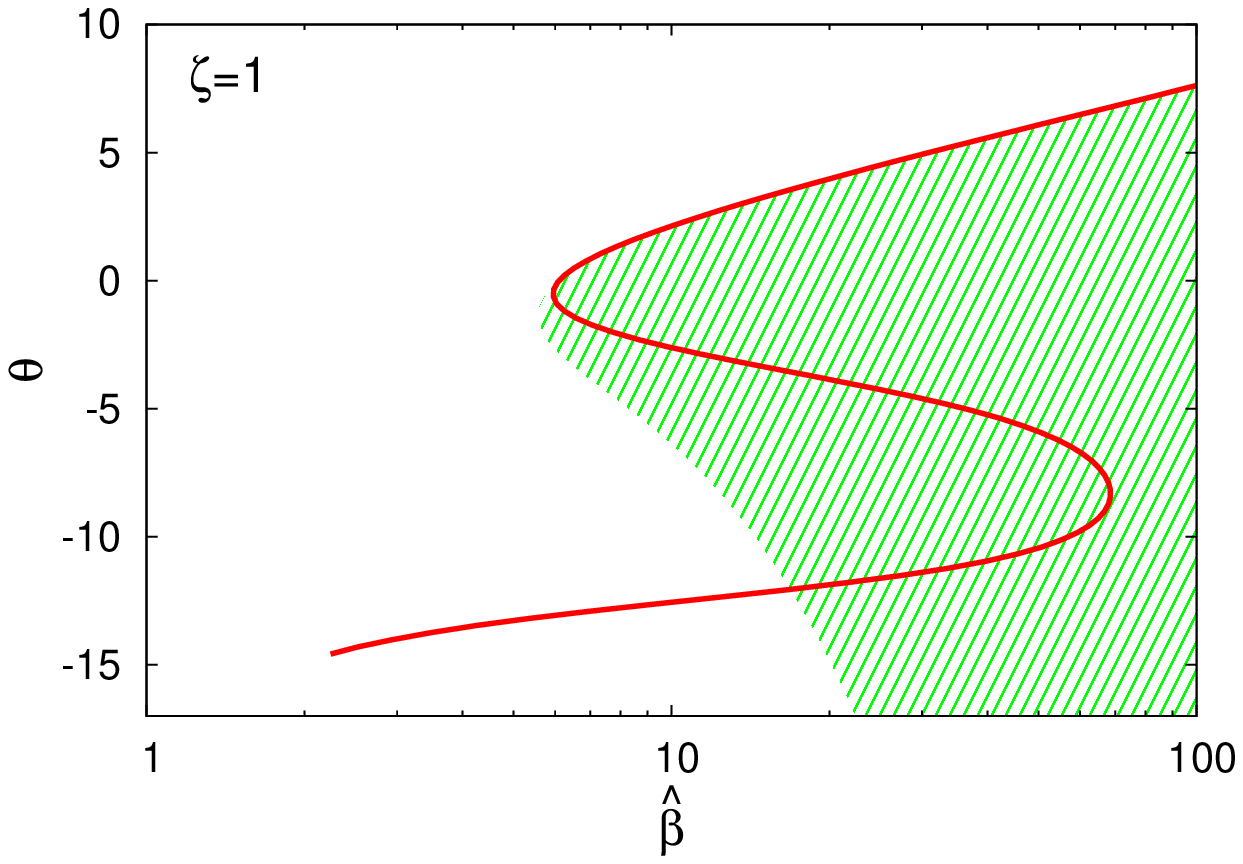}}
\epsfxsize=5.8 cm \epsfysize=5.5 cm {\epsfbox{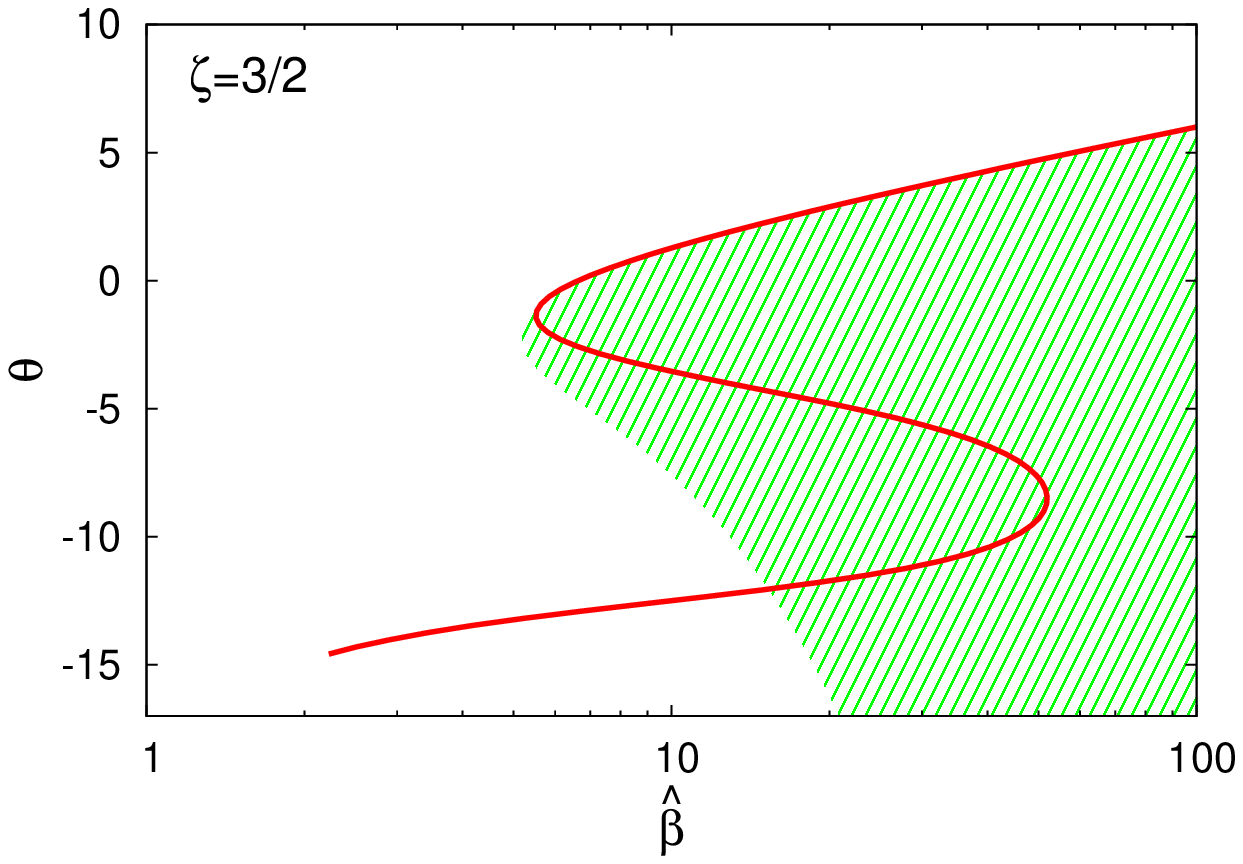}}
\end{center}
\caption{
Thermodynamic phase diagram for the ultra-local models with $\zeta=1/2, 1$ and $3/2$.
The shaded area is the region of initial inverse temperature $\hat{\beta}$ and density
$\theta$ where the thermodynamic equilibrium is inhomogeneous.
The solid line is the cosmological trajectory $(\hat{\beta}_{\rm coll}(z),\theta_{\rm coll}(z))$.
}
\label{fig:thermo-cosmo}
\end{figure*}

We can now study the non-linear dynamics of the cosmological scales
$x_{\rm coll}(z)$ that enter the non-linear regime found in Fig.~\ref{fig:radius-velocity}.
More precisely, we use a thermodynamic approach to investigate whether these
regions develop a fragmentation process and show strong small-scale inhomogeneities \cite{Chavanis2005,Balian2007}.
Because we are interested in the evolution at high redshift, $z \geq z_{\alpha}$,
when the fifth force is dominant, we neglect the Newtonian gravity and we consider the
thermodynamic equilibrium of systems defined by the energy $E$ and entropy $S$
given by
\beqa
E & = & \int d^3 x d^3 v \; f(\vx,\vv) \left( \frac{v^2}{2} + c^2 \ln A[\rho(\vx)] \right) ,
\hspace{0.5cm}
\label{energy-thermo} \\
S & = & -\int d^3 x d^3 v \; f(\vx,\vv) \; \ln \frac{f(\vx,\vv)}{f_0} .
\label{entropy-thermo}
\eeqa
Here $f(\vx,\vv)$ is the phase-space distribution function, $f_0$ is an irrelevant normalization
constant, and we used the fact that the fifth-force potential $\ln A$ is a function of the local
density.
Then, assuming that the scales that turn non-linear because of the fifth force at high redshift
reach a statistical equilibrium through the rapidly changing effects of the fluctuating
potential, in a fashion somewhat similar to the violent relaxation that takes
place for gravitational systems \cite{Lynden-Bell1967},
we investigate the properties of this thermodynamic equilibrium.

Contrary to the usual gravitational case, the potential $\ln A$ is both bounded and
short-ranged , so that we cannot build infinitely large negative (or positive) potential
energies and a stable thermodynamic equilibrium always exists, and it is possible to work
with either micro-canonical, canonical or grand-canonical ensembles.
In this respect, a thermodynamic analysis is better suited for such systems than for
standard 3D gravitational systems, where the potential energy is unbounded
from below and stable equilibria do not always exist, and different statistical
ensembles are not equivalent \cite{Padmanabhan1990}.

We work in the grand-canonical ensemble, where the dark matter particles are confined
in a box of size $x$ (the scale $x_{\rm coll}(z)$ that is turning non-linear at redshift $z$),
with a mean temperature $T=1/\beta$ and chemical potential $\mu$ that are set by
the initial velocity scale $c_{\rm coll}(z)$ and mean density $\bar\rho(z)$.
The analysis of the thermodynamic equilibria and phase transitions is described in details
in the companion paper \cite{Brax:2016vpd}.
The main result is that at high temperature, $T>T_c$ and $\beta < \beta_c$,
the thermodynamic equilibrium is homogeneous, whereas at low temperature,
$T<T_c$ and $\beta > \beta_c$, the equilibrium is inhomogeneous.
Indeed, at high temperature the system is dominated by its kinetic energy and the
potential energy associated with the fifth force (which is bounded) is negligible,
so that we recover a perfect gas without interactions, whereas at low temperature
the fifth-force potential becomes important and leads to strong inhomogeneities as it
corresponds to an attractive force.
In terms of the rescaled dimensionless variables $\theta$ and $\hat\beta$,
\beq
\theta = \ln\left(\frac{\rho}{\rho_{\alpha}}\right) , \;\;\;
\hat\beta = \alpha c^2 \beta ,
\label{theta-beta-def}
\eeq
this leads to the phase diagram shown in Fig.~\ref{fig:thermo-cosmo}.
The equilibrium is inhomogeneous inside the shaded region, which is limited at low
$\hat\beta$ by the inverse critical temperature $\hat\beta_c$,
with $\hat\beta_c \simeq \{ 6.85, 5.58, 5.14 \}$ for
$\zeta = \{ 1/2, 1, 3/2 \}$.
The upper and lower limits of the domain are the curves $\theta_+(\hat\beta)$ and
$\theta_-(\hat\beta)$, which obey the low-temperature asymptotes
\beq
\hat\beta \rightarrow \infty : \;\;\;
\theta_+ \sim \frac{1+\zeta}{\zeta} \ln \hat\beta , \;\;\;
\theta_- \sim - \hat\beta .
\label{theta+_theta-}
\eeq
Then, if the average initial temperature and density $(1/\hat\beta,\theta)$ fall outside the
shaded domain the system remains homogeneous.
If they fall inside the shaded domain the system becomes inhomogeneous and splits
over two domains with density $\theta_-$ and $\theta_+$, with a proportion such that the
total mass is conserved.
Because of the ultra-local property [i.e. $\ln A$ is a local function through $\rho(\vx)$],
the equilibrium factorizes over space $\vx$ so that the two domains at density
$\theta_{\pm}$ are not necessarily connected and can take any shape.

The solid curves in Fig.~\ref{fig:thermo-cosmo} are the cosmological trajectories
associated with the scale and velocity $\{x_{\rm coll}(z),c_{\rm coll}(z)\}$ displayed
in Fig.~\ref{fig:radius-velocity}, which correspond to
\beq
\theta_{\rm coll}(z) = \ln \left( \frac{\bar\rho(z)}{\rho_{\alpha}} \right) , \;\;\;
\hat\beta_{\rm coll}(z) = \frac{\alpha c^2}{c^2_{\rm coll}(z)} .
\label{beta-coll-def}
\eeq
This trajectory moves downward to lower densities with cosmic time, following $\bar\rho(z)$.
In agreement with the lower panel of Fig.~\ref{fig:radius-velocity},
the inverse temperature $\hat\beta_{\rm coll}$ first decreases until $a_{\alpha}$,
as the velocity $c_{\rm coll}(z)$ grows.
Next, $\hat\beta_{\rm coll}$ increases while $c_{\rm coll}(z)$ decreases along with the
fifth force, until we recover the $\Lambda\rm -CDM$ behavior at late times
and $\hat\beta_{\rm coll}$ decreases again thereafter.
We are interested in the first era, $a<a_{\alpha}$, and we find that the cosmological
trajectory is almost indistinguishable from the upper boundary $\theta_+(\hat\beta)$
of the inhomogeneous thermodynamic phase.
Indeed, at early times we have $c_{\rm coll} \simeq c_s$, hence
$\hat\beta_{\rm coll} \simeq \alpha/\epsilon_1$ from Eq.(\ref{cs-cN-def}).
Using Eq.(\ref{eps1-high-rho-squared}) we have at high densities, which also correspond
to $a<a_{\alpha}$, $\epsilon_1 \sim \alpha (\rho/\rho_{\alpha})^{-\zeta/(1+\zeta)}
= \alpha e^{-\zeta\theta/(1+\zeta)}$,
hence
\beq
a \ll a_{\alpha} : \;\;\; \theta_{\rm coll} \sim \frac{1+\zeta}{\zeta} \ln \hat\beta_{\rm coll} ,
\label{theta-beta-coll}
\eeq
and we recover the asymptote (\ref{theta+_theta-}) of $\theta_+(\hat\beta)$.

If $\theta_{\rm coll} > \theta_+$ we are in the homogeneous phase and the system
remains at the initial density $\bar\rho$.
If $\theta_{\rm coll} \lesssim \theta_+$ we are in the inhomogeneous phase and the system
splits over regions of densities $\theta_+$ and $\theta_-$. However, as we remain close
to $\theta_+$ most of the volume is at the density $\theta_+ \simeq \theta_{\rm coll}$
and only a small fraction of the volume is at the low density $\theta_-$.
Neglecting these small regions, we can consider that in both cases the system remains
approximately homogeneous.
This means that, according to this thermodynamic analysis, the cosmological density
field does not develop strong inhomogeneities that are set by the cutoff scale
of the theory when it enters the fifth-force non-linear regime.
Therefore, density gradients remain set by the large-scale cosmological density
gradients and the analysis of the linear growing modes and of the spherical collapse
presented in previous sections are valid.
On small non-linear scales and at late times, where Newtonian gravity
becomes dominant, we recover the usual gravitational instability that we neglected in
this analysis and structure formation proceeds as in the standard $\Lambda$-CDM
case.

\subsection{Halo centers}
\label{sec:halo-centers}

\begin{figure*}
\begin{center}
\epsfxsize=5.8 cm \epsfysize=5.5 cm {\epsfbox{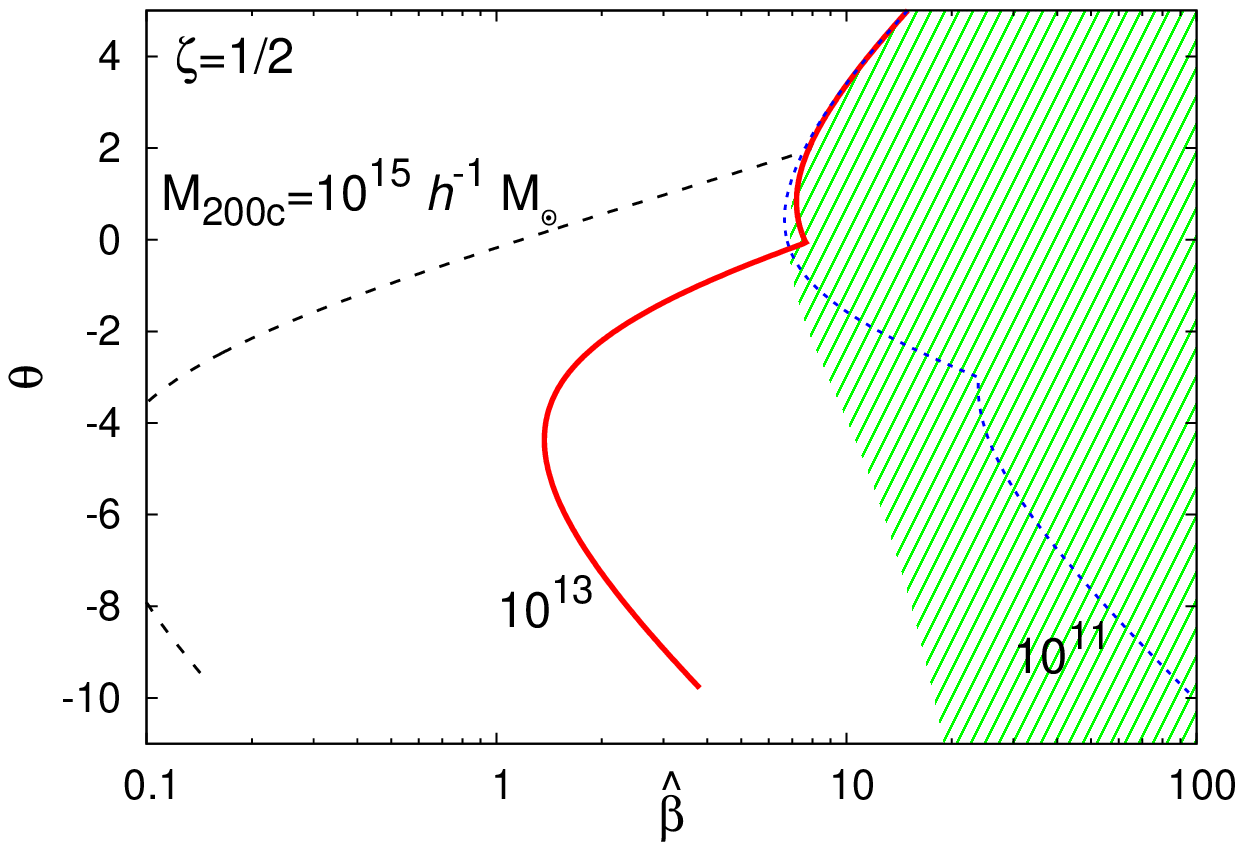}}
\epsfxsize=5.8 cm \epsfysize=5.5 cm {\epsfbox{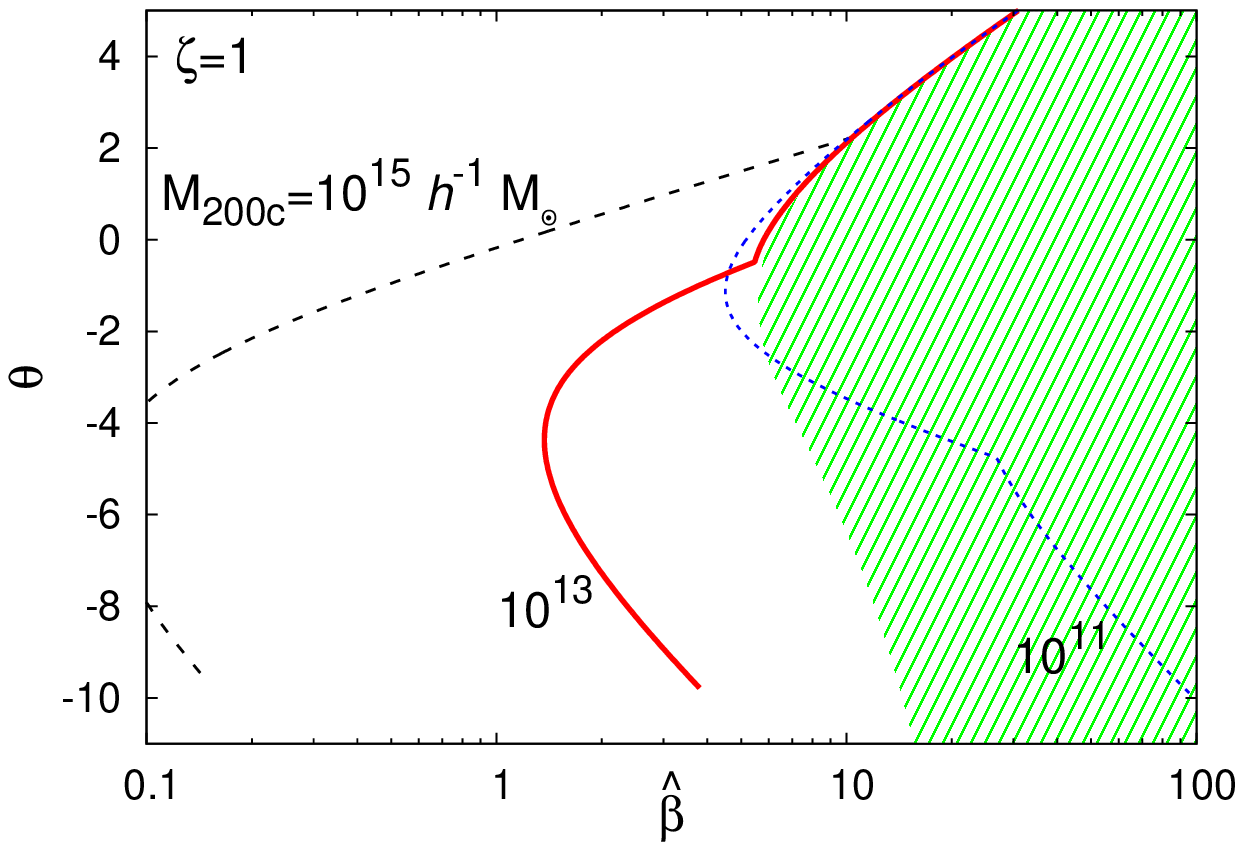}}
\epsfxsize=5.8 cm \epsfysize=5.5 cm {\epsfbox{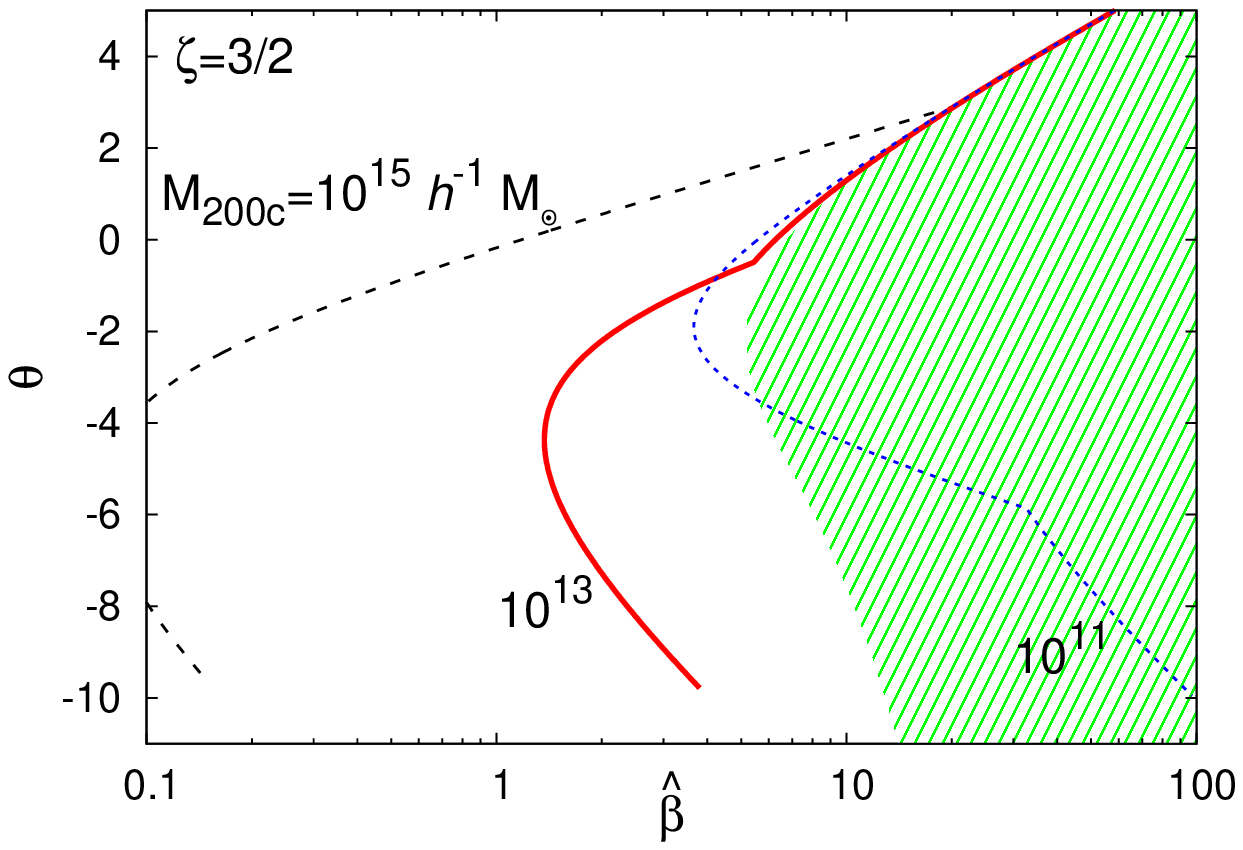}}
\end{center}
\caption{
Radial trajectory $(\hat\beta_r,\theta_r)$ over the thermodynamic phase diagram inside
NFW halos of mass $M=10^{15}, 10^{13}$ and $10^{11} h^{-1} M_{\odot}$. at $z=0$.
}
\label{fig:thermo-halos}
\end{figure*}

It is interesting to apply the thermodynamic analysis presented above to the
inner radii of clusters and galaxies.
Indeed, we have seen in section~\ref{sec:radial-profile-eta} that
the fifth force becomes large inside spherical halos and the ratio
$F_A/F_{\rm N}$ actually diverges at the center for shallow density
profiles, see Fig.~\ref{fig:ratio-forces} and Eq.(\ref{eta-small-r-gammap}).
However, this analysis was based on dimensional and scaling arguments and it fails
if the density field becomes strongly inhomogeneous so that the typical density
inside the halo is very different from the global averaged density.
The thermodynamic analysis used to derive the phase space diagram shown in
Fig.~\ref{fig:thermo-cosmo} neglected Newtonian gravity. However, we can also apply
its conclusions to a regime dominated by Newtonian gravity where at radius $r$ inside
the halo the structures built by gravity and the density gradients are on scale $r$.
Then, we can ask whether at this radius $r$ fifth-force effects may lead to a fragmentation
of the system on much smaller scales $\ell \ll r$. To study this small-scale behavior
we can neglect the larger-scale gravitational gradients $r$ and discard gravitational
forces.

Within a radius $r$ inside the halo the averaged reduced density and inverse temperature
are
\beq
\theta_r = \ln \left( \frac{\rho_<(r)}{\rho_{\alpha}} \right) , \;\;\;
\hat\beta_r = \frac{\alpha c^2}{{\rm Max}(c_s^2,v_{\rm N}^2)} ,
\label{theta-beta-halo-def}
\eeq
where $v_{\rm N}$ is the Newtonian circular velocity and $c_s$ is the fifth-force velocity
scale defined in Eq.(\ref{vN-cs-def}).
As seen in Eq.(\ref{eta-def-Delta}), the maximum ${\rm Max}(c_s^2,v_{\rm N}^2)$
shifts from one velocity scale to the other when the associated force becomes dominant.
Here we choose the non-analytic interpolation ${\rm Max}(c_s^2,v_{\rm N}^2)$
instead of the smooth interpolation $c_s^2+v_{\rm N}^2$ that we used
in Eq.(\ref{c-coll-def}) for the cosmological analysis for illustrative convenience.
Indeed, the discontinuous changes of slope in Fig.~\ref{fig:thermo-halos}
show at once the location of the transitions $|\eta|=1$ between the
fifth-force and Newtonian gravity regimes.

When the density grows at small radii as a power law, $\rho \propto r^{-\gamma_p}$,
we have seen in Eq.(\ref{eta-small-r-gammap}) that the fifth-force to gravity ratio
$\eta$ behaves as $\eta \sim r^{\gamma_p ( 1+2\zeta)/(1+\zeta) -2}$ with
\beq
v_{\rm N}^2 \sim r^{2-\gamma_p} ,  \;\;\; c_s^2 \sim r^{\gamma_p\zeta/(1+\zeta)} ,
\label{cs-vN-r}
\eeq
at high density $\rho \gg \rho_{\alpha}$, where we used
Eq.(\ref{dlnAdlnrho-high-density}).
This gives in the Newtonian gravity and fifth-force regimes
\beq
| \eta | < 1 : \;\;\; \theta_r \sim \frac{\gamma_p}{2-\gamma_p} \ln\hat\beta_r ,
\label{theta-r-beta-r-low-eta}
\eeq
\beq
| \eta | > 1 : \;\;\; \theta_r \sim \frac{1+\zeta}{\zeta} \ln\hat\beta_r .
\label{theta-r-beta-r-high-eta}
\eeq
For $\gamma_p>2$ we are in the Newtonian regime $v^2_{\rm N} \rightarrow \infty$,
$\hat\beta_r \rightarrow 0$, so that we are in the homogeneous phase of the
thermodynamic phase diagram as $\hat\beta_r < \hat\beta_c$.
For $(2+2\zeta)/(1+2\zeta) < \gamma_p < 2$ Newtonian gravity still dominates at small
radii and we have the asymptote (\ref{theta-r-beta-r-low-eta}) with
$\gamma_p/(2-\gamma_p) > (1+\zeta)/\zeta$, so that the radial trajectory
$(\hat\beta_r,\theta_r)$ moves farther above from the upper bound
$\theta_+$ of Eq.(\ref{theta+_theta-}) of the inhomogeneous phase and small radii
are within the homogeneous phase.
For $\gamma_p < (2+2\zeta)/(1+2\zeta)$ we are in the fifth-force regime
and we obtain $\theta_r \sim \theta_+$, so that the radial trajectory
$(\hat\beta_r,\theta_r)$ follows the upper boundary of the inhomogeneous phase domain.
This means that the dimensional analysis of section~\ref{sec:radial-profile-eta}
is valid as the fifth force does not push towards a fragmentation of the system
down to very small scales.

These asymptotic results apply to the small-radius limit $r\rightarrow 0$.
In Fig.~\ref{fig:thermo-halos} we show the full radial trajectories
$(\hat\beta_r,\theta_r)$ over the thermodynamic phase diagram, from $R_{200\rm c}$
inward, for the NFW halos that were displayed in Fig.~\ref{fig:ratio-forces} at $z=0$.
As we move inside the halo, towards smaller radii $r$, the density $\theta_r$ grows
and the trajectory moves upward in the figure.
The turn-around of $\hat\beta_r$ at $\theta_r \simeq -4$ corresponds to the NFW
radius $r_s$ where the local slope of the density goes through $\gamma=2$
and the circular velocity is maximum.
At smaller radii, $r \ll r_s$, the NFW profile goes to $\rho \propto r^{-1}$, hence
$\gamma_p = 1$. In agreement with the asymptotic analysis above, this implies
that we move farther into the fifth-force regime and we follow the upper boundary
$\theta_+$ of the inhomogeneous phase domain, so that the dimensional analysis
of section~\ref{sec:radial-profile-eta} is valid.
This also leads to an increasingly dominant fifth force at small radii and
characteristic velocities that are higher than the Newtonian circular velocity.
This may rule out these ultra-local scenarios.
However, on small scales the baryonic component is non-negligible and it actually
dominates on kpc scales inside galaxies. Since the baryons do not feel the fifth force
this could keep these models consistent with observations.
On the other hand, for low-mass halos, $M \lesssim10^{11} h^{-1} M_{\odot}$ at $z=0$,
we find that a significant part of the halo is within the inhomogeneous thermodynamic
phase.
This may leave some signature as a possible fragmentation of the system
on these intermediate scales into higher-density structures.
This process would next lead to a screening of the fifth force, because of the
ultra-local character of the fifth force.
Indeed, because it is set by the local density gradients, the fragmentation
of the system leads to a disappearance of large-scale collective effects and the
fifth force behaves like a surface tension at the boundaries of different domains.
Such a process may also happen in the case of massive halos at earlier stages
of their formation, which could effectively screen the fifth force
whereas the simple static analysis leads to a dominant fifth force at small radii.
However, a more precise analysis to follow such evolutionary tracks and check
the final outcomes of the systems requires numerical studies that are beyond the scope
of this paper.

\section{Dependence on the $\alpha$ parameter}
\label{sec:x-dependence}

\begin{figure*}
\begin{center}
\epsfxsize=8 cm \epsfysize=6 cm {\epsfbox{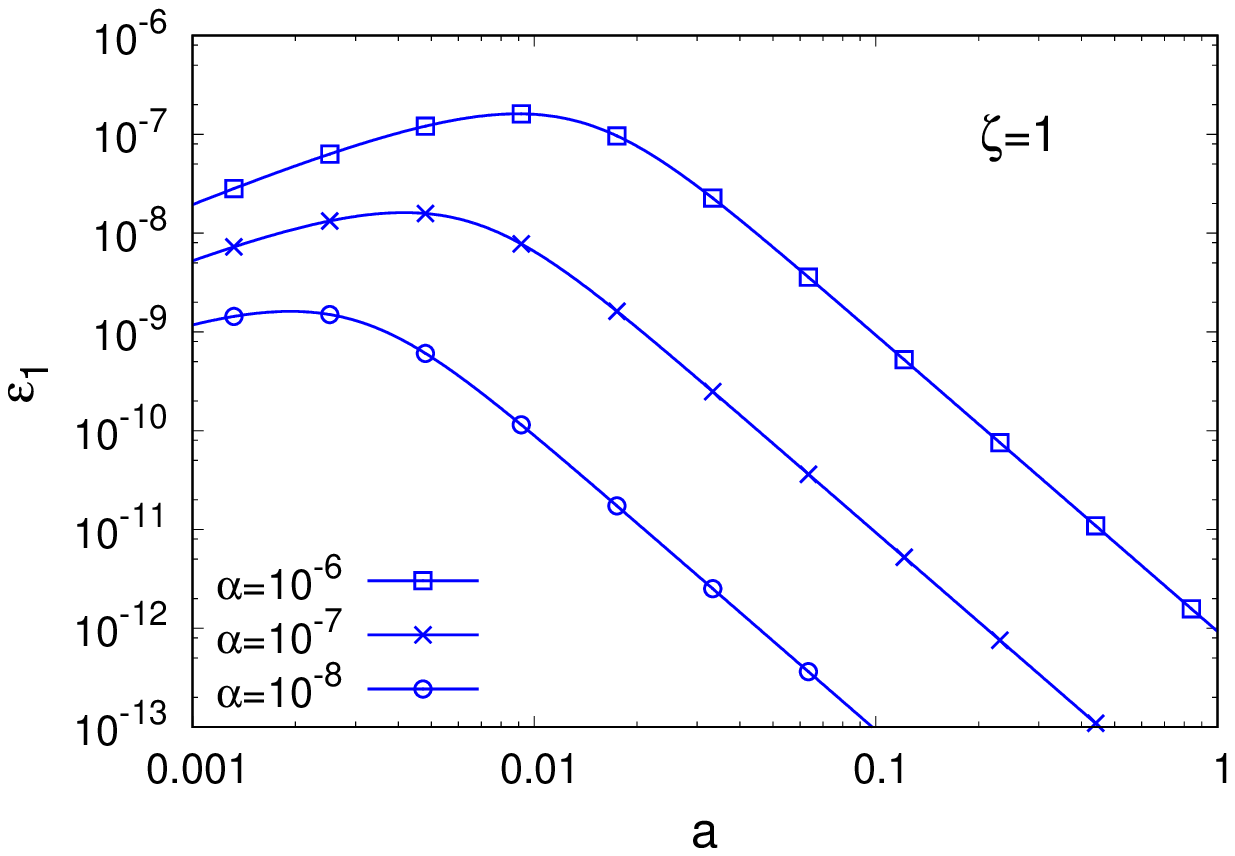}}
\epsfxsize=8 cm \epsfysize=6 cm {\epsfbox{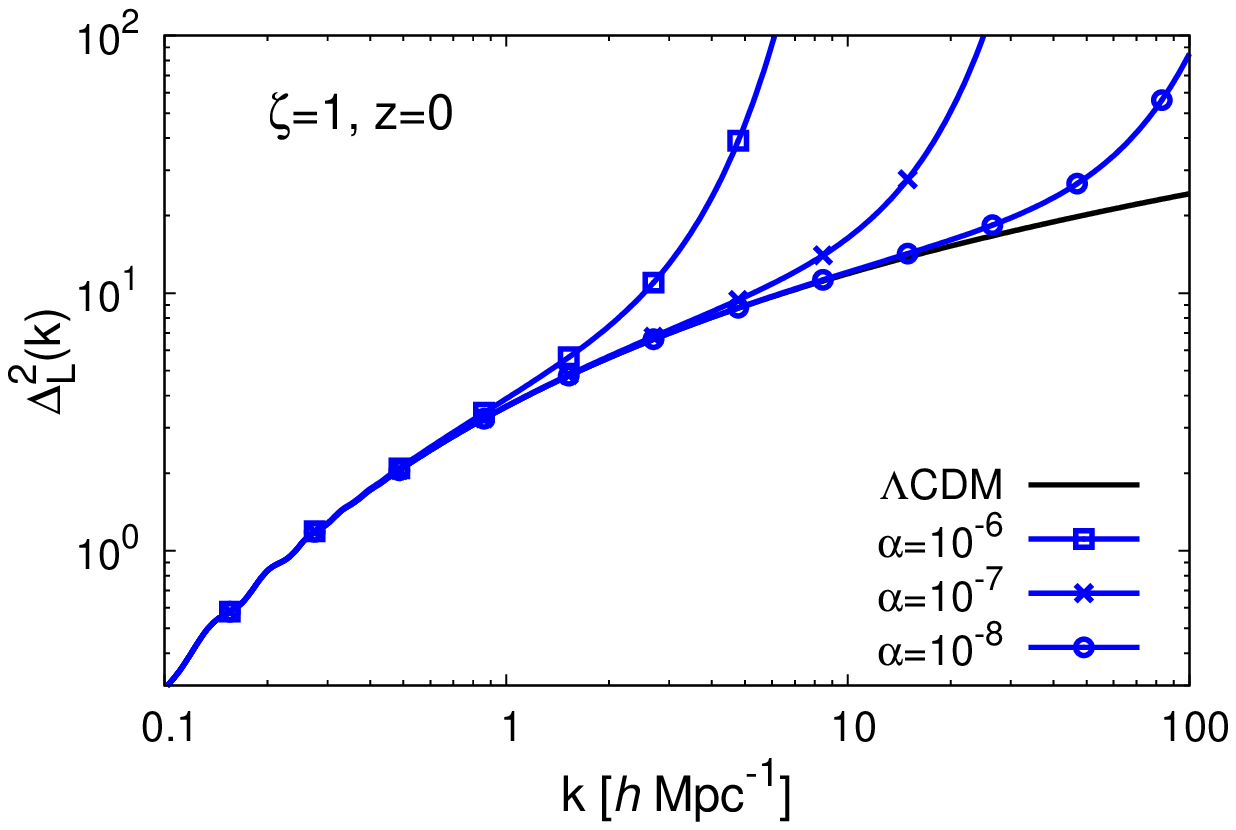}}\\
\epsfxsize=8 cm \epsfysize=6 cm {\epsfbox{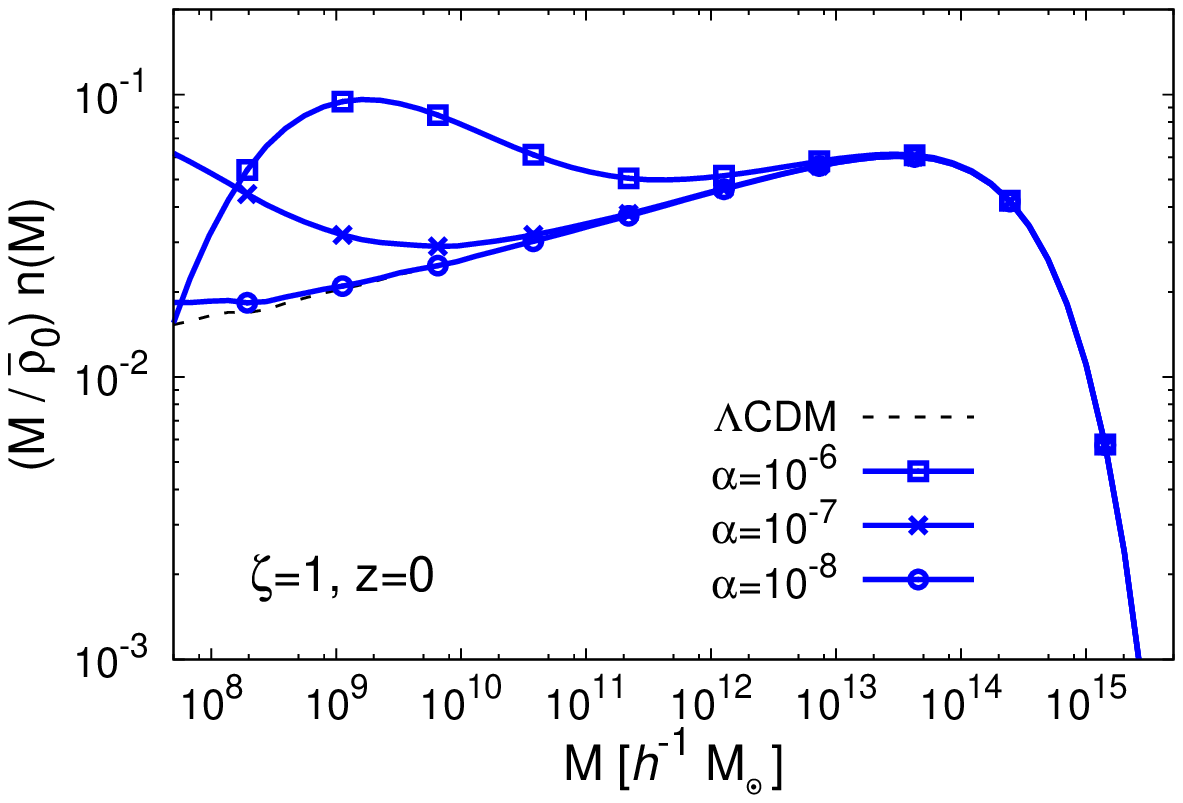}}
\epsfxsize=8 cm \epsfysize=6 cm {\epsfbox{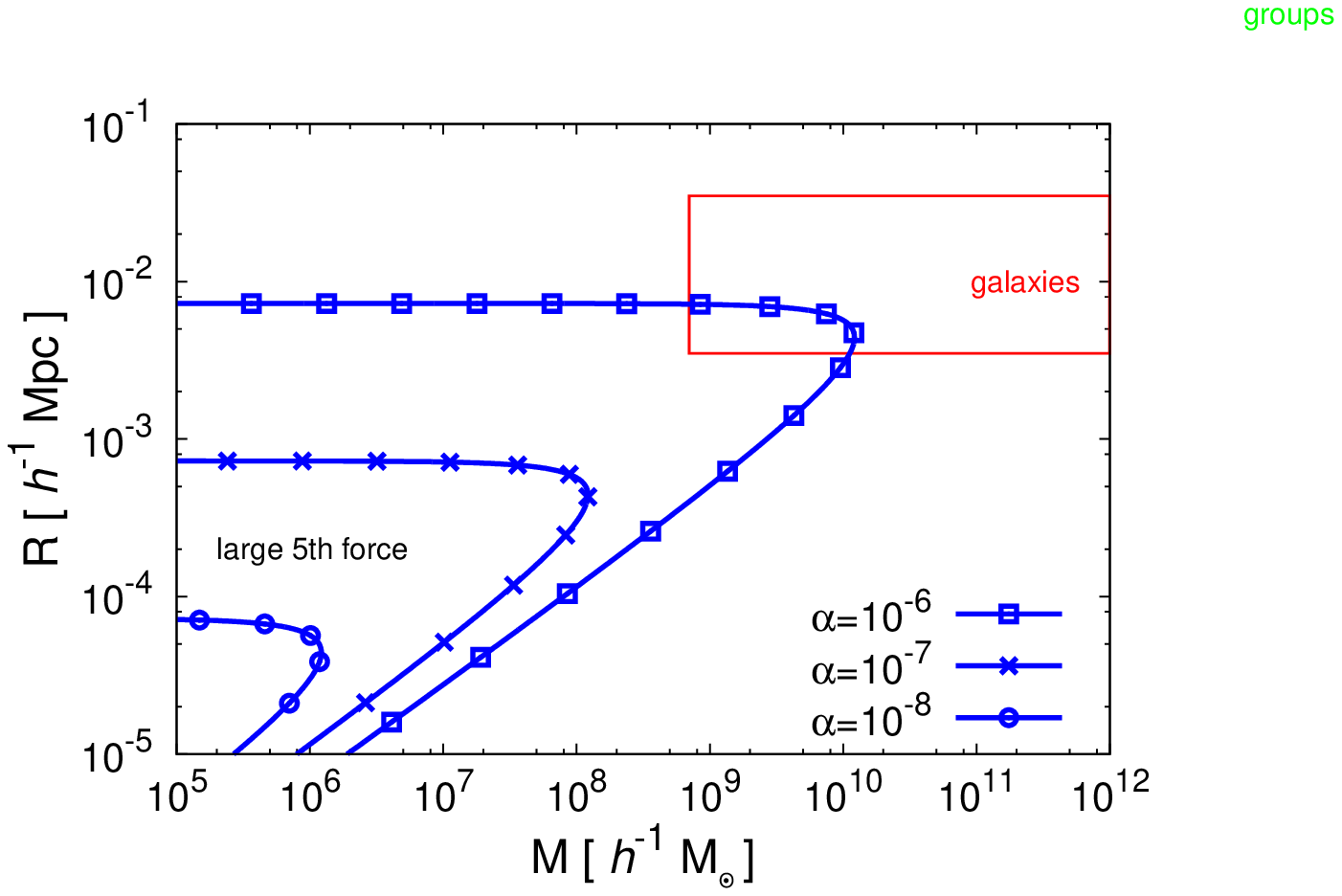}}
\end{center}
\caption{
Dependence on the parameter $\alpha$ of the deviations from the $\Lambda$-CDM predictions.
We plot models with $\zeta=1$ and $\alpha=10^{-6}$, $10^{-7}$ and $10^{-8}$.
{\it Upper left panel:} $\epsilon_1(a)$ as a function of the scale factor, as in Fig.~\ref{fig:eps1}.
{\it Upper right panel:} logarithmic linear power spectrum $\Delta_L^2(k,z)$ at redshift $z=0$,
as in Fig.~\ref{fig:Pk_SUSY}.
{\it Lower left panel:} halo mass function as in the lower panel of Fig.~\ref{fig:halo-mass}.
{\it Lower right panel:} domain in the mass-radius plane where the fifth force is greater than Newtonian gravity, as in Fig.~\ref{fig:eta_R_M}.
}
\label{fig:x}
\end{figure*}

In this section we investigate how the results obtained in the previous sections change when
we vary the parameter $\alpha$.
As a matter of example, we consider the model with $\zeta=1$ and we show our results in
Fig.~\ref{fig:x}, where we compare the case $\alpha=10^{-6}$ considered in the previous
sections with the two cases $\alpha=10^{-7}$ and $\alpha=10^{-8}$.

In agreement with the discussion in section~\ref{sec:cosmo-per}, as $\alpha$ decreases
the maximum amplitude of $\epsilon_1$ decreases as $\epsilon_1(a_{\alpha}) \sim \alpha$
while the associated scale factor decreases as $a_{\alpha} \sim \alpha^{1/3}$.
This implies that the effect of the fifth force is shifted to higher redshift with a lower amplitude,
whence a smaller impact of the scalar field on the matter power spectrum, $P(k,z)$, and
on the halo mass function, as we can check in the upper right and lower left panels
in Fig.~\ref{fig:x}.
The area in the $(M,R)$ plane where the fifth force is greater than Newtonian gravity also
shrinks as $\alpha$ decreases, as we can see in the lower right panel.
This is because $R_{\alpha} \propto \alpha$, which moves the upper branch down towards
small radii, whereas the lower branch slowly moves upward because at fixed mass
we have $R(M) \sim \alpha^{-1/(4\zeta+1)}$.
Therefore, galaxies are no longer sensitive to the modification of gravity if
$\alpha \lesssim 5 \times 10^{-7}$.

\section{Conclusion}
\label{sec:conclusion}

We have considered in this paper supersymmetric chameleon models with
a very large mass, $1/m_{\rm eff} \ll 10^{-4} {\rm mm}$, and coupling
$\beta \gg 1$.
This makes the range of the fifth force very small and leads to an equivalence
between these supersymmetric chameleon models and the ultra-local models
studied in a companion paper, for cosmological scales with $H \ll k/a \ll m_{\rm eff}$.
The background remains very close to the $\Lambda$-CDM cosmology in both sets
of models.
However, in contrast with the more general ultra-local models, in this supersymmetric
context only the dark matter is sensitive to the fifth force. Therefore, although the
ultra-local character of the models gives rise to an automatic screening mechanism
that ensures that we satisfy Solar System tests of gravity in that more general framework,
in the context studied in this paper this mechanism is not so critical as baryons, which
dominate on small scales and in the Solar System, never feel the fifth force (except through
its effects on the dark matter Newtonian potential) and follow General Relativity.

We have first described how to build such chameleon models in this supersymmetric
context. This involves several characteristic functions that enter the K\"ahler
potential $K$, which governs the kinetic terms of the model, the superpotential $W$
responsible for the interactions between the fields, and the coupling between the dark matter
and the dark energy.
This also introduces several energy scales that may be different.
We have shown in details how these models are equivalent to ultra-local models
for cosmological purposes. This leads to a great simplification as the latter involve
a single free function, $\ln A(\tilde\chi)$.
As in most dark energy and modified gravity models, we also need to introduce
a cosmological constant and the associated energy scale.
In addition, we need a small parameter $\alpha \lesssim 10^{-6}$, which however
appears as a ratio of several energy scales. This provides a natural setting to explain
why this quantity can be significantly different from unity.

Next, we have used the ultra-local models identification to study the cosmological
properties of these scenarios.
We have considered both the background dynamics and the evolution of linear
perturbations. Whereas the background remains very close to the $\Lambda$-CDM
evolution, within an accuracy of $10^{-6}$, the growth of cosmological structures
is significantly amplified on scales below $1 h^{-1} {\rm Mpc}$.
This fifth-force effect shows a fast increase at high $k$ as it corresponds
to a pressure-like term in the linearized equations of motion.
Another property that is peculiar to these models, as opposed to most dark energy
or modified gravity models, is that the fifth force is the  greatest at a high redshift
$z_{\alpha} \sim \alpha^{-1/3} \sim 100$ and for galaxies (among cosmological
structures).

We have also considered the modifications to the spherical collapse of cosmological
structures. The faster growth of structures at $z \sim z_{\alpha}$ leads to
an acceleration of the collapse at these early times and to a lower linear density
threshold $\delta_L^{\Lambda \rm-CDM}$ required to reach a non-linear density
contrast of $200$ today, especially on smaller scales where the fifth force is greater.
This leads to a higher halo mass function at intermediate masses,
$10^8 \lesssim M < 10^{14} h^{-1} M_{\odot}$, as compared with the
$\Lambda$-CDM cosmology.
Next, we have considered the behavior of the fifth force inside spherical halos.
We find that the fifth force increasingly dominates at smaller radii in halos with a shallow
density profile, $\gamma_p \lesssim 1$, as for NFW profiles.
On the other hand, the fifth force is negligible on cluster scales and of the same order
as Newtonian gravity on galaxy scales.
This suggests that galaxies could be the best probes of such models.

To investigate the non-linear fifth force regime, and to check that the previous cosmological
analysis is not violated by small-scale non-linear effects, we have used the thermodynamic
analysis developed in the companion paper.
Again, we find that for these supersymmetric chameleon models the cosmological scales
that turn non-linear at high redshift because of the fifth force are at the boundary of the
inhomogeneous domain in the thermodynamic phase diagram. This suggests that they
do not develop strong small-scale inhomogeneities and that the standard mean field
cosmological analysis is valid.
The same behavior is found at small radii in spherical halos, which again suggests that
the spherically averaged analysis applies. However, for low-mass halos,
$M \lesssim 10^{11} h^{-1} M_{\odot}$ at $z=0$, intermediate radii fall within the
inhomogeneous phase.
This could lead to some fragmentation of the system with the formation of intermediate
mass clumps. On the other hand, this same process leads to a self-screening of the
fifth force as isolated clumps no longer interact through the fifth force because of its
ultra-local character.
Finally, we have considered the dependence of our results on the value of the parameter
$\alpha$.
We find that for $\alpha \ll 10^{-7}$ the deviations from the $\Lambda$-CDM cosmology
are likely to be negligible (contrary to the models studied in the companion paper)
because they have a lower amplitude and are pushed to lower scales where baryons
are dominant.

Thus, we find that although such models follow the $\Lambda$-CDM behavior at the
background level they display a non-standard behavior for the dark matter perturbations
on small scales, below $1 h^{-1} {\rm Mpc}$. At the level of the preliminary analysis
presented in this paper they appear to remain globally consistent with observational
constraints. However, the effects of the fifth force deep inside halos, on kpc scales,
may provide strong constraints and rule out this models.
In particular, the thermodynamic analysis presented in this paper may not be sufficient
as the systems may not reach this equilibrium because of incomplete relaxation.
To go beyond the analytic approaches used in this paper and to make an accurate
comparison with data on galaxy scales requires numerical simulations, which we leave
to future work.

\begin{acknowledgments}

This work is supported in part by the French Agence Nationale de la Recherche
under Grant ANR-12-BS05-0002. This project has received funding from the European Union’s Horizon 2020 research and innovation programme under the Marie Skłodowska-Curie grant agreement No 690575.

\end{acknowledgments}

\bibliography{ref1finalsusy}

\end{document}